\documentclass[leqno,a4paper,11pt]{article}
\RequirePackage{amsmath}
\RequirePackage{amssymb}
\RequirePackage{amsthm}
\usepackage[english]{babel}
\usepackage[latin1]{inputenc}
\usepackage{rotating}
\usepackage{enumerate}
\usepackage{ulem} 
\usepackage{float}
 \usepackage{verbatim} 
\usepackage{lscape}
\usepackage{colordvi}
\usepackage{fancyhdr}
\usepackage{times,graphicx,epsfig}
\usepackage{psfrag} 
\usepackage[it,small]{subfigure}
\usepackage{tikz}
\usetikzlibrary{arrows,shapes}
\usepackage{pgflibrarysnakes} 
\usepackage{multirow}
\usepackage{multicol}
\tikzstyle{every picture}+=[remember picture]
\usepackage{appendix}

\DeclareMathAlphabet{\mathpzc}{OT1}{pzc}{m}{it}

\renewcommand{\emph}{\textit}
\newcommand{\aout}[1]{
\tikz[baseline]
 \node [cross out,draw,anchor=text] {$#1$};}

\newcommand{\QC}{quasi-cyclic }
\newcommand{\WV}{weights vector}
\newcommand{\GR}{Gr\"{o}bner}

\newcommand{\PPP}{\prod_{1\leq h\leq \alpha} p_h^1}

\newcommand{\e}{\varepsilon}
\newcommand{\ep}{\epsilon}
\newcommand{\HH}{$ \mathbf{H} $ }
\newcommand{\HHin}{ \mathbf{H}  }
\newcommand{\ZZ}{ $\mathbb{Z} $  }
\newcommand{\ZZin}{ \mathbb{Z}  }
\newcommand{\figur}{Figure}
\newcommand{\etal}{\emph{et al.}}
\newcommand{\GG}{$ \mathbf{G} $ }

\newcommand{\NNin}{\mathbb{N}}
\newcommand{\FF}{\mathbb {F}}
\newcommand{\M}{\mathcal{M}}
\newcommand{\C}{\mathcal{C}}

\newtheorem{theorem}{Theorem}[section]
\newtheorem{proposition}[theorem]{Proposition}
\newtheorem{fact}[theorem]{Fact}
\newtheorem{lemma}[theorem]{Lemma}

\newtheorem{remark}[theorem]{Remark}
\newtheorem{definition}[theorem]{Definition}
\newtheorem{example}[theorem]{Example}

\newcounter{newname} 

\newcounter{CtrLabelii}

\newenvironment{doubleenum}
{\begin{list}{\labelenumi\the\value{CtrLabelii}}{\usecounter{CtrLabelii}}}
{\end{list}}

\title{{\tt Quasi-cyclic LDPC codes with high girth}}

\begin{document}



\def\d{\partial}
\def\a{\alpha}
\def\b{\beta}
\def\g{\gamma}
\def\e{\varepsilon}
\def\L{\lambda}
\def\N{\nu}



\maketitle

\noindent
{\bf Christian Spagnol}
{\tt (christians@rennes.ucc.ie)}
\\
\noindent
Department of Electronic engineering, UCC Cork, Ireland.
\\

\noindent
{\bf Marta Rossi}
{\tt (marta.rossi@posso.dm.unipi.it)}
\\
\noindent
Department of Mathematics and Appl., University of Milan-Bicocca, Italy.
\\

\noindent
{\bf Massimiliano Sala}
{\tt (msala@bcri.ucc.ie)}
\\
\noindent
Department of Mathematics university of Trento, Italy /Boole Centre for Research in Informatics, UCC Cork, Ireland.
\\

\vskip 1cm

\noindent
{\bf Abstract.}
We study a class of quasi-cyclic LDPC codes. We provide precise conditions
guaranteeing high girth in their Tanner graph.
Experimentally, the codes we propose perform no worse than random
LDPC codes with their same parameters, which is a significant
achievement for algebraic codes.
 \\



\section{Introduction}

The LDPC codes are codes that approach optimal decoding performances,
with an acceptable decoding computational cost (\cite{Gallager:63,Hu:EIOTSPAFDLDPCC,MacKay_Neal:95}).
In this paper we present a class of quasi-cyclic LDPC codes and we show
that we are able to guarantee some relevant properties of the codes.
Experimentally, their decoding performance is comparable with
the performance obtained by random LDPC codes.

Traditionally, coding theory is divided into two main research areas:
algebraic coding theory and probabilistic coding theory.
The former (\cite{h24}) deals with codes endowed with a nice algebraic
structure, which allows both to study their internal properties
and to have efficient encoding-decoding techniques, the latter deals
with convolutional codes, which are very difficult
to study and randomly constructed, but having superior decoding performance.
However, the rediscovery of the LDPC codes by MacKay (\cite{MacKay_Neal:95})
triggered a radical change in coding theory: now we have linear block
codes that may reach decoding performance close to the upper bound
given by the Shannon limit (\cite{Chung&Forney&Richardson&Urbanke01}).
Dozens of papers have appeared since MacKay's paper, some of them trying
to endow some structure (either algebraic or geometrical) on LDPC
codes. However this has seldom been successful, because the structure
brings a regularity in the parity-check matrix of the code,
which naturally pushes towards the creation of many dangerous small cycles
in their Tanner graph.
It the object of this paper to propose a family of LDPC codes, possessing
an algebraic structure but not suffering from the performance limitations
common to other similar families.
The family of \QC~LDPC codes are of great interest for the possibility of
exploiting the structure of the parity check matrix to achieve very fast and efficient encoding and decoding.
 Unfortunately the BER/SNR performance of this class
of codes is known to be worst than  random generated LDPC codes in particular for medium/long length.

The object of this contribution is to study the family of \QC~LDPC codes,
and provide precise conditions guaranteeing high girth in the Tanner
graph. 
Although several families of quasi-cyclic LDPC codes have
been proposed, no general study on their girth properties have
been published.
Previous researches presented in  the literature focuses on studies of
the properties of particular classes of \QC~LDPC codes constructed from circulant  matrices
obtained from a monomial.  It is our purpose to fill this gap, by formally
classify all cases when cycles of length less than $10$ may
arise in the general case.
Therefore, in this contribution matrices obtained from polynomial
are also considered.
The study is restricted to polynomials composed of two or less
monomials, the reason for such limitation is the fact that circulant
obtained from polynomial composed by three or more monomials internal
cycles of length h $6$ always exist.
Hence such polynomial are of no interest if codes with higher girth is wanted.

From the  classification obtained, it is obvious how to identify
necessary and sufficient conditions for any quasi-cyclic LDPC
code to have girth at least $10$.
Various  constructions are presented that perform no worse than random LDPC codes with the same parameters, which is a significant achievement for 
algebraic codes. 

The remainder of this paper is structured as follows:
\begin{itemize}
\item Section \ref{sec:prel} provides notations, recall some
      relevant well-known facts and prove some simple preliminary statements;
\item Section \ref{sec:cycles_gen} deals with theclassification of the
cases when    cycles up to length $8$ may arise for a rather general code family;
\item Section \ref{sec:QC_case} improves results from
      Section \ref{sec:cycles_gen} for the generic \QC~case;
\item In section \ref{sec:polysTOcycles} the existence of short cycles
is linked with condition on the polynopmial representation of the
      circulant matrices;
\item In sections~\ref{sec:Bresnan_code},~\ref{sec:extendedBresnan}
      and ~\ref{sec:qc_g10}
      varius subclass of \QC~codes are studied in details and detailed
      conditions to avoid short cycles are given. The performances of
      such codes is compared with other codes,
\item Finnaly in section \ref{sec:concl} comments, conclusions,
      and outline some further research are presented.
\end{itemize}

\section{Preliminaries and notation}\label{sec:prel}

In this section some known facts are recalled, some notation are given and some simple 
statements that will be useful later on are proved.

\subsection{LDPC codes and Tanner graphs}
\label{Tanner}

The parity-check matrix $H=(h_{i,j})$ of any binary $[n,k,d]$ linear code $C$
may be represented by a graph $\Gamma$, known as the Tanner graph
(\cite{TanSriFuj01,Tanner:81}).
The Tanner graph is formed by two types of nodes: the ``bit nodes''
and the ``check nodes''.
Bit nodes correspond to matrix columns and check nodes correspond
to matrix rows, so that there are $r = n - k$ check nodes and $n$ bit nodes.
We connect the check node $i$ to the bit node $j$ if and only if
the entry $h_{i,j}=1$. There is no edges connecting two check nodes or
two bit nodes (this kind of graph is called a {\em bipartite} graph).
In other words, $H$ is the adjacency matrix of $\Gamma$.

\begin{example}
An example of a binary LDPC code and relative Tanner graph can be seen in \figur~\ref{fig:tanner_cycle}

\begin{figure}[t!h]
\begin{minipage}[b]{0.5\linewidth}\centering
\tiny{
 \begin{equation*}\vspace*{-2cm}
        \HHin =\left[
          \begin{array}{cccccccccc}
            1 & 0 & 1 & 0 & 1 & 0 & 0 & 0 & 0 & 0\\
            0 & 1 & 0 & 1 & 0 & 1 & 0 & 0 & 0 & 0\\
            0 & 0 & 0 & 0 & 0 & 0 & 1 & 1 & 1 & 0\\
            0 & 0 & 1 & 0 & 1 & 0 & 0 & 0 & 0 & 1\\
            0 & 0 & 0 & 1 & 0 & 0 & 1 & 0 & 0 & 0\\
            0 & 0 & 0 & 0 & 0 & 0 & 0 & 1 & 1 & 1\\
          \end{array}\right]
\end{equation*}
\vspace*{0.75cm}
}
\end{minipage}
\begin{minipage}[b]{0.5\linewidth}\centering
    \psfrag{C0}{\scriptsize{$c_0$}}    \psfrag{C1}{\scriptsize{$c_1$}}    \psfrag{C2}{\scriptsize{$c_2$}}    \psfrag{C3}{\scriptsize{$c_3$}}
    \psfrag{C4}{\scriptsize{$c_2$}}    \psfrag{C5}{\scriptsize{$c_3$}}    \psfrag{b0}{\scriptsize{$b_0$}}    \psfrag{b1}{\scriptsize{$b_1$}}
    \psfrag{b2}{\scriptsize{$b_2$}}    \psfrag{b3}{\scriptsize{$b_3$}}    \psfrag{b4}{\scriptsize{$b_4$}}    \psfrag{b5}{\scriptsize{$b_5$}}
    \psfrag{b6}{\scriptsize{$b_6$}}    \psfrag{b7}{\scriptsize{$b_7$}}    \psfrag{b8}{\scriptsize{$b_8$}}    \psfrag{b9}{\scriptsize{$b_9$}}
      \psfrag{CheckNodes}{\small{Check Nodes}} \psfrag{BitNodes}{\small{Variable Nodes}}
\includegraphics[width=0.75\textwidth]{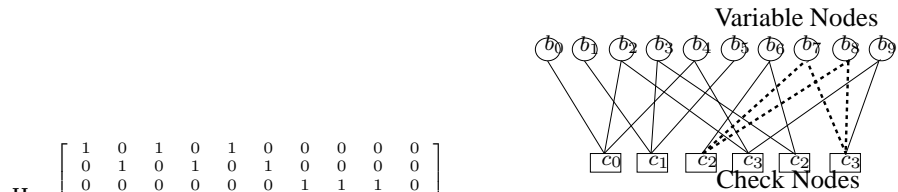}
\end{minipage}
    \caption[Cycle on Tanner Graph]{Parity check matrix H and the associated
      Tanner Graph,  the presence of a cycle is highlighted}\label{fig:tanner_cycle}
    \end{figure}

\end{example}
Now we introduce LDPC codes - Low-Density Parity-Check codes - a
class of linear error correcting codes. Historically, these codes were
discovered by Gallager in 1963 in his PhD thesis \cite{Gallager:63}.
These codes were largely ignored, because of some implementation issues.
In the 1990's they were rediscovered by MacKay \cite{MacKay_Neal:95}
and now the research continues vigorously, with dozens of papers
published every year (\cite{ryan}).
\begin{definition}
An LDPC code is a linear block code for which the parity
check matrix has a low density of non-zero entries.

A $(c,s)$-regular LDPC code is a linear code whose
parity check matrix $H$ contains exactly $c$ ones per column and $s$ ones
per row.
\end{definition}
We do not specify what we mean by low density because it depends on
the context. For example, for a typical $(3,6)$-regular binary code
(rate $1/2$) of block length $n$, there are only three ones in each column of
$H$ and so the fraction of ones in this matrix is $6/n$.

The decoding algorithm for these codes is usually called the ``sum-product
algorithm''. We summarize some properties of these codes:
\begin{itemize}
\item the LDPC codes have excellent decoding performance, near to the channel
  capacity (\cite{Gallager:63,MacKay_Neal:95, Richardson&Shokrollahi&Urbanke01density}),
\item the sum-product algorithm is based on the probabilities received from the
      channel and it may be idealized as a belief-propagation algorithm
      (see \cite{belief}),
      with information being passed and updated by the bit nodes to
      the check nodes and vice-versa, in a loop;
\item the information propagation is heavily hindered by the presence of
      small cycles in the Tanner graph (\cite{tian2});
\item the best LDPC codes are irregular and are created by some random-walk
      optimization algorithm (\cite{Chung&Forney&Richardson&Urbanke01, Richardson&Shokrollahi&Urbanke01density,Richardson&Urbanke01capacity,Richardson_UrbankeMacKay:EFF_Enc01,Tian03});
\item there is no known algebraic class of LDPC codes that has performance
      comparable to the best known LDPC codes.
\end{itemize}
There are two serious issues for a code not possessing an algebraic structure:
\begin{itemize}
\item the encoding process is computationally expensive,
\item it is very difficult to study its properties.
\end{itemize}
There are many family of LDPC codes that have been proposed endowed with
an algebraic (or geometrical) structure, but none of them has clearly shown,
at present, a decoding performance comparable with the random LDPC codes
(see \cite{alg1, Johnson:03,Johnson:CFIDFPG,KIM:04,Fossorier:01,Kou02,OsGrefSmar02,Rosenthal:01}).

\subsection{Girth}\label{subsec:girth}

\begin{definition}
In a graph, a {\bf cycle} is a path that starts from a vertex $v$ and ends
in $v$.
The {\bf girth} of a graph is the smallest of its cycles.
\end{definition}

 Obviously the girth of a bipartite graph is always even.
The girth is considered one of the important parameters of a LDPC code, it is commonly accepted that the presence of short cycles in the graph is one of the main
parameters affecting the coding gain achievable by the LDPC code~\cite{MacKay:GECCBOVSM99}.
The dependency of the performances of a LDPC code on its girth
distribution, in particular  when small cycles have been avoided, is still under debate
since  mathematical proof has not yet been obtained.
 In contrast with
the deteriorating  effect of cycles on the performance of the LDPC
 codes, 
Etzion~\etal~\cite{Etzion98,Etzion99}  proved how cycle-free graphs cannot support good codes.
Still, simulations and applications have shown that the belief
propagation algorithm is generally very effective, even in the presence of cycles in the
graph~\cite{MacKay:96,Chung&Forney&Richardson&Urbanke01}. 
Nevertheless it is commonly accepted that the presence
of short cycles in the graph is one of the main causes of reducing the coding gain achieved
by the LDPC code~\cite{MacKay:GECCBOVSM99}, and so the girth is considered one of the significant parameters of
a code.

For Tanner graphs of $(c,s)$-regular LDPC codes, it is possible
to give upper bounds on the girth, see~\cite{Rosenthal:01}.
\begin{theorem}[]\label{th:gith_bound}
Consider a $(c,s)$-regular $[N, K , d]$ LDPC code.
Let $R = N - K$ and $\alpha=(c-1)(s-1)$.
If the girth $g \equiv  2 \mod( 4)$, then
$             g\leq 4 \log_\alpha r + 2$,
otherwise
$            g\leq 4 \log_\alpha r + 4$.
\end{theorem}
When $r=404$ and $\alpha=(3-1)(6-1)=10$ (as for our simulation),
we get in the worst case
$\quad      g \leq 4 \log_{10} (404)+4 = 14$,
but in practice it is very difficult to find such codes with $g\geq 10$,
so that ensuring $g\geq 8$ is already interesting.

One of the most promising families of LDPC codes with a nice structure
was proposed by Rosenthal and Vontobel (\cite{Rosenthal:01}).
These are based on Tanner graphs built starting from Ramanujan graphs and
hence are guaranteed to have a very high girth.
Unfortunately, their decoding performance have been questioned (\cite{MacKay:WOMARMLDPCC})
and it is not evident how an efficient encoding could be implemented.

There is a family of LDPC codes, which has been proposed by Fossorier
(\cite{Fossorier:04}), which is particularly interesting for us, because
they are quasi-cyclic and so their structure is quite similar to ours.
With Fossorier's codes, it is easy to get a girth as high as $8$ or
$10$.
However, the construction by Fossorier does not provide codes whose Tanner
graph has a girth higher than $12$, as shown by Fossorier himself
in the same paper.

Another interesting construction has been provided in \cite{OsGrefSmar02,Os03}.
They can get very high girth, but there are some open problems, in particular
on how to get an efficient encoding.

To simplify the search for cycle in a Tanner graph
representing a \HH parity check  matrix, a novel and convenient
definition of cycles of length $l_c$ in an arbitrary 
binary matrix is given here. 

\begin{definition}\label{def:linked}
Let $s,N,M$ be natural numbers with $s\geq2$ and $N,M\geq3$, and define $l_c=2s$. Let $B$ be any $N\times M$ matrix over \ZZ$_2$. 
A sub-set $V$ of  $l_c$ entries  of $B$ is called {\bf linked} if the entries lay in $s$ columns and $s$ rows.
\end{definition}
\begin{definition}\label{def:cycle}
Let $B$ be any $N\times M$ matrix over \ZZ$_2$. 
A sub-set $V$ of $l_c$ entries  ($l_c=2s$) of $B$ is called a {\bf $2s$-cycle} if :
\begin{itemize}
\item $V$ is linked, and 
\item for any $r$ such that $ 2\leq  r < s$ there is no  linked sub-set $W \subset  V$ of $2r$ entries. 
\end{itemize}
\end{definition}

\begin{figure}[htbp]
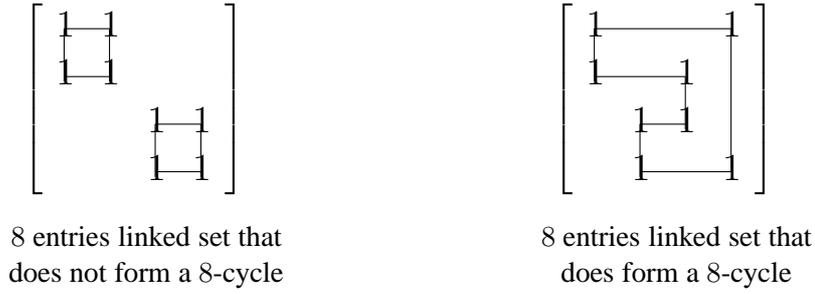
\label{fig:linked}
 \centering 
\caption[Linked sets]{Example of linked sets with $s=4,t=8$}
\end{figure}

In order to clarify the meaning of the definitions two examples are shown
in \figur~\ref{fig:linked}

\unitlength=1mm
\newcommand{\downLine}{
  \begin{picture}(0,0)(0,0)
    \put(1,-6){\tikz\draw[thin] (0mm,0mm) -- (0mm,-7mm);}
\end{picture}}
\newcommand{\downLineLong}{
  \begin{picture}(0,0)(0,0)
    \put(1,-18){\tikz\draw[thin] (0mm,0mm) -- (0mm,-19mm);}
\end{picture}}
\newcommand{\orLine}{
  \begin{picture}(0,0)(0,0)
    \put(1,1){\tikz\draw[thin] (0mm,0mm) -- (6mm,0mm);}
\end{picture}}
\newcommand{\orLineLong}{
  \begin{picture}(0,0)(0,0)
    \put(1,1){\tikz\draw[thin] (0mm,0mm) -- (12mm,0mm);}
\end{picture}}
\newcommand{\orLineVeryLong}{
  \begin{picture}(0,0)(0,0)
    \put(1,1){\tikz\draw[thin] (0mm,0mm) -- (18mm,0mm);}
\end{picture}}

\begin{table}[ht]
\begin{minipage}[b]{0.5\linewidth}\centering
  \begin{equation*}
   \qquad \qquad 
\Large{
    \left[
     \begin{array}{cccc}
       {\downLine}{\orLine}1 & {\downLine}1 &  &   \\
       {\orLine}1 & 1 &  &  \\
       &  & {\downLine}{\orLine}1 & {\downLine}1  \\
       &  & {\orLine}1 & 1 
      \end{array}
    \right]
   }
 \qquad \qquad
  \end{equation*}
$8$ entries linked set that \\does not  form a $8$-cycle
\end{minipage}
\hspace{0.5cm}
\begin{minipage}[b]{0.5\linewidth}
\centering
 \begin{equation*}
  \qquad \qquad      
   \Large{
    \left[
     \begin{array}{cccc}
      {\downLine} {\orLineVeryLong}1 &  &   & {\downLineLong}1\\
      {\orLineLong}1 &   & {\downLine}1 &  \\
      & {\downLine} {\orLine} 1 & 1 &  \\
      & {\orLineLong} 1 &   &1 
     \end{array}
    \right]  
  } 
\qquad\qquad
  \end{equation*}
$8$ entries linked set that \\ does form a $8$-cycle 
\end{minipage}
\end{table}

%
Matrix (a) contains $8$ entries, they form a linked set since
   they lie in $4$ rows and $4$ columns but they do not form a
   $8$-cycle since they can be grouped  in two smaller linked sets of
   $4$ elements each.
Matrix (b) represents  a linked set of entries that form a $8$-cycle,
since the $8$ entries lie in  $4$ rows and $4$ columns but there are no
   smaller linked sets.

Note that if a matrix column contains a point of a $2s$-cycle $V$ then it 
contains exactly two points of $V$. The same is true for the rows.
Moreover a linked sub-set $V$ of $2s$ entries either is a $2s$-cycle or it contains at least a $2r$-cycle  with $2 \leq r < s$.
\subsection{Circulant matrices}\label{subsec:circulants}
Binary circulant matrices are important  as they form
the ``bricks'' with which  parity-check matrices for \QC~LDPC codes are ``built''.
\begin{definition}
Let $m\geq 6$. Let $C$ be an $m\times m$ matrix over $\FF_2$.
$C$ is circulant if its rows are obtained by successive shifts
(to the right). The matrix $C$ is weight-$l$ if the weight of any row
is $l$. 
In case of circulant of weight-$2$, that are used extensively in
this work, the polynomial representation of the first row, $p(x)\in \FF_2[x]$,
is called the {\bf polynomial} of $C$. 
Let $p(x)=x^a+x^b$, with $a<b$.
$s(p)=\min(b-a,a+m-b)$ is called the
{\bf separation} of $p$ (or of $C$).
\end{definition}
Consider, for example, the following weight-$2$ circulant matrix.
\begin{equation}
  C=\left(
  \begin{array}{ccccc}
    1&1&0&0&0 \\
    0&1&1&0&0 \\
    0&0&1&1&0 \\
    0&0&0&1&1 \\
    1&0&0&0&1 \\
  \end{array} \right)
\end{equation}
Its polynomial is $p=x+1$, with parameters $m=5$, $b=1$, $a=0$, $s(p)=1$.

Note, similar matrices can be described as an superimposition of two
permutation matrices used by other authors
(e.g.~\cite{Milenkovic06Proto}), in such case the exponents are
related to the power of the single permutation matrices, and the
analysis presented in this chapter can be rewritten with such notation.

When the natural number $m$ is greater or equal to $3$ the equations:
\begin{equation}
a \equiv b \mod(m),\qquad p(x) \equiv q(x) \mod(x^m+1)\,,
\end{equation}
will be abbreviated with, respectively, :
\begin{equation}
\quad a \equiv b, \qquad p(x) \equiv q(x) \,,
\end{equation}
where the polynomial congruence is in \ZZ$_2[x]$.

Let $p=x^a+x^b$, sometimes some statements where the role of $a$ and
$b$ may be exchanged are needed. To provide a concise formulation in these cases,
the notation $\epsilon(p)$ in congruences modulo $m$ is introduced.
Let $f$ be any function $f:\NNin \mapsto \ZZin$, then
\begin{itemize}
\item $f(\ep(p)) \equiv l$,
      means ``$f(a) \equiv l$ {\em or} $f(b) \equiv l$'',
\item $f(\ep(p)) \not \equiv l$,
      means ``$f(a) \not \equiv l$ {\em and} $f(b) \not \equiv l$''.
\end{itemize}
In the case of a circulant matrix of weight-$1$,  $\epsilon(p)$ refers to the exponent of the monomial.

This notation is extended to the case when more polynomials, $p_1 \ldots p_s$,
are involved in a function, as follow. 
Let $p_1=x^{a_1}+x^{b_1},\,\ldots \,, p_s=x^{a_s}+x^{b_s}$.
Let $f$ be any function $f:\NNin^s \mapsto \ZZin$. Then
\begin{itemize}
\item $f(\ep(p_1),\,\ldots \,,\ep(p_s)) \equiv l$,      means 
  that there is a combination $(z_1,\ldots,z_s)$   where  $z_i\in\{a_i,b_i\}$ such that      $f(z_1,\ldots, z_s) \equiv l$ and
\item $f(\ep(p_1),\ldots ,\ep(p_s)) \not\equiv l$,      means 
  that for every possible combinations $(z_1,\ldots,z_s)$  where $z_i\in\{a_i,b_i\}$ then 
      $f(z_1,\ldots, z_s) \not\equiv l$.
\end{itemize}

\begin{remark}
To avoid ambiguity, statements of kind: 
$       f(\ep(p_1),\ldots \ep(p_r)) \equiv     f(\ep(p_{r+1}),\ldots,\ep(p_s))\quad $or \\
 $      f(\ep(p_1),\ldots \ep(p_r)) \not\equiv f(\ep(p_{r+1}),\ldots,\ep(p_s)) \,.$\\
will never be used.
\end{remark}
Some simple facts on weight-$2$ circulant matrices are collected here.
These follow directly from the circularity of the matrix.
%
\begin{proposition}\label{prop:cas}
Let $C=\{c_{i,j}\}$ be an $(m\times m)$ weight-$2$ circulant matrix
over \ZZ$_2[x]$ with polynomial $p(x)$. Then:
\begin{enumerate}
\item $c_{i,j}=1$ if and only if $j \equiv i + \ep(p)$,
\item $c_{x,y}=c_{t,y}=1$ (with $x\not=t$) if and only if
      $t-x \equiv \pm s(p)$,
\item $c_{x,y}=c_{x,z}=1$ (with $y\not=z$) if and only if
      $z-y \equiv \pm s(p)$,
\item $c_{x,y}=c_{t,y}=1$ and $c_{x,z}=c_{w,z}=1$
      (with $y\not=z$, $x\not=t$, $x\not=w$)
      if and only if  $x-t \equiv \pm s(p)$ and
      $w-x \equiv x-t \equiv \pm s(p)$.
\end{enumerate}
\end{proposition}


%

%
\begin{lemma} \label{lem:sp}
Let $p$ and $q$ be two polynomials in \ZZ$_2[x]$ with
degree at most $m-1$. Then
$$
     s(p)=s(q) \quad \Leftrightarrow \quad s(p) \equiv \pm s(q)
$$
\begin{proof}
Note that $s(p)$ and $s(q)$ are not greater
than $m/2$ and positive.
\end{proof}
\end{lemma}

\begin{definition}
\label{shift}
Let $m\geq 3$. Let $p$ and $q$ be two polynomials in \ZZ$_2[x]$ with
degree at most $m-1$. We say that $p$ is a {\bf shift} of $q$
 if there is $0\leq \mu\leq m-1$ s.t.
$$
   p \, \equiv \, x^\mu q  \,.
$$
In this case we write $p\sim q$.
\end{definition}
Observe that $\sim$ is an equivalence relation in the set formed by
all polynomials over \ZZ$_2$ with degree less than $m$.
\begin{lemma} \label{sp}
Let $p$ and $q$ be two polynomials in \ZZ$_2[x]$ with
degree at most $m-1$. Then
$$
     s(p)=s(q) \quad \Leftrightarrow \quad s(p) \equiv \pm s(q)
$$
\begin{proof}
Note that $s(p)$ and $s(q)$ are not greater
than $m/2$ and positive.
\end{proof}
\end{lemma}

There is a link between the separation of a polynomial and
its roots.
\begin{fact}
\label{roots}
Let $m\geq 3$. Let $p$ and $q$ be two weight-$2$ polynomials in \ZZ$_2[x]$ with
degree at most $m-1$. Then
$$  p\sim q \quad \Leftrightarrow \quad s(p)=s(q)  \,.  $$

Moreover, if $p$ and $q$ are both maximal or if they are both minimal,
then the non-zero roots of $p$ and $q$ are the same
(and with the same multiplicity) if and only if $p\sim q$.
\begin{proof}
Let $p=x^{a_p}+x^{b_p}$ and $q=x^{a_q}+x^{b_q}$, with $a_p < b_p$ and
$a_q < b_q$.

${\bf p\sim q \Rightarrow s(p)=s(q)}$.\\
If $p\sim q$, then    $p \equiv  x^\mu q$, for some
$0\leq \mu \leq m-1$.
That is, $p + x^\mu q = \lambda (x^m+1)$, for some $\lambda\in \ZZin_2[x]$.
But $\deg(p + x^\mu q)\leq 2m-2$, since
$\deg(p)\leq m-1$, $\d q\leq m-1$ and $\mu\leq m-1$.
Also, either $\d \left(\lambda(x^m+1) \right)\geq m$ or $\lambda=0$.\\
There are three cases:
\begin{enumerate}
\item $\deg( x^\mu q)\leq m-1$. \\
      Then $\lambda=0$ and $p=x^\mu q$, which means
      $$
      b_p-a_p \equiv \pm s(p), \quad s(p)= s(x^\mu q) =
      s(x^{a_q+\mu}+x^{b_q+\mu})\,,
      $$
      $$
      s(x^{a_q+\mu}+x^{b_q+\mu})  \equiv \pm \big( (b_q+\mu)-(a_q+\mu))
      \equiv \pm (b_q-a_q) \equiv \pm s(q) \,,
      $$
      so that $s(p) \equiv \pm s(q)$ and hence $s(p)=s(q)$ (Lemma \ref{sp}).

\item $\deg(x^\mu q) \geq m$ and we have $x^\mu q=x^{a_q+\mu}+x^{b_q+\mu}$,
      with $a_q+\mu \leq m-1$ and $b_q+\mu \geq m$.\\
      Then $x^{b_q+\mu}=x^{b_q+\mu-m}(x^m+1)+x^{b_q+\mu-m}$, so that
      $\lambda=x^{b_q+\mu-m}$ and either
      $
      a_p=b_q+\mu-m, b_p=a_q+\mu$, or
      $b_p=b_q+\mu-m, a_p=a_q+\mu$,
      which implies $s(p)\equiv \pm \left((a_q+\mu)-(b_q+\mu-m) \right) \equiv$
      $ \pm (a_q-b_q+m) \equiv$ $\equiv \pm (b_q-a_q) \equiv \pm s(q)$
      and hence $s(p)=s(q)$ (Lemma \ref{sp}).
\item $\deg(x^\mu q) \geq m$ and we have $x^\mu q=x^{a_q+\mu}+x^{b_q+\mu}$,
      with $a_q+\mu \geq m$ and $b_q+\mu \geq m$.\\
      This case is the same as case 1), with the role of $p$ and $q$ exchanged.
      Since $\sim$ is an equivalence relation, we do not have to deal with it.
\end{enumerate}

${\bf s(p)=s(q) \Rightarrow p\sim q}$.\\
If $s(p)=s(q)$, there are four cases:
\begin{itemize}
\item $s(p)=b_p-a_p$ and $s(q)=b_q-a_q$. Then $b_p-a_p=b_q-a_q$. We may
      assume $b_q\geq b_p$, so that $b_q-b_p=a_q-a_p$, i. e.
      $q=x^{b_q-b_p} p$.
\item $s(p)=m-a_p+b_p$ and $s(q)=m-a_q+b_q$. Again $b_p-a_p=b_q-a_q$ and so
      we may argument as before.
\item $s(p)=b_p-a_p$ and $s(q)=m-b_q+a_q$. It is enough to take
      $\mu=b_p-a_q=a_p+m-b_q$:
      $x^{\mu} (x^{a_q}+x^{b_q})=x^{b_p}+x^{a_p+m} \equiv x^{b_p}+x^{a_p}$.
\item $s(p)=m-b_p+a_p$ and $s(q)=b_q-a_q$. Same argument.

\end{itemize}

We now suppose both $p$ and $q$ minimal and we want to show that $p\sim q$
if and only if they have the same non-zero roots (with the same
multiplicity). The case when they are both maximal is analogous and
will not be shown.\\
It is obvious that two polynomials $p$ and $q$ have the same non-zero roots
with the same multiplicity if and only if $p=x^i q$, for some $i$.
Assuming both $p$ and $q$ minimal, we have
$
  p\sim q \quad \Leftrightarrow \quad p=x^\mu q
$,
and so our desired logical equivalence follows.
\end{proof}
\end{fact}

\section{Cycle configurations for generic matrices}\label{sec:cycles_gen}
In this section some notations, facts and lemmas useful to identify which cycles can exist in a given matrix are presented.
A rather general class of matrices is studied.
The general results obtained here will be specialized to the \QC~case
in following subsections.

For the remainder of this chapter, if not differently specified, $m,\alpha,\beta\gamma$ are natural numbers with 
 $m\geq 3$, $\alpha,\beta,\gamma\geq 1$,
\begin{definition}\label{def:bigmatrix}
Given a matrix $B$ over \ZZ$_2$, $B$ is said to be in $ \M_{m,\alpha,\beta,\gamma}$
if it may be written as
$$
B=\, \left[\begin{array}{ccc}
A_{1,1}      & \cdots & A_{1,\alpha\beta}      \\
\vdots       & \vdots &    \vdots              \\
A_{\alpha\gamma,1} & \cdots & A_{\alpha\gamma,\alpha\beta}
\end{array}\right]\,
$$
where the $A_{i,j}$'s are binary square matrices of dimension $m$.
This decomposition is referred to as the {\bf standard decomposition}
of $B$ in $\M_{m,\alpha,\beta,\gamma}$.
Any matrix $A_{i,j}$ is called a {\bf decomposition sub-matrix}(d.s.).
For any $i$ in $\{1,\ldots,\alpha\gamma\}$, that the set
$\{ A_{i,j}\mid 1\leq j\leq\alpha\beta\}$ is called a {\bf decomposition row}(d.r.) of $B$.
Similarly, for any $j$ in $\{1,\ldots,\alpha\beta\}$, the set
$\{ A_{i,j}\mid 1\leq i\leq\alpha\gamma\}$ is called a {\bf decomposition column}(d.c.) of $B$.
\end{definition}
%

The d.s.'s $\{A_{i,j}\}$ are defined unambiguously and
the uniqueness of the standard decomposition in $\M_{m,\alpha,\beta,\gamma}$
 is obvious.
The term  ``standard decomposition'' will be used rather than
``standard decomposition in $\M_{m,\alpha,\beta,\gamma}$''.
If $B\in \M_{m,\alpha,\beta,\gamma}$, $B$ can also be viewed as:
\begin{equation}\label{matrixB}
B= \left[\begin{array}{ccc}
  L_{1,1}      & \cdots & L_{1,\beta}      \\
  \vdots       & \vdots &    \vdots              \\
  L_{\gamma,1} & \cdots & L_{\gamma,\beta}
  \end{array}\right],
\;
L_{r,s}= \left[\begin{array}{ccc}
  A_{(r-1)\alpha+1,(s-1)\alpha+1} & \cdots & A_{(r-1)\alpha+1,s\alpha} \\
  \vdots                          & \vdots &    \vdots                   \\
  A_{r\alpha,(s-1)\alpha+1}       & \cdots & A_{r\alpha,s\alpha}
  \end{array}\right]
,
\end{equation}
where $L_{r,s}$ is a square matrix of sub-matrices, with dimension $\alpha m$,
($1\leq r\leq\gamma$, $1\leq s\leq\beta$).

\begin{remark}
If $H\in \M_{m,\alpha,\beta,\gamma}$ has  full rank and $\beta>\gamma$,
then it represent a binary linear code
with dimension $(\beta-\gamma)m\alpha$ and length $\beta m\alpha$.
The information rate is
$$
   \frac{K}{N}=\frac{(\beta-\gamma)m\alpha}{\beta m\alpha}=
   \frac{\beta-\gamma}{\beta}\,.
$$
If $H\in \M_{m,\alpha,\beta,\gamma}$ is not full rank the rate of the
code is lower and the value presented above can be considered
as \emph{designed rate}.
Note that in some cases adding redundant rows, hence not having full,
can be used as a method to improve performances tanks to the extra
checks that a codeword has to satisfy.
\end{remark}
\noindent
It is clear that any rate can be achieved with this code construction by
choosing $\beta$ and $\gamma$ appropriately.

\begin{definition}\label{def:first}
Let $B\in \M_{m,\alpha,\beta,\gamma}$ and let
$\{A_{i,j}\}_{1\leq i\leq \alpha\gamma, 1\leq j\leq \alpha\beta}$
 form its standard decomposition. Define the  sets of indexes $\mathcal{I}=\{i_1,\ldots,i_r\}\subset\{1,\ldots,\alpha\gamma\}$ and
$\mathcal{J}=\{j_1,\ldots,j_s\}\subset \{1,\ldots,\alpha\beta\}$.
With the notation $B_{\mathcal{I},\mathcal{J}}$ the sub-matrix of $B$ defined by the d.r.'s in $\mathcal{I}$ and
the d.c.'s in $J$ is denoted. The sub-matrix $B_{\mathcal{I},\mathcal{J}}$ is said to be of type $(r,s)$ and 
 $B_{\mathcal{I},\mathcal{J}}$ is called a {\bf decomposition minor}(d.m.).
Given two d.m.'s  $B_{\mathcal{I},\mathcal{J}}$ and $C_{\mathcal{I'},\mathcal{J'}}$,
they are considered \textbf{equivalent} if
it is possible to obtain one from the other by d.r. permutations or
by d.c. permutations or by both.
An equivalence class is called a {\bf configuration} of type $(r,s)$.
\end{definition}

Foe example the following two matrices are decomposition minors of
matrix~$B$ in~\ref{matrixB} and are equivalent configurations since
$D'$ can be obtained from $D$ by switching the first and second rows
and then the first and second columns.
$$
D= \left[\begin{array}{ccc}
  L_{1,2} & L_{1,4} & L_{1,6}      \\
  L_{2,2} & L_{2,4} & L_{2,6}        \\
  L_{4,2} & L_{4,4} & L_{4,6}
  \end{array}\right],
\;
D'= \left[\begin{array}{ccc}
  L_{2,4} & L_{2,2} & L_{2,6}      \\
  L_{1,4} & L_{1,2} & L_{1,6}        \\
  L_{4,4} & L_{4,2} & L_{4,6}
  \end{array}\right],
$$
Note that the relation defined on d.m.'s is actually an equivalence
relation, so that ``an equivalence class'' makes sense.

The following lemma will be useful later on.
\begin{lemma}\label{lem:id}
Let $I$ be the $m\times m$ identity matrix over \ZZ$_2$.
Let $B\in \M_{m,\alpha,\beta,\gamma}$ and let $\{A_{i,j}\}$ form its
standard decomposition. Suppose that there are $i$ and $j$ s.t. $A_{i,j}=I$.
Let $B_{x,y}$ be an entry of $B$ included in $A_{i,j}$.
Then
$$
  B_{x,y}=1 \qquad \Leftrightarrow \qquad x\equiv y
$$
\begin{proof}
It is know that $x=x'+(i-1)m$ and $y=y'+(j-1)m$, with $1\leq x'\leq m$ and
$1\leq y'\leq m$. The pair $(x',y')$ represents the components inside
$A_{i,j}$. But $A_{i,j}$ is the identity, so that
$$
  B_{x,y}=1 \quad \Leftrightarrow \quad x'=y' \quad
                  \Leftrightarrow \quad x\equiv y
$$
\end{proof}
\end{lemma}

Using this notation and the definitions of cycles on a matrix given
previously~\ref{def:cycle}, the following lemmas are obvious.
\begin{lemma}\label{lem:equiv}
Let $B_{I,J}$ and $C_{I',J'}$ be equivalent
d.m.'s of a matrix $B\in \M_{m,\alpha,\beta,\gamma}$.
Then $B_{I,J}$ (strictly) contains a $2s$-cycle if and only if $C_{I,J}$ does.
\end{lemma}
%
%
Lemma~\ref{lem:equiv} allows us to talk about ``configuration of
cycles''(see Def.~\ref{def:first}), meaning equivalent decomposition
minors that contain cycles of  the same type.

All the possible (d.m.) configurations with $2s$-cycles are classified next.
\begin{lemma}\label{lem:class-2s-cycle}
With the notation introduced above, let $B\in \M_{m,\alpha,\beta,\gamma}$.
Then the only configurations of $B$ that may (strictly) contain a $2s$-cycle are
of type\footnote{For
brevity any configuration that is the transpose of another is omitted.}:
\begin{itemize}
\item \label{item:2s-1,1}
$
\begin{array}{cc}
      (1,1) \qquad\qquad & \left|A_{i j}\right|\,,\\
\end{array}
$

\item \label{item:2s-1,2}
$
\begin{array}{cc}
      (1,2)\qquad\qquad & \left|A_{i,j_1} A_{i,j_2} \right|\,,\\
       \vdots\qquad\qquad  & \vdots
\end{array}
$
\item \label{item:2s-1,s}
$
\begin{array}{cc}
      (1,s) \qquad \qquad& \left|A_{i,j_1} \cdots  A_{i,j_s} \right|\,,\\
\end{array}
$
\item \label{item:2s-2,2}
$
\begin{array}{cc}
      (2,2)\qquad\qquad & 
      \left|\begin{array}{cc}
        A_{i_1,j_1} & A_{i_1,j_2} \\
        A_{i_2,j_1} & A_{i_2,j_2} \\
        \end{array}\right|\,,\\
 \vdots\qquad\qquad  & \vdots\\
\end{array}
$
\item \label{item:2s-2,s}
$
\begin{array}{cc}
      (2,s) \qquad\qquad&
      \left|\begin{array}{ccc}
        A_{i_1,j_1} & \cdots & A_{i_1,j_s}\\
        A_{i_2,j_1} & \cdots & A_{i_2,j_s} \\
        \end{array}\right|\,,\\
\vdots\qquad\qquad & \vdots\\
\end{array}
$
\item \label{item:2s-s-1,s-1}
$
\begin{array}{cc}
      (s-1,s-1)\ \ &
      \left|\begin{array}{ccc}
        A_{i_1,j_1} & \cdots & A_{i_1,j_{s-1}}\\
	\vdots&&\vdots\\\
        A_{i_s-1,j_1} & \cdots & A_{i_s-1,j_{s-1}} \\
        \end{array}\right|\,,\\
\end{array}
$
\item \label{item:2s-s-1,s}
$
\begin{array}{cc}

      (s-1,s) &
      \qquad\   \left|\begin{array}{ccc}
        A_{i_1,j_1} & \cdots & A_{i_1,j_s}\\
	\vdots&&\vdots\\
        A_{i_s-1,j_1} & \cdots & A_{i_s-1,j_s} \\
        \end{array}\right|\,,\\
\end{array}
$
\item \label{item:2s-s,s}
$
\begin{array}{cc}
      (s,s)&
      \qquad \qquad\   \left|\begin{array}{ccc}
        A_{i_1,j_1} & \cdots & A_{i_1,j_s}\\
	\vdots&&\vdots\\
        A_{i_s,j_1} & \cdots & A_{i_s,j_s} \\
        \end{array}\right|\,.\\
\end{array}
$
\end{itemize}
\begin{proof}
Since a $2s$-cycle $V$ needs $s$ matrix rows, it needs at most $s$ d.r's, and analogously for the d.c.'s. So configuration $(s,s)$ is the largest that can occur. Since matrix rows (and matrix columns) can be grouped in d.r.'s (d.c.'s), any d.m. of $(s,s)$ is possible. This
 proves the claim.
\end{proof}
\end{lemma}

To any d.m. configuration, one or more cycle configurations may be associated.
To proceed it is necessary to characterize the cycle configurations arising from
the previous lemma.
In order to do so, a convenient notation for a cycle configuration is presented.
Any d.m. configuration in the statement of Lemma~\ref{lem:class-2s-cycle} is a sub-configuration of an $(s,s)$ configuration,
 which can generically be represented as
$$
  \left|\begin{array}{cccc}
        T_{1,1} & \cdots & T_{1,s}\\
	\vdots &&\vdots\\
	T_{s,1} & \cdots& T_{s,s} \\
  \end{array}\right|\,,
$$
where any $T_{i,j}$ is an $A_{h,k}$, for some $h$ and $k$.
A numerical representation for cycles is adopted as follows.
Let $V$ be any $2s$-cycle contained in the $(s,s)$ configuration as above.
For any matrix $T_{i,j}$
With $t_{i,j}\geq 0$ the number of points of $V$
contained in $T_{i,j}$ is denoted.
With d.r. and d.c. permutations, it can be supposes that
\begin{equation}\label{eq:s-cycle-repr}
   t_{1,1}\geq t_{i,j},\, 1\leq i,j \leq s, \quad
   t_{1,2}\geq t_{1,3}\geq \ldots \geq t_{1,s},\, t_{2,1}\geq t_{3,1}\geq \ldots \geq t_{s,1} \,.
\end{equation}
A cycle presentation fulfilling conditions (\ref{eq:s-cycle-repr}) will be
called a {\bf (1)-presentation}.

It is obvious that for a $2s$-cycle  it must be:
\begin{equation}
\sum_{1\leq i,j \leq s}  t_{i,j}=2s \,.
\end{equation}

For example, a $4$-cycle $V$,  a $6$-cycles $\tilde V$, and a $8$-cycle $\bar V$
may be represented as
$$
V= \left|\begin{array}{ccc}
         2 & 2 \\
         0 & 0 \\
  \end{array}\right|,
\qquad
\tilde V=\left|\begin{array}{ccc}
        2 & 2 & 0 \\
        0 & 2 & 0 \\
        0 & 0 & 0 \\
  \end{array}\right|,
\qquad
\bar V= \left|\begin{array}{cccc}
         2 & 2 & 0 &0\\
         0 & 1 & 1 &0\\
         0 & 1 & 1 &0\\
	 0 & 0 & 0 &0\\
  \end{array}\right|,
$$
and all these representations are (1)-presentations.
To clarify the implications of the representation consider the
following cycle configurations:
$$
W= \left|\begin{array}{ccc}
         2 & 1 \\
         0 & 3 \\
  \end{array}\right|,
\qquad
\tilde W=\left|\begin{array}{ccc}
        2 & 0 & 1 \\
        2 & 0 & 1 \\
        0 & 0 & 0 \\
  \end{array}\right|,
\qquad
$$
they do not form valid (1)-representations since in $W$ $t_{1,1}$ is
not the maximum of the elements and in $\tilde W$ the elements of the
first row are not ordered.

Note that cycle $\tilde V$ allows another (1)-presentation that is 
 a column permutation of it:
$$
\tilde V= \left|\begin{array}{ccc}
         2 & 2 & 0\\
         2 & 0 & 0\\
         0 & 0 & 0\\
  \end{array}\right|.
$$

For convenience, in a cycle presentation d.r.'s and d.c.'s containing only zeros will be dropped.
The previous presentations may be written as follows
$$
V= \left|\begin{array}{cc}
         2 & 2 \\
  \end{array}\right|,
\qquad
\tilde V=\left|\begin{array}{ccc}
        2 & 2 \\
        0 & 2 \\
  \end{array}\right|,
\qquad
\bar V= \left|\begin{array}{ccc}
         2 & 2 & 0\\
         0 & 1 & 1\\
         0 & 1 & 1\\
  \end{array}\right| \,.
$$

The notation presented above leaves some ambiguity. In fact, for
example, $t_{1,1}=2$ does not specify whether the two points in
$T_{1,1}$ lie in the same cycle column or cycle row or in neither.

\begin{remark}[Transpose]\label{remark:g8-transpose}
Let $D$ be a d.m. of a matrix $B\in \M_{m,\alpha,\beta,\gamma}$ and $D^T$
its transpose.
It is clear that $D$ contains a $2s$-cycle if and only if
$D^T$ contains a $2s$-cycle. In the general case, it is not possible to obtain $D$ from $D^T$
by d.c. or d.r. operations, so $D$ and $D^T$ are {\em not} necessarily
equivalent according to the definition given.
However, if all cycle configurations associated to a given
configuration are classified, then, automatically, all cycle configurations
for its transpose are obtained
(by transposing all of its cycle configurations).
\end{remark}

Two lemmas are provided next, these are applied in the analysis of
the $2s$-cycle configurations arising in Lemma~\ref{lem:class-2s-cycle}.
\begin{definition} \label{def:r&c-weigth}
The \emph{row weight} of a d.r. in a cycle configuration is defined as the sum of the $t_{i,j}$'s in the d.r, and similarly for the d.c.'s.
 \end{definition}
\begin{lemma}\label{lem:ganzo}
In any cycle configuration, column weights and row weights are even.
\begin{proof}
By transposing, the statement for d.c.'s is true if and only if it is
true for d.r.'s. 
It is here shown for d.r.'s. 
Given a d.r., if the relevant $t_{i,j}$'s sum to an odd number,
then one point in  not in a cycle row with another.
But, by definition, any point in a 2s-cycle shares one cycle row
with one and only one other point in the cycle, hence there cannot be
d.r. with odd weight.
\end{proof}
\end{lemma}

Lemma~\ref{lem:ganzo} will be applied many times.
To simplify the relevant notation, the phrase ``Lemma~\ref{lem:ganzo} r-1'' will be used meaning
``Lemma~\ref{lem:ganzo} applied to the first d.r.''. Similarly with
notations like ``r-2'', ``r-3''.
The same notation is used the columns (e.g. ``c-2'').

\begin{lemma}[Isolation] \label{lem:isolation}
In any d.m. configuration $D$ containing a $2s$-cycle, the following situation cannot occur with $2\leq r <s$:
$$D=
  \left|\begin{array}{cc}
 D_1  & 0\\
  0& D_2 \\
\end{array}\right|
$$
where:\\
\begin{center}
$
\begin{array}{cc}
D_1=
  \left| \begin{array}{cccc}
        T_{1,1} & \cdots & T_{1,r}\\
	\vdots &&\vdots\\
	T_{r,1} & \cdots& T_{r,r} \\
  \end{array}\right|
,&
D_2=
   \left| \begin{array}{cccc}
        T_{r+1,r+1} & \cdots & T_{r+1,s}\\
	\vdots &&\vdots\\
	T_{s,r+1} & \cdots& T_{s,s} \\
\end{array}\right|\\
\end{array}
$
\end{center}

\begin{proof}
It is known, by definition~\ref{def:cycle},  how in any row (and column) of a cycle $V$ there must be exactly two cycle points.
If one cycle point lies in a d.r. of the $D_1$ then also the second cycle point of that d.r. must lie in  $D_1$.
Hence, for every d.r. in $D_1$ there are two cycle points, the same is true for the d.c.'s. 
So in $D_1$  there are $2r$ points that lie in $r$ d.r.'s and $r$ d.c.'s , hence they are {\bf linked},
but $r < s$ and so this contradicts the definition of cycle
given (Definitions~\ref{def:cycle}).
\end{proof}
\end{lemma}

To ease the reading in  situations where lemma~\ref{lem:isolation} is not
satisfy for a certain d.m. $D_1$ phrases of the type ''$D_1$ is
isolated'' are used throughout the chapter.

%
%
%

\subsection{Possible configurations}\label{subsec:possible_conf}
The aim is now to determine \emph{all} the
possible $2s$-cycle configuration (using (1)-presentations) 
that can arise in a  matrix.
Some new terminology and lemmas are introduced first.

\begin{definition}\label{def:weight_vector}
The {\bf weights vector} of a $2s$-cycle configuration is defined as the vector containing all the $t_{ij}$'s in such configuration.
\end{definition}
For example the cycle configurations presented previously:
$$
V= \left|\begin{array}{cc}
         2 & 2 \\
  \end{array}\right|,
\qquad
\tilde V=\left|\begin{array}{ccc}
        2 & 2 \\
        0 & 2 \\
  \end{array}\right|,
\qquad
\bar V= \left|\begin{array}{ccc}
         2 & 2 & 0\\
         0 & 1 & 1\\
         0 & 1 & 1\\
  \end{array}\right| \,.
$$
have weights vectors
$$
\begin{array}{ccc}
[2,2],&[2,2,2],&[2,2,1,1,1,1]. \\
\end{array}
$$

\begin{theorem}\label{th:possible_WV}
The \WV s that can make a $2s$-cycle configuration are all the possible combinations of numbers, that 
sum to $2s$ and that do not contain exactly two odd values.
\begin{proof}
The sum of the elements  in a  \WV~must be $2s$ for the definition.  Hence all the possible vectors that sum to $2s$ are candidate to be 
\WV s of a $2s$-cycle configuration. The vectors that have two odd
weights can be discharge since in such cases it is impossible to have 
even row weight and column weight for each d.r. and d.c.. 
 In fact, if both sub-matrices with odd  weights are in the same
 d.r. the two d.c.'s where they lie have odd weight, and vice-versa.
\end{proof}
\end{theorem}

The following theorem specifies which of these \WV s can form a cycle configuration of chosen dimension.
\begin{theorem}\label{th:r,c_VW}
An $(r,c)$ $2s$-cycle configuration can be obtained only from \WV s that have the following characteristics:
\begin{itemize}
\item they contain \textbf{at least} $r+c-1$ elements,
\item they contain \textbf{no more} than $\min (cr, 2s)$ elements,
\item the value of \textbf{the biggest element is no more than} $2s-2(c-1)$
\end{itemize}
\begin{proof}
Suppose $r\leq c$, for remark~\ref{remark:g8-transpose} the case $r > c$
can be reduce to this by transposing the configuration. 
Any d.r. must have at least one element with $t_{ij}\neq0$ otherwise
it would be a $(r-1,c)$ 
cycle configuration, the same is true for the d.c.'s. 

The value $t_{1,1}$ must not be $0$ for the definition of (1)-presentation, then for
Lemma~\ref{lem:isolation} there must be at least another d.m. with $t_{ij}\neq 0$ 
in the same d.r. or d.c. Suppose, without loss of generality, that $t_{1,2}\neq 0$. 
Applying Lemma~\ref{lem:isolation} to the sub-matrix $[T_{1,1},\,T_{1,2}]$ 
implies that there must be another d.m. with $t_{i,j}\neq0$ in the first
d.r. or in one of the first  two d.c.'s, otherwise
$[T_{1,1},\,T_{1,2}]$  would be isolated. 
Assuming $t_{1,3}\neq 0$ another sub-matrix, $[T_{1,1},\,T_{1,2},\,T_{1,3}]$, is obtained.
Repeating the process until all $t_{1,x}\neq0$,  $r-1$
d.r.'s that must have at least one $t_{i,j}\neq0$ element are left, 
hence a total of at least $c+r-1$ weights are needed. 
Note that the same result will be obtained if instead of ''filling''
the d.c.'s first the  d.r.'s are filled, or any combination. 
It is obvious that any \WV~cannot have more that $2s$ values.
Moreover, the sums of the elements in a \WV~is $2s$ but there cannot be more elements that d.m. hence
the max number of elements in a \WV~is $min(2s,rc)$.
Using the (1)-presentation then $t_{1,1} \geq t_{i,j}$ for every
$i,j$, the remaining d.c. must not be zero and have at least row
weight $2$.
Since the sum of all row weights must be $2s$, the max value of $t_{1,1}$ is  $2s-2(c-1)$.
\end{proof}
\end{theorem}
Using theorems~\ref{th:possible_WV} and~\ref{th:r,c_VW}  it is possible to determine the set
of possible \WV s 
that can be used to form a $(r,s)$ $2s$-cycle configuration.  Obtaining the configurations from the \WV s is 
a mater of placing the weights in such a way that they satisfy Lemma~\ref{lem:ganzo} and Lemma~\ref{lem:isolation}.

Next, an example is presented.
The aim is to find which configurations of type $(4,5)$ can have
$10$-cycles. The process starts by looking for all possible vectors of number that sum to $10$.
$$
\begin{array}{l}
\left[10\right],\\
\aout{\left[9,1\right]},\\
\left[8,2\right],\aout{\left[8,1,1\right]},\\
\aout{\left[7,3\right]},\aout{\left[7,2,1\right]},\left[7,1,1,1\right],\\
\left[6,4\right],\aout{\left[6,3,1\right]},\left[6,2,2\right],\aout{\left[6,2,1,1\right]},\left[6,1,1,1,1\right],\\
\aout{\left[5,5\right]},\aout{\left[5,4,1\right]},\aout{\left[5,3,2\right]},\left[5,3,1,1\right],\aout{\left[5,2,2,1\right]},\left[5,2,1,1,1\right],\left[5,1,1,1,1,1\right],\\
\left[4,4,2\right],\aout{\left[4,4,1,1\right]},\aout{\left[4,3,3\right]}, \aout{\left[4,3,2,1\right]},\left[4,3,1,1,1\right],\left[4,2,2,2\right],\aout{\left[4,2,2,1,1\right]},\\
\qquad\left[4,2,1,1,1,1\right],\left[4,1,1,1,1,1,1\right],\\
\left[3,3,3,1\right],\aout{\left[3,3,2,2\right]},\left[3,3,2,1,1\right],\left[3,3,1,1,1,1\right],\aout{\left[3,2,2,2,1\right]},\left[3,2,2,1,1,1\right],\\
\qquad\left[3,2,1,1,1,1,1\right],\left[3,1,1,1,1,1,1,1\right],\\
\left[2,2,2,2,2\right],\aout{\left[2,2,2,2,1,1\right]},\left[2,2,2,1,1,1,1\right],\left[2,2,1,1,1,1,1,1\right],\\
\qquad\left[2,1,1,1,1,1,1,1,1\right],\left[1,1,1,1,1,1,1,1,1,1\right].
\end{array}
$$
The crossed out vectors are not to be considered since they contain exactly two odd  numbers.
Applying Theorem~\ref{th:r,c_VW} it is possible to eliminate  \WV s that do not have length between $8(=5+4-1)$ 
and $10(=\min(10,20))$ and that do not  have max value less or equal to $2(=10-2(5-1))$.
The remaining candidates are :
$$
\left[2,2,1,1,1,1,1,1\right],\left[2,1,1,1,1,1,1,1,1\right],\left[1,1,1,1,1,1,1,1,1,1\right].
$$
It is now necessary to place the values of the \WV s inside the $(4,5)$ d.m..
It is now proved how the only possible configurations  are :
\begin{equation*}
\begin{array}{cc}
1.\quad
\left|\begin{array}{ccccc}
        2 & 1 & 1 & 0 &0 \\
        0 & 1 & 0 & 1 &0 \\
        0 & 0 & 1 & 0 &1 \\
        0 & 0 & 0 & 1 &1 \\
        \end{array}\right|,
\quad2\quad
\left|\begin{array}{ccccc}
        1 & 1 & 1 & 1 &0 \\
        1 & 0 & 0 & 0 &1 \\
        0 & 1 & 1 & 0 &0 \\
        0 & 0 & 0 & 1 &1 \\
        \end{array}\right|,
\end{array}
\end{equation*}
This can be proved with some easy considerations.
There are twenty $T_{i,j}$ with $ 1\leq i\leq4, 1\leq j\leq 5$ .
It must be $\sum_{i,j} t_{i,j}=10$, and
$t_{1,1} \geq t_{i,j},$ $\forall i,\, \forall j\, s.t.\, i\leq4, j\leq 4$ .
There are five columns and for any of this column the weight must be
even and not zero, (if there is a zero column then it falls in a
smaller configuration), 
the only possibility to have a total sum of ten and column weights even
is that  all columns have weight $2$.
Moreover, there are four rows and for any of these rows the weight must
be even and not zero, (if there is a zero column then it falls in a
smaller configuration),  
the only possibility for the total sum to be ten and row weights even is
to have one row with weight $4$ and the remaining with weight $2$.
\begin{itemize}
\item Considering the \WV~$\left[2,2,1,1,1,1,1,1\right]$.\\
An element $t_{i,j}=2$ cannot  be on a d.r. of weight $2$ because both
the d.c.and d.r. would be completed 
but this cannot be otherwise it would be isolated ( Lemma~\ref{lem:isolation}).
So, both the elements $t_{i,j}=2$ must be in the same row, the only
row with weight~$4$. In this
case such a row has weight~$4$ and 
cannot have other non zero elements in it, but also the two columns
are completed, since they have weight~$2$. 
This situation does not satisfy  Lemma~\ref{lem:isolation}, hence this \WV~does not lead to any cycle configuration of this size.
\item Considering the \WV~$\left[2,1,1,1,1,1,1,1,1\right]$.\\
Element $t_{1,1}=2$ since it is the max.  Lemma~\ref{lem:isolation}
implies that there must be at least another non zero element in row
one, 
but since only weight of value one are present and the row weight must
be even then there must be two one elements in row one. Supposed that
$t_{1,2}=t_{1,3}=1$. Considering now Lemma~\ref{lem:ganzo} applied to the
remaining rows and columns 
configuration~$1$, or a column/row permutation of it, is found.
\item  Considering the \WV~$\left[1,1,1,1,1,1,1,1,1,1\right]$.\\
It can be supposed that the first row has weight $4$ hence it must
 have four elements $t_{1,j}=1$
 that can be  $t_{1,1}=t_{1,2}=t_{1,3}=t_{1,4}=1$. 
Applying  Lemma~\ref{lem:ganzo} to all remaining rows and columns configuration~$2$,
or a column/row permutation of it, is obtained.
\end{itemize} 
And this prove the claim.

\subsection{4-cycle configurations for generic H matrices}\label{subsec:4cycle_gen}\
In this section all the possible configurations that 
can give cycles of length  $4$ in the case of a generic matrix are found and listed.
\begin{theorem}
\label{the:gencase4}
Let $B\in \M_{m,\alpha,\beta,\gamma}$.
The only possible $4$-cycle configuration, in  (1)-presentation,  are
as follows \footnotemark:
\begin{enumerate}
\item \label{item:g4-1,1}
      $(1,1)$, $$\left| 4 \right|,$$
\item \label{item:g4-1,2}
      $(1,2)$, $$\left|2\, 2 \right|,$$
\item \label{item:g4-2,2}
      $(2,2)$,
      $$\left|\begin{array}{cc}
        1 & 1 \\
        1 & 1 \\
        \end{array}\right|.$$
\end{enumerate}
\begin{proof}
Following from Theorem~\ref{th:possible_WV} the only possible \WV s for a 4-cycles are:
$$
\begin{array}{ccc}
[4],&[2,2],&[1,1,1,1]. \\
\end{array}
$$
From them it is straightforward to find the configurations listed in
the statement.
\end{proof}
\end{theorem}

\subsection{6-cycle configurations for generic H matrices}\label{subsec:6cycle_gen}
In this section all the possible configurations that 
can give cycles of length  $6$  in the general case are found and listed. 

%
\begin{theorem}\label{the:gencase6}
Let $B\in \M_{m,\alpha,\beta,\gamma}$.
The only possible $6$-cycle configurations, in  (1)-presentation, are
as follows\footnotemark[\value{footnote}]
\footnotetext{For
brevity any configuration that is the transpose of another is omitted.}:
%
\begin{enumerate}
\item \label{item:g6-1,1}
      $(1,1)$, $$\left| 6 \right|,$$
\item \label{item:g6-1,2}
      $(1,2)$,
      $$\left|\begin{array}{cc}
        4 & 2 \\
       \end{array}\right|,$$
\item \label{item:g6-1,3}
      $(1,3)$,
      $$\left|\begin{array}{ccc}
        2 & 2 & 2 \\
       \end{array}\right|,$$
\item \label{item:g6-2,2}
      $(2,2)$, \\ 
      $$ \begin{array}{cc}
        \setcounter{newname}{1}
        \theenumi.\arabic{newname}
        \left|\begin{array}{cc}
        2 & 2 \\
        0 & 2 \\
        \end{array}\right|,&
        \addtocounter{newname}{1}
        \theenumi.\arabic{newname},
           \left|\begin{array}{cc}
        3 & 1 \\
        1 & 1 \\
        \end{array}\right|,
\end{array}
$$
\item \label{item:g6-2,3}
      $(2,3)$,
      $$\left|\begin{array}{ccc}
        2 & 1 & 1 \\
        0 & 1 & 1 \\
        \end{array}\right|,$$


\item \label{item:g6-3,3}
      $(3,3)$,
      $$\left|\begin{array}{ccc}
        1 & 1 & 0 \\
        1 & 0 & 1 \\
        0 & 1 & 1 \\
        \end{array}\right|.$$
\end{enumerate}

\begin{proof}

Lemma~\ref{lem:class-2s-cycle} proves that these are all the
dimensions that $6$-cycle configurations can have. It is now necessary
to prove that the listed configurations are the \emph{all and only} possible configuration that can have $6$-cycles.
To do so it is necessary to prove that for each dimension all the
possible  configurations associated it are listed in the statement.

Following from Theorem~\ref{th:possible_WV} the only possible \WV s for a 6-cycles are:
$$
\begin{array}{l}
\left[6\right],\\
\aout{\left[5,1\right]},\\
\left[4,2\right],\aout{\left[4,1,1\right]},\\
\aout{\left[3,3\right]},\aout{\left[3,2,1\right]},\left[3,1,1,1\right],\\
\left[2,2,2\right],\aout{\left[2,2,1,1\right]},\left[2,1,1,1,1\right],\\
\left[1,1,1,1,1,1\right],\\
\end{array}
$$

The \WV s are used to simplify the process of determining the valid
configurations.
The study of each case is presented next.
\hrule
\vskip 2mm
\noindent
{\bf Configuration (1,1)}. 
A type $(1,1)$ configuration can be generated only by \WV~$\left[6\right]$ and it corresponds to case~\ref{item:g6-1,1}.\\
\hrule
\vskip 2mm
\noindent
{\bf Configuration  (1,2)}. 
A type $(1,2)$ configuration can be generated only by \WV~
$\left[4,2\right]$ and it corresponds 
to  case~\ref{item:g6-1,2}.\\
\hrule
\vskip 2mm
\noindent
{\bf Configuration  (1,3)}. 
A type $(1,3)$ configuration can be generated only by  \WV~
$\left[2,2,2\right]$ and it 
corresponds to case~\ref{item:g6-1,3}.
\hrule
\vskip 2mm
\noindent
{\bf Configuration (2,2)}. 
Applying Theorem~\ref{th:r,c_VW}, configurations of type $(2,2)$ must have one of the following \WV s:
$$
\begin{array}{l}
\left[2,2,2\right],\left[3,1,1,1\right].\\
\end{array}
$$
Applying Lemma~\ref{lem:ganzo} it is straightforward that these
 two \WV s result
 in configurations~\ref{item:g6-1,3}.1 and~\ref{item:g6-1,3}.2.\\
\hrule
\vskip 2mm
\noindent
{\bf Configuration (2,3)}. 
Applying Theorem~\ref{th:r,c_VW}, configurations of type $(2,3)$ must have one of  the following \WV s:
$$
\begin{array}{l}
\left[2,1,1,1,1\right],\left[1,1,1,1,1,1\right].\\
\end{array}
$$
\begin{itemize}
\item $\left[1,1,1,1,1,1\right]$ is not a possible choice because it
would require to have an odd (3) row weight since there are only two d.r.'s
\item $\left[2,1,1,1,1\right]$ results in  configuration~\ref{item:g6-2,3}.
This can be proved with some short considerations. The value $t_{1,1}=2$ since
it is the max value and  $t_{2,1}=0$ because otherwise  column weight
of c-1 would be odd. It follows that all the other must be $1$ and
this proves the claim.
\end{itemize}
\hrule
\vskip 2mm
\noindent
{\bf Configuration (3,3)}. 
Applying Theorem~\ref{th:r,c_VW}, configurations of type $(3,3)$ must have one of the following \WV s:
$$
\begin{array}{l}
\left[2,1,1,1,1\right],\left[1,1,1,1,1,1\right]\\
\end{array}
$$
\begin{itemize}
\item $\left[2,1,1,1,1\right]$ is not a possible choice. In fact,
there are three d.r.'s and for any of this d.r.'s  the row weight
must be even and not zero 
(if there is a zero row then it fall in a smaller configuration). The
sum of all d.r.'s must be six, hence 
all d.r's must have row weight 2. The same is true for the d.c.'s.  If
any  $t_{i,j}=2$ then that element 
is isolated (Lemma~\ref{lem:isolation}) but this is not allowed hence this \WV~cannot generate valid configurations.
\item $\left[1,1,1,1,1,1\right]$ results in configuration~\ref{item:g6-3,3}.
To prove this it is sufficient to consider that as discussed in the
previous configuration  each d.r. and d.c. mush have weight two.
Any possible way to put two 1-elements in each d.r. and d.c.,
satisfying Lemma~\ref{lem:isolation}, 
results in configuration~\ref{item:g6-3,3} or in a column/row permutation of it.
\end{itemize}
Thanks to Remark~\ref{remark:g8-transpose}, it is not necessary to prove the transposed configurations
\end{proof}
\end{theorem}

\subsection{8-cycle configurations for generic \HH matrix}\label{subsec:8cycle_gen}
In this section all the possible configurations that 
can give cycles of length  $8$ in the case of general matrices are found and listed.  
\begin{theorem} \label{the:gencase8}
Let $B\in \M_{m,\alpha,\beta,\gamma}$.
The only possible $8$-cycle configurations, in (1)-presentations,  are as follows\footnote{For brevity any configuration that is the
transpose of another is omitted.}:
\begin{enumerate}
\item \label{item:g8-1.1} 
      $(1,1)$, $$\left| 8 \right|,$$
\item  \label{item:g8-1.2}    $(1,2)$,\\
 $$
 \begin{array}{cc}
        \setcounter{newname}{1}
        \theenumi.\arabic{newname} \label{item:g8-1.2.1}
      	\left|\begin{array}{cc}
        6 & 2 \\
       \end{array}\right|,&
        \addtocounter{newname}{1}
        \theenumi.\arabic{newname}
      \label{item:g8-1.2.2}
      	\left|\begin{array}{cc}
        4 & 4 \\
       \end{array}\right|,
    \end{array}
$$
\item\label{item:g8-1.3}
      $(1,3)$,
      $$\left|\begin{array}{ccc}
        4 & 2 & 2 \\
       \end{array}\right|,$$

\item\label{item:g8-1.4}
      $(1,4)$,
      $$\left|\begin{array}{cccc}
        2 & 2 & 2 & 2\\
       \end{array}\right|,$$

\item \label{item:g8-2.2}
      $(2,2)$,\\
   $$ \begin{array}{ccc}
        \setcounter{newname}{1}
        \theenumi.\arabic{newname}

	\label{item:g8-2.2.1}
      	\left|\begin{array}{cc}
        5 & 1 \\
        1 & 1 \\
        \end{array}\right|,&
        \addtocounter{newname}{1}
        \theenumi.\arabic{newname}

	\label{item:g8-2.2.2}
      	\left|\begin{array}{cc}
        4 & 2 \\
        2 & 0 \\
        \end{array}\right|,&
        \addtocounter{newname}{1}
        \theenumi.\arabic{newname}

	\label{item:g8-2.2.3}
      	\left|\begin{array}{cc}
        4 & 2 \\
        0 & 2 \\
        \end{array}\right|,\\ \\
        \addtocounter{newname}{1}
        \theenumi.\arabic{newname}
	\label{item:g8-2.2.4}
      	\left|\begin{array}{cc}
        3 & 1 \\
        3 & 1 \\
        \end{array}\right|,&
        \addtocounter{newname}{1}
        \theenumi.\arabic{newname}
	\label{item:g8-2.2.5}
      	\left|\begin{array}{cc}
        3 & 1 \\
        1 & 3 \\
        \end{array}\right|,&
        \addtocounter{newname}{1}
        \theenumi.\arabic{newname}
	\label{item:g8-2.2.6}
      	\left|\begin{array}{cc}
        2 & 2 \\
        2 & 2 \\
        \end{array}\right|,

	\end{array}
$$

\item \label{item:g8-2.3}
      $(2,3)$,\\
   $$ \begin{array}{ccc}
        \setcounter{newname}{1}
        \theenumi.\arabic{newname}
	\label{item:g8-2.3.1}
      	\left|\begin{array}{ccc}
        4 & 1 & 1 \\
        0 & 1 & 1 \\
        \end{array}\right|,&
        \addtocounter{newname}{1}
        \theenumi.\arabic{newname}
	\label{item:g8-2.3.2}
      	\left|\begin{array}{ccc}
        3 & 2 & 1 \\
        1 & 0 & 1 \\
        \end{array}\right|,&
        \addtocounter{newname}{1}
        \theenumi.\arabic{newname}
	\label{item:g8-2.3.3}
      	\left|\begin{array}{ccc}
        3 & 1 & 0 \\
        1 & 1 & 2 \\
        \end{array}\right|,\\ \\
        \addtocounter{newname}{1}
        \theenumi.\arabic{newname}
	\label{item:g8-2.3.4}
      	\left|\begin{array}{ccc}
        2 & 2 & 2 \\
        2 & 0 & 0 \\
        \end{array}\right|,&
        \addtocounter{newname}{1}
        \theenumi.\arabic{newname}
	\label{item:g8-2.3.5}
      	\left|\begin{array}{ccc}
        2 & 1 & 1 \\
        2 & 1 & 1 \\
        \end{array}\right|,&
        \addtocounter{newname}{1}
        \theenumi.\arabic{newname}
	\label{item:g8-2.3.6}
      	\left|\begin{array}{ccc}
        2 & 2 & 0 \\
        2 & 0 & 2 \\
        \end{array}\right|,
	\end{array}
$$
\item\label{item:g8-2.4}
      $(2,4)$,\\
   $$ 
   \begin{array}{ccc}
        \setcounter{newname}{1}
        \theenumi.\arabic{newname}
	\label{item:g8-2.4.1}
      	\left|\begin{array}{cccc}
        2 & 2 & 1 & 1\\
        0 & 0 & 1 & 1\\
        \end{array}\right|,&
        \addtocounter{newname}{1}
        \theenumi.\arabic{newname}
	\label{item:g8-2.4.2}
      	\left|\begin{array}{cccc}
        2 & 1 & 1 & 0\\
        0 & 1 & 1 & 2\\
        \end{array}\right|,&
        \addtocounter{newname}{1}
        \theenumi.\arabic{newname}

	\label{item:g8-2.4.3}
      	\left|\begin{array}{cccc}
        1 & 1 & 1 & 1\\
        1 & 1 & 1 & 1\\
        \end{array}\right|,
	\end{array}
$$

\item\label{item:g8-3.3}
      $(3,3)$,\\
   $$ \begin{array}{cccc}
        \setcounter{newname}{1}
        \theenumi.\arabic{newname}
	\label{item:g8-3.3.1}
      	\left|\begin{array}{ccc}
        3 & 1 & 0 \\
        1 & 0 & 1 \\
        0 & 1 & 1 \\
        \end{array}\right|,&
        \addtocounter{newname}{1}
        \theenumi.\arabic{newname}
	\label{item:g8-3.3.2}
      	\left|\begin{array}{ccc}
        2 & 1 & 1 \\
        2 & 0 & 0 \\
        0 & 1 & 1 \\
        \end{array}\right|,&
        \addtocounter{newname}{1}
        \theenumi.\arabic{newname}
	\label{item:g8-3,3.3}
      	\left|\begin{array}{ccc}
        2 & 1 & 1 \\
        1 & 1 & 0 \\
        1 & 0 & 1 \\
        \end{array}\right|,&
        \addtocounter{newname}{1}
        \theenumi.\arabic{newname}
	\label{item:g8-3.3.4}
      	\left|\begin{array}{ccc}
        2 & 0 & 0 \\
        1 & 1 & 0 \\
        1 & 1 & 2 \\
        \end{array}\right|.
	\end{array}
$$
\item\label{item:g8-3.4}
$(3,4)$,\\
   $$ \begin{array}{cc}
        \setcounter{newname}{1}
        \theenumi.\arabic{newname}
	\label{item:g8-3.4.1}
        \left|\begin{array}{cccc}
        2 & 1 & 1 & 0\\
        0 & 1 & 0 & 1\\
        0 & 0 & 1 & 1\\
        \end{array}\right|,&
        \addtocounter{newname}{1}
        \theenumi.\arabic{newname},
        \label{item:g8-3.4.2}
        \left|\begin{array}{cccc}
        1 & 1 & 1 & 1\\
        1 & 1 & 0 & 0\\
        0 & 0 & 1 & 1\\
        \end{array}\right|,
        \end{array}
$$
\item\label{item:g8-4.4}
      $(4,4)$,
     $$\left|\begin{array}{cccc}
        1 & 1 & 0 & 0\\
        1 & 0 & 1 & 0\\
        0 & 1 & 0 & 1\\
        0 & 0 & 1 & 1\\
        \end{array}\right|.$$

\end{enumerate}
\begin{proof}

The proof of this theorem is long and similar to the proof of
theorems~\ref{the:gencase4} and~\ref{the:gencase6}, with the addition
of some logical reasoning.
To improve readability it is here omitted, an interested reader can
find the full and detailed process in Appendix.

\end{proof}	

\end{theorem}

The previous three theorems list all the configurations that can
contain cycles of length less than ten. In particular such
configurations identify how many cycle points  lie in each sub-matrix
(d.m.).
To improve such result it is necessary to endow the
d.m. with some structure that allows to predict the positions of such
point. The next section considers the case where the d.m.'s are circulant matrices.

\section {The \QC~ case}
\label{sec:QC_case}
This section restricts the discussion to matrices which can be used as
parity-check matrices for LDPC \QC~codes, and gives some generic
definitions and lemmas.

A quasi-cyclic code of index $t$ is a linear block code C in which a cyclic shift of any codeword
in C by $t$ positions is also a codeword. The generator matrix \GG for these codes is a matrix
where every row is a $t$ circular shift of the previous row. It can be shown that every generator
matrix for a quasi-cyclic code can be decomposed in circulant matrices.
This property makes the encoding suitable for low cost hardware implementation. 
Major results on a Array codes, a class of \QC codes, have been presented by Fan~\cite{Fan00}
and later by Fossorier~\cite{Fossorier:04}. Array codes are composed by $J$ rows of $L$ circulant matrices.
The circulant matrices used was limited to have weight-1.
The condition given by Fan~\cite{Fan00} (later re-proposed in Theorem 2.1 in~\cite{Fossorier:04}) gives the conditions for the
existence of cycles in such class and is rewritten here using the notations presented previously.
We remind the reader that with our notations $\epsilon(p)$ is any of
the exponents of the polynomial $p$ and $s(p)$ is the separation of the
two monomials.

\begin{theorem}[Fan-2000]\label{th:fossorier}
A necessary and sufficient condition for the matrix \HH to have a $2s$-cycle is:
\begin{equation*}
\sum_{k=0}^{s-1} \bigl[ \epsilon(p_{1,k})-\epsilon(p_{2,k})\bigr] = 0 \mbox{ mod p}
\end{equation*}
were $\epsilon(p_{1,k})$ and $\epsilon(p_{2,k})$ are the exponents of the circulant matrices that contain the two cycle-points that lie in the same cycle-column $k$.
\end{theorem}
\begin{remark}
The theorem was given in the case of a particular case of \QC~codes
with only weight 1 circulants but the same theorem holds in the case when weight-2 circulant matrices $C$ are present.
The presence of  weight-2 circulant matrices allows cycle columns to lie in the same circulant.
In such case the difference $\epsilon(p_{1,k})-\epsilon(p_{2,k})$ is
equivalent to the $s(p_k)$ of such circulant.
Previous work considering weight-2 circulant matrices was presented by
Smarandache and Vontobel~\etal~\cite{Smarandache04BinomialQC} but focuses on the problem of minimum distance and
not girth properties.
\end{remark}
However, the construction by Fossorier cannot provide codes whose Tanner
graph has a girth higher than $12$, as shown by Fossorier himself
in the same paper and known from Fan.
To overtake such limitations and to complete the study 
of the cycles in \QC~codes 
 a wider class of \QC~LDPC codes are considered in this
thesis.
\begin{definition}
\label{def:QC_general} Let $B\in \M_{m,\alpha,\beta,\gamma}$ and let $\{A_{i,j}\}$
form its standard decomposition. The matrix $B$ is said to be in
$\C_{m,\alpha,\beta,\gamma}$ if any d.s. $A_{i,j}$ 
can only be either a
weight-$2$ circulant matrix, a weight-$1$ circulant matrix or an $m\times m$ zero
matrix.
\end{definition}
\noindent It is well-known that any \QC~code has a parity-check matrix \HH made of circulant
sub-matrices. However, if an LDPC code with good girth characteristics
is desired, it is necessary to avoid
sub-matrices that are weight-$t$ circulant matrix, with $t \geq 3$, since they contain internal
6-cycles~\cite{BondHuiSch01} and codes with higher girth are wanted. Matrices in $\C_{m,\alpha,\beta,\gamma}$ are the only
interesting parity-check matrices for good \QC~LDPC codes.
This family of \HH matrices is the focus of the study presented here.

In the case of \QC~LDPC codes a cycle configuration $\sf{c}$ in $\C_{m,\alpha,\beta,\gamma}$ will be
described as:
\begin{equation} \label{eq:co}
\sf{c}=\,
\left|\begin{array}{ccc}
        C-2 & J-1 & C-1 \\
        O   & J-1 & J-1 \\
        \end{array}\right| \,,
\end{equation}
where  $C-2$ is a weight-$2$ circulant matrix containing two points of the
cycle, $O$ is a zero matrix, $J-1$ is a weight-$1$  matrix containing one point of the
cycle, $C-1$ is a weight-$2$ circulant matrix containing one point,
and so on. 
Clearly, notations of the type $O-i$ are not used because it is obvious that
in a zero matrix there must be $i=0$.
In a  situation like (\ref{eq:co}) it will be written that $\sf{c}$ ``contains'' a matrix (d.s.) $C-2$,
three matrices $J-1$ and a matrix $C-1$. When no confusion can arise,
even looser notation will be adopted.
 For example, in (\ref{eq:co}) it may be said that $\sf{c}$ contains $C-2$,
that $\sf{c}$ contains $C-1$ and that $\sf{c}$ contains $J-1$.
Similarly, phrases of the type  ``$\sf{c}$ contains d.m.'s, d.s.'s,d.r.'s, d.c.'s'' will be used with the obvious meaning. 
For example, in (\ref{eq:co}) $\sf{c}$ contains the following d.m.'s
$$
\left|\begin{array}{ccc}
        C-2 & J-1 & C-1 \\
        \end{array}\right| \,,
\left|\begin{array}{ccc}
        O   & J-1 & J-1 \\
        \end{array}\right| \,,
\left|\begin{array}{ccc}
        C-2 & J-1 \\
        O   & J-1 \\
        \end{array}\right| \,.
$$
It is sometimes convenient to gather together cycle configurations possessing
a given number of cycle points.
To be more precise, the notation $\Delta-i$
means that $\Delta$ can be any allowable matrix containing $i$
points.
In other words, the notation
$$
\left|\begin{array}{ccc}
        C-2 & \Delta-2
        \end{array}\right| \,,
$$
is equivalent to the union of
$$
  \left|\begin{array}{ccc}
        C-2 & C-2
        \end{array}\right| \,
  \qquad \mbox{and} \qquad
  \left|\begin{array}{ccc}
        C-2 & J-2
  \end{array}\right| \,.
$$

Let $\sf{c}$ be a cycle configuration, the following easy lemmas can be used 
to show that $\sf{c}$ cannot exist in $\C_{m,\alpha,\beta,\gamma}$.
The expression ``to discard configuration $\sf{c}$'' is synonymous
with ``to show that configuration $\sf{c}$
cannot exist in $\C_{m,\alpha,\beta,\gamma}$''.

\begin{lemma} \label{lem:lemmino1}
Suppose $\sf{c}$ contains $J-i$, with $i\geq 2$. Then
\begin{enumerate}
\item $J-i$ can contain neither a cycle column nor a cycle row,
\item in $\sf{c}$, any d.r. and any d.c. containing $J-i$
      ($i\geq 1$) must contain another non-zero matrix,
\item $\sf{c}$ can be discarded if it contains only one d.r. or
      only one d.c. .
\end{enumerate}
\begin{proof}
Point one is obvious, because a $J$ is a shift of an identity matrix
hence there is only one non zero entry for each row and column.
Point two follows from point one and the fact that the cycle point
contained in the $J$ must be part of a linked set.
Part three follows from part two.
\end{proof}
\end{lemma}

\begin{lemma}\label{lem:lemmino5} 
If $\sf{c}$ correspond to a 2s-cycle with $2s\leq 8$, it  cannot
contain a weight-$1$ circulant matrix $J-i$ with $i\geq 3$.
\begin{proof}
Otherwise any cycle point in $J$ will need another point in the same
row and one in the same column, to form a linked set. For lemma~\ref{lem:lemmino1}
this points cannot be in the $J$ but this implies the existence of at
least nine points in the linked set, but this defies the definitions
of 2s-cycle with $2s\leq 8$
\end{proof}
\end{lemma}

The following two lemmas are obvious and are reported here only for clarity.
\begin{lemma}\label{lem:lemmino2}
Suppose $\sf{c}$ has a $(1,2)$ d.m. composed of two weight-$1$ circulant matrices\\ $|J-i\,J-1|$, $i\geq 1$.
Then in that d.m. there cannot be more than one cycle row. Similarly for a $(2,1)$ d.m.
in $\sf{c}$.
\end{lemma}
\begin{lemma}\label{lem:lemmino3} 
Suppose that  $\sf{c}$ contains a matrix $J-i$,
$i\geq 1$. 
If  $C-i$ is substituted to $J-i$, another possible cycle
configuration is obtained.
\end{lemma}
Passing from a  cycle configurations obtained with the procedure
outlined in the previous section to the respective configurations in the \QC~case
require the application of lemma~\ref{lem:lemmino1}, lemma~\ref{lem:lemmino2},
lemma~\ref{lem:lemmino3} 
and lemma~\ref{lem:lemmino5} to the original configuration.
For example, configuration:
$$\left|\begin{array}{ccccc}
        2 & 1 & 1 & 0 &0 \\
        0 & 1 & 0 & 1 &0 \\
        0 & 0 & 1 & 0 &1 \\
        0 & 0 & 0 & 1 &1 \\
        \end{array}\right|,
$$
 results in a cycle configuration for \QC~codes \\
$$\left|\begin{array}{ccccc}
        C-2 & \Delta-1 & \Delta-1 & 0 &0 \\
        0 & \Delta-1 & 0 & 1 &0 \\
        0 & 0 & \Delta-1 & 0 &\Delta-1 \\
        0 & 0 & 0 & \Delta-1 &\Delta-1 \\
        \end{array}\right|.
$$
In fact cycle configuration:\\
$$\left|\begin{array}{ccccc}
        J-2 & \Delta-1 & \Delta-1 & 0 &0 \\
        0 & \Delta-1 & 0 & 1 &0 \\
        0 & 0 & \Delta-1 & 0 &\Delta-1 \\
        0 & 0 & 0 & \Delta-1 &\Delta-1 \\
        \end{array}\right|,
$$
can be discard because in $J-2$ there is no cycle column (lemma~\ref{lem:lemmino1}).

In the following subsections 
the theorems presented previously (\ref{the:gencase4},~\ref{the:gencase6},~\ref{the:gencase8}) are
specialized for the \QC~case. All the cycle configurations
that can appear in such case are listed. For every circulant in a 
configuration the possible weights that such a circulant can 
have is discussed.

\subsection{$4$-cycle configurations for \QC~H matrices}\label{subsec:4cycle_QC}
\begin{theorem}\label{the:QC_g4} Let be $M \in \C_{m,\alpha,\beta,\gamma}$. The configurations in $M$
that may contain  cycles of length $4$,  are the following: 
\footnote{For
brevity any configuration that is the transpose of another is omitted.}
\begin{enumerate}
    \item \label{item:QC4-1,1}$$
            \left| C-4 \right|,
          $$
    \item \label{item:QC4-1,2}$$
            \left| C-2 \quad C-2 \right|,
          $$
    \item\label{item:QC4-2,1} $$
        \left|\begin{array}{cc}
            \Delta-1 & \Delta-1\\
            \Delta-1 & \Delta-1 \\
           \end{array}\right|  \,.
         $$
\end{enumerate}
\begin{proof}

All the configurations, for the generic case, that appeared in
theorem~\ref{the:gencase4}.
For each configuration it is proved how only configurations listed
in the theorem are valid in the  case of a \QC~LDPC code.
%
%
\vskip 3mm
\hrule
\vskip 2mm
\noindent
{\bf Configuration~\ref{item:g4-1,1}} gives \\
$$
\left| C-4 \right| \,.
$$
Since cycle configuration $|J-4|$ may be discard (Lemma~\ref{lem:lemmino5}).
\vskip 3mm
\hrule
\noindent
{\bf Configuration~\ref{item:g4-1,2}} gives \\
$$
\left|\begin{array}{cc}
        C-2 & C-2 \\
       \end{array}\right| \,.
$$

Cycle configurations $|C-2\, J-2|$ and $|J-2\, J-2|$ may be discard
because in $J-2$ there is no cycle column (Lemma~\ref{lem:lemmino1}-1).

\vskip 2mm
\hrule
\noindent
{\bf Configuration~\ref{item:g4-2,2}} gives \\
$$
\left|\begin{array}{cc}
        \Delta-1 & \Delta-1\\
        \Delta-1 & \Delta-1 \\
       \end{array}\right|  \,.
$$
Note how, thanks to Lemma~\ref{lem:lemmino3}, it is necessary only to
show that the following configuration is possible:
$$
\left|\begin{array}{cc}
        J-1 & J-1\\
        J-1 & J-1 \\
       \end{array}\right| \,.
$$
and this is obvious.
For Remark~\ref{remark:g8-transpose} it is not necessary to study the cycle configurations that are transposed of the one considered.
Hence it has been proved that the listed configurations are the only valid ones.
\end{proof}
\end{theorem}
To better explain the meaning of $\Delta$ and what it implies 
 the (non-equivalent)
cycle configurations present in the statement of
Theorem~\ref{the:QC_g4} case~\ref{item:QC4-2,1} , are reported:
$$
\left|\begin{array}{cc}
        J-1 & J-1\\
        J-1 & J-1 \\
       \end{array}\right| \,,
\left|\begin{array}{cc}
        C-1 & J-1\\
        J-1 & J-1 \\
       \end{array}\right| \,,
\left|\begin{array}{cc}
        C-1 & C-1\\
        J-1 & J-1 \\
       \end{array}\right| \,,
$$
$$
\left|\begin{array}{cc}
        C-1 & J-1\\
        C-1 & J-1 \\
       \end{array}\right| \,,
\left|\begin{array}{cc}
        C-1 & C-1\\
        C-1 & J-1 \\
       \end{array}\right| \,,
\left|\begin{array}{cc}
        C-1 & C-1\\
        C-1 & C-1 \\
       \end{array}\right| \,.
$$

\subsection{$6$-cycle configurations for \QC~H matrices}\label{subsec:6cycle_QC}
\begin{theorem}\label{the:QC_g6} Let be $M \in \C_{m,\alpha,\beta,\gamma}$. The configurations in $M$
that may contain a cycle of length  $6$,  are the
following \footnote{For
brevity any cycle configuration that is the transpose of another is omitted.}

\begin{enumerate}
    \item \label{item:QC6-1,1}$$
            \left| C-6 \right|,
          $$
    \item \label{item:QC6-1,2}$$
            \left| C-4 \quad C-2 \right|,
          $$
    \item \label{item:QC6-1,3}$$
            \left| C-2 \quad C-2 \quad C-2  \right| \,.
          $$
    \item \label{item:QC6-2,2a}$$
        \left|\begin{array}{cc}
        C-2 & \Delta-2 \\
         O  & C-2 \\
       \end{array}\right|\,,
$$
    \item \label{item:QC6-2,2b}$$
    \left|\begin{array}{cc}
        C-3 & J-1\\
        J-1 & J-1 \\
       \end{array}\right|
    $$
\item\label{item:QC6-2,3} $$
    \left|\begin{array}{ccc}
        C-2 & \Delta-1 & \Delta-1 \\
         O  & \Delta-1    & \Delta-1 \\
       \end{array}\right|,
    $$
    \item \label{item:QC6-3,3}$$
    \left|\begin{array}{ccc}
        \Delta-1 & \Delta-1 &    O     \\
        \Delta-1 &    O     & \Delta-1 \\
            O    & \Delta-1 & \Delta-1 \\
       \end{array}\right|\,.
    $$
\end{enumerate}

\begin{proof}

All the configurations that appear  in Theorem~\ref{the:gencase6} are
considered and it is proved how there exist no other 
cycle configurations from the one listed in the statement.
The process also proves that no unnecessary cycle configurations are listed.
%
%
\vskip 1mm
\noindent
{\bf Configuration~\ref{item:g6-1,1}} gives\\
$$
\left| C-6 \right| \,.
$$
Since cycle configuration $|J-6|$ can be discard (Lemma~\ref{lem:lemmino5}).
%
%
\vskip 3mm
\hrule
\noindent
{\bf Configuration~\ref{item:g6-1,2}} gives\\
$$
\left|\begin{array}{cc}
        C-4 & C-2 \\
       \end{array}\right| \,.
$$
Other cycle configurations $|C-4\, J-2|$, $|J-4\, J-2|$ and
$|J-4\, C-2|$ may be discard, because in $J-2$ and in $J-4$
there is no cycle column.
\vskip 3mm
\hrule
\noindent
{\bf Configuration~\ref{item:g6-1,3}} gives\\
$$
\left|\begin{array}{ccc}
        C-2 & C-2 & C-2\\
       \end{array}\right| \,.
$$

Other cycle configurations may be discard because they contain $J-2$ and in $J-2$  there is no cycle column (Lemma~\ref{lem:lemmino1}-1).

\vskip 3mm
\hrule
\noindent
{\bf Configuration~\ref{item:g6-2,2}} \\
\begin{enumerate}
\item Configuration~\ref{item:g6-2,2}.1 gives
$$
\left|\begin{array}{cc}
        C-2 & C-2 \\
         O  & C-2 \\
       \end{array}\right|\,,
 \qquad
\left|\begin{array}{cc}
        C-2 & J-2 \\
         O  & C-2 \\
       \end{array}\right|\,.
$$
Other cycle configurations may be discard because they contain $J-2$ 
as the only non-zero matrix in a d.r. or d.c. (Lemma~\ref{lem:lemmino1}-2).

\item Configuration~\ref{item:g6-2,2}.2 gives
$$
\left|\begin{array}{cc}
        C-3      & \Delta-1\\
        \Delta-1 & \Delta-1 \\
       \end{array}\right|\,.
$$
In fact any cycle configuration of kind
$$
\left|\begin{array}{cc}
        J-3      & \Delta-1\\
        \Delta-1  & \Delta-1 \\
       \end{array}\right|
$$
can be discarded (lemma~\ref{lem:lemmino5}).
On the other hand, cycle configuration
$$
\left|\begin{array}{cc}
        C-3 & J-1\\
        J-1 & J-1 \\
       \end{array}\right|
$$
is obviously acceptable and so lemma~\ref{lem:lemmino3} can be applied.
\end{enumerate}
\vskip 3mm
\hrule
\noindent
{\bf Configuration~\ref{item:g6-2,3}} gives\\ 

$$
\left|\begin{array}{ccc}
        C-2 & \Delta-1 & \Delta-1 \\
         O  & \Delta-1    & \Delta-1 \\
       \end{array}\right|.
$$

In fact , the following cycle configuration is obviously possible
$$
\left|\begin{array}{ccc}
        C-2 & J-1 & J-1 \\
         O  & J-1 & J-1 \\
       \end{array}\right|.
$$
Hence lemma~\ref{lem:lemmino3} can be apply.

To discard all cycle configurations of type
$$
\left|\begin{array}{ccc}
        J-2 & \Delta-1 & \Delta-1 \\
         O  & \Delta-1 & \Delta-1 \\
       \end{array}\right|\,,
$$
it is enough to apply Lemma~\ref{lem:lemmino1}-2 to d.c. 1.

\vskip 3mm
\hrule
\noindent
{\bf Configuration~\ref{item:g6-3,3}} gives\\
$$
\left|\begin{array}{ccc}
        \Delta-1 & \Delta-1 &    O     \\
        \Delta-1 &    O     & \Delta-1 \\
            O    & \Delta-1 & \Delta-1 \\
       \end{array}\right|\,.
$$
Clearly, the following cycle configuration is possible
$$
\left|\begin{array}{ccc}
        J-1 & J-1 & O   \\
        J-1 &  O  & J-1 \\
         O  & J-1 & J-1 \\
       \end{array}\right|\,,
$$
hence Lemma~\ref{lem:lemmino3} can be applied.

For remark~\ref{remark:g8-transpose} it is not necessary to study the cycle configurations that are transposed of the one considered.
Hence it has been proved that the listed configurations are the only valid ones.
\end{proof}
\end{theorem}

\subsection{$8$-cycle configurations for \QC~H matrices}\label{subsec:8cycle_QC}
\begin{theorem}\label{the:QC_g8}Let be $M \in \C_{m,\alpha,\beta,\gamma}$. The configurations in $M$
that may contain a cycles of length $8$,  are the
following \footnote{For
brevity any cycle configuration that is the transpose of another is omitted.}
\begin{enumerate}
    \item $$
            \left| C-8 \right|,
          $$
    \item $$
            \left| C-6 \quad C-2 \right|,
          $$
    \item $$
            \left| C-4 \quad C-4 \right|,
          $$

    \item $$
            \left| C-4 \quad C-2 \quad C-2  \right| \,.
          $$
    \item $$
            \left| C-2 \quad C-2 \quad C-2 \quad C-2 \right| \,.
          $$
    \item $$
        \left|\begin{array}{cc}
        C-5      & \Delta-1 \\
        \Delta-1 & \Delta-1 \\
       \end{array}\right|\,,
        $$
    \item $$
        \left|\begin{array}{cc}
        C-4      & \Delta-2 \\
        0        & \C-2 \\
       \end{array}\right|\,,
        $$
    \item $$
        \left|\begin{array}{cc}
        C-4 & C-2 \\
        C-2 & 0 \\
       \end{array}\right|\,,
        $$
    \item $$
        \left|\begin{array}{cc}
        C-3      & C-3 \\
        \Delta-1 & \Delta-1 \\
       \end{array}\right|\,,
        $$
    \item $$
        \left|\begin{array}{cc}
        C-3      & \Delta-1 \\
        \Delta-1 & C-3 \\
       \end{array}\right|\,,
        $$
    \item $$
        \left|\begin{array}{cc}
        \Delta-2      & \Delta-2 \\
        \Delta-2 & \Delta-2 \\
       \end{array}\right|\,,
        $$
\item $$
    \left|\begin{array}{ccc}
        C-4 & \Delta-1 & \Delta-1 \\
         O  & \Delta-1    & \Delta-1 \\
       \end{array}\right|,
    $$
\item $$
    \left|\begin{array}{ccc}
        C-3     & C-2   & \Delta-1 \\
      \Delta-1  &  O    & \Delta-1 \\
       \end{array}\right|,
    $$
\item $$
    \left|\begin{array}{ccc}
        C-3     & O   & \Delta-1 \\
      \Delta-1  &  C-2    & \Delta-1 \\
       \end{array}\right|,
    $$
\item $$
    \left|\begin{array}{ccc}
       \Delta-2 & C-2   & C-2 \\
         C-2    & O & O \\
       \end{array}\right|,
    $$
\item $$
    \left|\begin{array}{ccc}
       \Delta-2 & \Delta-1   & \Delta-1 \\
       \Delta-2 & \Delta-1   & \Delta-1 \\
       \end{array}\right|,
    $$
\item $$
    \left|\begin{array}{ccc}
       C-2 & \Delta-2   &  O\\
       O        & \Delta-2   & C-2 \\
       \end{array}\right|,
    $$

\item $$
    \left|\begin{array}{cccc}
       C-2 & C-2  & \Delta-1 & \Delta-1 \\
       O   &  O   & \Delta-1 & \Delta-1 \\
       \end{array}\right|,
    $$
\item $$
    \left|\begin{array}{cccc}
       C-2 &  O  & \Delta-1 & \Delta-1 \\
       O   & C-2 & \Delta-1 & \Delta-1 \\
       \end{array}\right|,
    $$
\item $$
    \left|\begin{array}{cccc}
       \Delta-1 & \Delta-1  & \Delta-1 & \Delta-1 \\
       \Delta-1 & \Delta-1  & \Delta-1 & \Delta-1 \\
       \end{array}\right|,
    $$

    \item $$
    \left|\begin{array}{ccc}
           C-3 & \Delta-1 &    O     \\
        \Delta-1 &    O     & \Delta-1 \\
            O    & \Delta-1 & \Delta-1 \\
       \end{array}\right|\,.
    $$
    \item $$
    \left|\begin{array}{ccc}
          \Delta-2 & \Delta-1 &    \Delta-1     \\
        C-2 &    O     & O \\
            O    & \Delta-1 & \Delta-1 \\
       \end{array}\right|\,.
    $$
    \item $$
    \left|\begin{array}{ccc}
        \Delta-2 & \Delta-1 & \Delta-1     \\
        \Delta-1 & \Delta-1 & O \\
        \Delta-1 &  O       & \Delta-1 \\
       \end{array}\right|\,.
    $$
    \item $$
    \left|\begin{array}{ccc}
            C-2 &    O     & O \\
         \Delta-1 & \Delta-1 & O\\
          \Delta-1 & \Delta-1 &    C-2     \\
       \end{array}\right|\,.
    $$

    \item $$
    \left|\begin{array}{cccc}
        C-2 & \Delta-1 & \Delta-1 & O\\
           O     & \Delta-1 &     O    & \Delta-1\\
           O     &   O      & \Delta-1 & \Delta-1\\
       \end{array}\right|\,.
    $$
\item $$
    \left|\begin{array}{cccc}
        \Delta-1 & \Delta-1 & \Delta-1 & \Delta-1\\
        \Delta-1 & \Delta-1 &     O    & 0 \\
           O     &   O      & \Delta-1 & \Delta-1\\
       \end{array}\right|\,.
    $$
    \item $$
    \left|\begin{array}{ccccc}
        \Delta-1 & \Delta-1 &     O    & O\\
        \Delta-1 &    O     & \Delta-1 & O\\
           O     & \Delta-1 &     O    & \Delta-1\\
           O     &   O      & \Delta-1 & \Delta-1\\
       \end{array}\right|\,.
    $$

\end{enumerate}
\begin{proof}
Once more for brevity the full proof is omitted here and can be found
in Appendix.
\end{proof}
\end{theorem}

\section{Relations between polynomials and the existence of cycles}\label{sec:polysTOcycles}
This subsection gives an  interpretation to the cycle configurations
presented previously  in terms of the polynomials associated to the circulant matrices involved.
First a general statement on the girth of the Tanner graph associated
to a binary weight-2 circulant matrix is given.
The following proposition has first been presented
in~\cite{martinaLDPC} and~\cite{Marta_tesi:04}.
Here only the main result is reported.\\

The notation introduced  in subsection~\ref{subsec:circulants} is
presented again here for clarity:
\begin{itemize}
\item $\epsilon(p)$ : any of the exponents ($a,b$) of the polynomial  $p(x)=x^a+x^b$,
\item $s(p)$ :  the separation ($\min(b-a,a+m-b)$)  of the polynomial $p$.
\end{itemize}

\begin{proposition}\label{prop:girth-circulant}
Let $m\geq 3$.
Let $M=(M_{i,j})$ be a circulant binary $m\times m$ matrix, generated by a
weight-$2$ polynomial $p$. Let $s(p)$ be the separation of $p$
and $g$ be the girth of the Tanner graph of $M$.
Then
\begin{equation*}
    g=2 \frac{m}{\gcd(m,s(p))}
\end{equation*}

\begin{proof}

Let $G$ be the Tanner graph of $M$.  Then each check node is connected
exactly to two bit nodes, and vice-versa.
We denote by $c_1,\ldots,c_m$ and $b_1,\ldots,b_m$, respectively, the
check nodes and the bit nodes of $G$.
Without loss of generality, we may suppose that $M_{1,1}=1$.
By circularity, $M_{j,j}=1$ for $1\leq j\leq m$, which implies
that $c_j$ is connected to $b_j$, for $1\leq j\leq m$.

By definition of separation, we have either $s=b-a$ or
$s=m-(b-a)$.
By circularity we may assume $s=b-a$, so that $M_{1,s+1}=1$ and
$b-a \leq m/2$.
But then we have $M_{j,s+j}=1$, for
$1\leq j\leq m-s$, and $M_{j,s+j-m}=1$, for $m-s+1\leq j\leq m$.
Therefore, check node $c_j$ is also connected to:
either bit node $b_{s+j}$,
which happens when $1\leq j\leq m-s$,
or bit node $b_{s+j-m}$,
 which happens when  $m-s+1\leq j\leq m$.
No other non-zero entry is present in $M$ and so no other connection exists
among nodes in $G$.

For any $i$, we want to find the length $g_i$ of the minimum length cycle
containing $c_i$. The girth of $G$ will be $g=\min_{1\leq i\leq m} g_i$.
By symmetry of $G$, $g_i$ does not depend on $i$, in particular $g=g_1$.\\
We now determine $g_1$, constructing a path as follows:
\begin{itemize}
\item we start from $c_1$. From $c_1$, we may go either to $b_1$ or to
      $b_{s+1}$. We choose to go to $b_1$. Clearly, the cycle will be closing
      when we will find ourself in $b_{s+1}$, as the next step
      will be $c_1$. From now on, the path allows
      no more choices. We will use arrows to shorten our notation.
\item $c_1 \rightarrow b_1$, $b_1 \rightarrow c_{m-s+1}$.
      If $m-s+1=1$, the cycle will be closed and $g_1=2$: this is impossible
      since $s<m$.
\item $c_{m-s+1} \rightarrow b_{m-s+1}$, $b_{m-s+1} \rightarrow c_{m-2s+1}$
      ($m-2s+1\geq 1$).\\
      If $m-2s+1=1$, the cycle is closed and $g_1=4$.

\item We perform steps of type
      $$ c_{m-(l-1)s+1} \rightarrow b_{m-(l-1)s+1},\quad
         b_{m-(l-1)s+1} \rightarrow c_{m-l s+1}\,,  $$
      until either there is an $l\geq 1$ s.t. $m-ls+1=1$ or there is no
      such $l$.
      We analyze the two cases separately.\\
      There is an $l$ s.t. $m=ls$. This is equivalent to $s|m$.
          This is also equivalent to $g_1=2l$ because we have formed
              a length-$2l$ cycle and we have encountered no smaller cycles.
          \\
          Or there is no $l$ s.t. $m-ls+1=1$.
          In this case let $\bar l=| m/s|$. Then $g \geq 2 \bar l$,
              i.e. $g > 2 m/s$.
          The next two steps will be
          $$c_{m-{\bar l}s+1} \rightarrow b_{m-{\bar l}s+1}, \quad
            b_{m-{\bar l}s+1} \rightarrow c_{2m-({\bar l}+1)s+1}\,,$$
              i.e. we have to ``wrap on the graph''.

\item We perform steps of type
      $$ c_{2m-(l-1)s+1} \rightarrow b_{2m-(l-1)s+1},\quad
         b_{2m-(l-1)s+1} \rightarrow c_{2m-l s+1}\,,  $$
      until either there is an $l\geq |m/s|$ s.t. $2m-ls+1=1$ or there is no
      such $l$. We analyze the two cases separately.\\
      There is an $l> |m/s|$ s.t. $2m=ls$ (but no $l\geq 1$ is s.t.
              $m=ls$).
          Obviously this is equivalent to $s|2m$ and not $s| m$.
         
          This is also equivalent to $g_1=2l$ because we have formed
              a length-$2l$ cycle and we have encountered no smaller cycles.
\\
     Or there is no $l$ s.t. $2m-ls=0$ (but no $l\geq 1$ is s.t.
          $m=ls$).
          This is equivalent to $s$ not dividing $2m$
          (and $s$ not dividing $m$).
          This is also equivalent to $g> |2m/s|$.
          In this case let $\bar l=|2 m/s|$.
          The next two steps will be
          $$c_{2m-{\bar l}s+1} \rightarrow b_{2m-{\bar l}s+1}, \quad
            b_{2m-{\bar l}s+1} \rightarrow c_{3m-({\bar l}+1)s+1}\,,$$
              i.e. we have to ``wrap on the graph'' again.

\item In the general case, after $2l$ steps we reach $c_{q m-ls+1}$,
      where $s$ does not\\
      divide any $z m$ for $1\leq z\leq q-1$,
      by a generalization of the above arguments.
      At some stage we must meet again $c_1$, so that
      $c_{q m-ls+1}=c_1$. In other words, the girth is $2l$, with
      $l=q m/s$, if $q$ is s.t.
      $$
           s | q m, \quad s \not| z m,\, 1\leq z\leq q-1 \,.
      $$
      This means that $q m$ is the smallest multiple of $m$ that is also
      a multiple of $s$, i.e. $q m$ is the minimum common multiple of
      $m$ and $s$. \\
      Let $[,]$ denote minimum common multiple.
      For any two integers $a$ and
      $b$, we have $[a,b]/a=b/\gcd(a,b)$, so that
      $$
        \frac{q m}{s}= \frac{[m,s]}{s}=\frac{m}{\gcd(m,s)} \,.
      $$
\end{itemize}
\end{proof}
\end{proposition}

\begin{remark}[Minimum dimension of the circulants]
Theorem~\ref{prop:girth-circulant} indirectly gives a  minimum dimension
that  the  circulant matrices must have to allow high girth.
In particular if \HH matrices  with girth $10$ (or higher)  are desired the circulant
matrices must be at least $[5\times 5] $.
To prove this is sufficient to consider that to have $g \geq 10$ it
must be $\frac{m}{\gcd(m,s(p))} \geq 5$, hence $m \geq 5$.
\end{remark}

\begin{figure}[hbtp]
\psfrag{Output}{\scriptsize{Output}}
\psfrag{Circulant dimension m}{Circulant dimension $m$}
\psfrag{Achievable girth}{Achievable girth}
\centering
\includegraphics[width=0.8\columnwidth]{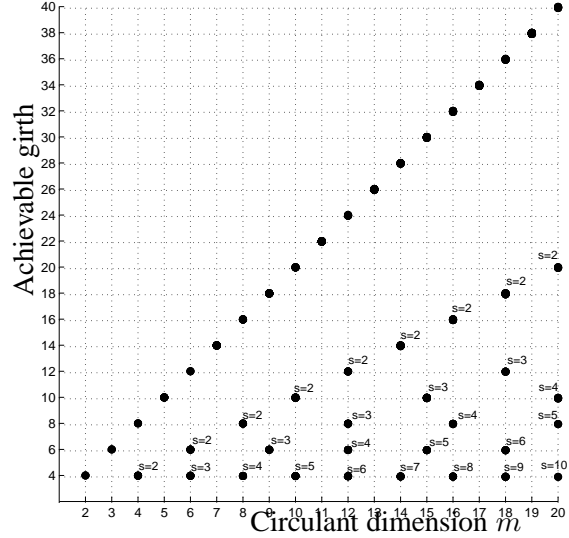}
 \caption{\label{fig:achievableGirth} Girth values that can be
 obtained with a single circulant given its dimension}
\end{figure}

A plot of the  girth that can be obtained given a certain circulant
dimension, changing the value of the separation, is presented
in \figur~\ref{fig:achievableGirth}. 
It is evident that the maximum girt ($g=2m$) is also achievable,
e.g. by choosing the separation to be one. Of more interest is to
consider that which  values of $m$ and $s$  must be  avoided to guarantee a good girth.

The following theorems present which conditions, on separations and
exponents,  must be satisfied for a certain cycle, in a cycle
configurations, to exist.
Once this set of conditions is known matrices
with high girth can be generated by choosing the polynomials of the
circulants in such a way that none of the conditions are satisfied.

\subsection{Conditions for the 4-cycles}\label{subsec:polysTOcycles-4}

\begin{theorem}
\label{the:condition4} Let be $M \in \C_{m,\alpha,\beta,\gamma}$. The configurations in $M$
that may contain a cycles of length $4$ and associated conditions,  are the following

\begin{enumerate}
    \item\label{item:g4-1}
        \begin{align*}
           &\left| C-4 \right|, & s(p)=m/2
        \end{align*}
    \item \label{item:g4-2}
    \begin{align*}
            &\left| C^1-2 \quad C^2-2 \right|,  & s(p^1)=s(p^2)
          \end{align*}
    \item \label{item:g4-3}
        \begin{align*}
       & \left|\begin{array}{cc}
            \Delta^1-1 & \Delta^2-1\\
            \Delta^3-1 & \Delta^4-1 \\
           \end{array}\right|, &\epsilon(p^1)-\epsilon(p^2)-\epsilon(p^3)+\epsilon(p^4)=0
         \end{align*}
\end{enumerate}

\begin{proof}

It is of interest to note that the conditions associated with every 
configuration can be proved applying theorem~\ref{th:fossorier}, that
is the generalization of Fossorier theorem for  weight-2 circulants. 

It has been chosen to follow a different methodology that
slightly reduces the notation and allow to use a graphical 
representation that easier to understand.
The use of separations instead of the difference of exponents
makes the conditions clearer and more evident; it also reduces the
number of variables involved making them easier to check.
The proof is divided in shorter lemmas each considering a case of the
main theorem.
An $4$-cycle is formed by two  cycle columns called $y,z$ and two
cycle rows called $x,t$. 
In every $4$-cycle there are $4$ cycle points, they take the name of
the column and row where they lie in
$(x,y),(x,z),(t,y),(t,z)$. The notation is used to compute
the conditions.
To allow the reader to better understand the meaning of the formulae
for each lemma a graphical representation of the cycle is presented.
For example  \figur~\ref{fig:circulant_4-1} presents an example of
$4$-cycle on two weight-$2$ circulants.
The two rectangular blocks represent the two circulants and the dashed 
diagonal lines the position of the ones in the matrices. The cycle
columns and  cycle rows are marked  by the dash-doted lines and the cycle is
outlined with a wider line.

\begin{lemma}\label{lem:g4-1} 
There is a $4$-cycle in case~\ref{item:g4-1}  if and
only if
$$
   2|m \mbox{ and } \quad s(p)=m/2 \,.
$$
\begin{proof}
There is a $4$-cycle if and only if $g=4$, because smaller girths are not possible.
Applying Proposition~\ref{prop:girth-circulant} to the case $g=4$ and $M=C$, there is a $4$-cycle if and only if
$$
   4= g=2 \frac{m}{\gcd(m,s)}
$$
i.e. $m= 2 \gcd(m,s)$. In particular, $m$ is even and $m/2 | s$, but $s\leq
m/2$, so that $s=m/2$. On the other hand, if $m$ is even and $s=m/2$ then
$\gcd(m,s)=s=m/2$.
\end{proof}
\end{lemma}
\begin{lemma}\label{lem:g4-2} 
There is a $4$-cycle in case~\ref{item:g4-2} if and
only if
$$
s(p^1) = s(p^2)
$$
\begin{proof}
It can be assumed that there is a $4$-cycle if and only if column $y$ lies in $C^1$ and
column $z$ lies in $C^2$ (\figur~\ref{fig:circulant_4-1}).

\begin{figure}[htbp]
\begin{center}
\psfrag{x}{x} \psfrag{t}{t} \psfrag{y}{y} \psfrag{z}{z}
\includegraphics[width=0.8\columnwidth]{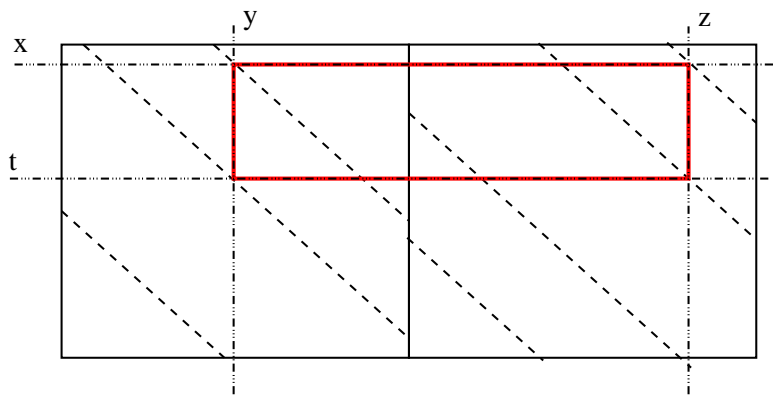}
\end{center}
\caption[$4$-cycle on two weight-$2$ circulants]{Example of $4$-cycle on two weight-$2$ circulants.}
\label{fig:circulant_4-1}
\end{figure}

Applying Prop.~\ref{prop:cas}-2 to column $y$ yields
\begin{equation} \label{eq:dim_g4-1-1}
x-t \equiv \pm s^1 \,.
\end{equation}
Applying Proposition~\ref{prop:cas}-2 to cycle column $z$ yields
\begin{equation} \label{eq:dim_g4-1-2}
x-t \equiv \pm s^2 \,.
\end{equation}
From (\ref{eq:dim_g4-1-1}) and (\ref{eq:dim_g4-1-2}), 
$ s^1 \equiv \pm s^2$ is obtained and hence $s^1 =
s^2$ (Lemma~\ref{lem:sp}).
\end{proof}
\end{lemma}
\begin{lemma}\label{lem:g4-3} 
There is a $4$-cycle in case~\ref{item:g4-3} if
and only~if
$$
 \epsilon(p^1)-\epsilon(p^2)-\epsilon(p^3)+\epsilon(p^4)=0
$$
\begin{proof}
It can be assumed that there is a $4$-cycle if and only if, simultaneously, 
cycle point $(x,y)$ lies in $\Delta^1$,
cycle point $(x,z)$ lies in $\Delta^2$,
cycle point $(t,y)$ lies in $\Delta^3$ and
cycle point $(t,z)$ lies in $\Delta^4$ (\figur~\ref{fig:circulant_4-2}).

\begin{figure}[htbp]
\begin{center}
\psfrag{x}{x} \psfrag{t}{t} \psfrag{y}{y} \psfrag{z}{z}
\includegraphics[width=0.8\columnwidth]{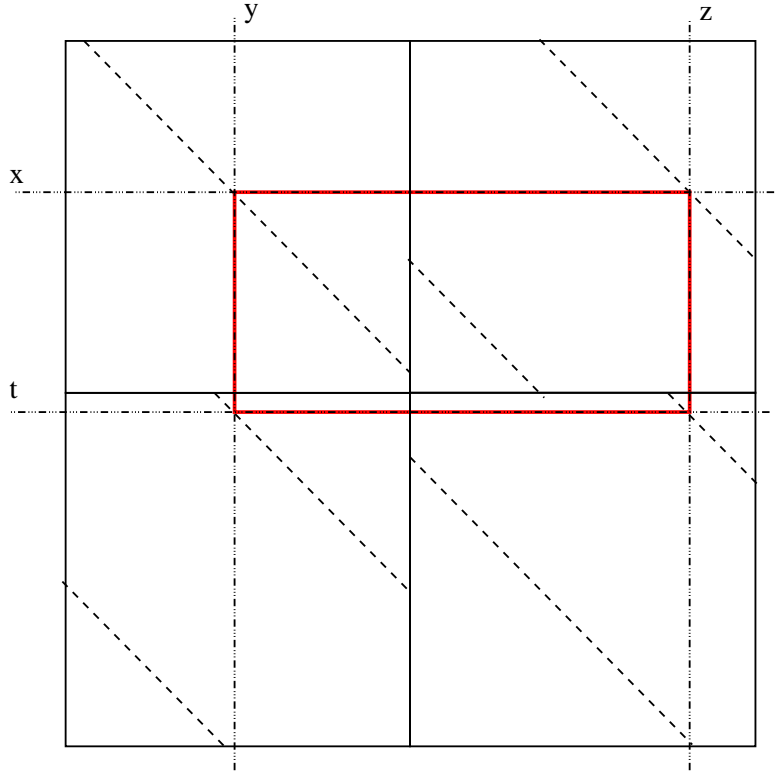}
\end{center}
\caption[$4$-cycle on four  weight-$1$ circulants]{Example of $4$-cycle on four  weight-$1$ circulants.}
\label{fig:circulant_4-2}
\end{figure}

Since cycle point $(x,y)$ lies in $\Delta^1$, using Proposition~\ref{prop:cas}-1:
\begin{equation} \label{eq:dim_g4-3-1}
y \equiv x +\ep(p^1) \,.
 \end{equation}
Since cycle point $(x,z)$ lies in $\Delta^2$, using Proposition~\ref{prop:cas}-1:
\begin{equation} \label{eq:dim_g4-3-2}
z \equiv x +\ep(p^2) \,.
 \end{equation}
Since cycle point $(t,y)$ lies in $\Delta^3$, using Lemma~\ref{lem:id}:
\begin{equation} \label{eq:dim_g4-3-3}
y \equiv t +\ep(p^3)\,.
 \end{equation}
Since cycle point $(t,z)$ lies in $\Delta^4$, using Lemma~\ref{lem:id}:
\begin{equation} \label{eq:dim_g4-3-4}
z \equiv t +\ep(p^4) \,.
 \end{equation}
The desired result is obtained from (\ref{eq:dim_g4-3-1}), (\ref{eq:dim_g4-3-2}), (\ref{eq:dim_g4-3-3})
 and (\ref{eq:dim_g4-3-4}).

\end{proof}
\end{lemma}

It has hence been proved that the conditions listed on the statement
considers all the possible $4$-cycles that can exist 
on the studied \QC~matrices.
\end{proof}
\end{theorem}
For example consider the polynomial $p(x)=1+x^3$ with $m=6$ for such
 polynomial $s(p)=3$ hence $s(p)=m/2$ and for condition $1$  a
 $4$-cycle exist. 
  \unitlength=1mm
\newcommand{\downLineOne}{
  \begin{picture}(0,0)(0,0)
    \put(1,-17){\tikz\draw[red,very thick] (0mm,0mm) -- (0mm,-19mm);}
\end{picture}}
\newcommand{\orLineOne}{
  \begin{picture}(0,0)(0,0)
    \put(1,2){\tikz\draw[red,very thick] (0mm,0mm) -- (17mm,0mm);}
\end{picture}}
 \begin{equation*}
          \left[
            \begin{array}{cccccc}
              \downLineOne \orLineOne 1 & 0 & 0 &  \downLineOne 1 & 0 &0\\
              0 & 1 & 0 &  0 & 1 &0\\
              0 & 0 & 1 &  0 & 0 &1\\
              \orLineOne  1 & 0 & 0 &  1 & 0 &0\\
              0 & 1 & 0 &  0 & 1 &0\\
              0 & 0 & 1 &  0 & 0 &1
            \end{array}\right]
      \end{equation*}

\subsection{Conditions for the 6-cycles}\label{subsec:polysTOcycles-6}

\begin{theorem}
\label{the:condition6} Let be $M \in \C_{m,\alpha,\beta,\gamma}$. The configurations in $M$
that may contain a cycles of length $6$,  are the following

    \begin{enumerate}
    \item \label{item:g6-1}
    \begin{align*}
           & \left| C-6 \right|,& s(p)=m/3
          \end{align*}
    \item \label{item:g6-2}
    \begin{align*}
            &\left| C^1-4 \quad C^2-2 \right|,&  s(p^2)\equiv \pm2s(p^1)
          \end{align*}
   \item \label{item:g6-3}

    \begin{align*}
           \left| C^1-2 \quad C^2-2 \quad C^3-2  \right|,
							  &\qquad  s(p^1)\pm s(p^2)\pm s(p^3) \equiv0,\\
          \end{align*}

  \item \label{item:g6-5}
    \begin{gather*}
        \left|\begin{array}{cc}
        C^1-2 & \Delta^2-2 \\
         O  & C^3-2 \\
       \end{array}\right|,\\
    \epsilon(p^2)-\epsilon(p^2) \equiv \pm s(p^1) \pm s(p^3),
   \end{gather*}

    \item\label{item:g6-6}
   \begin{gather*}
    \left|\begin{array}{cc}
        C^1-3 & \Delta^2-1\\
        \Delta^3-1 & \Delta^4-1 \\
       \end{array}\right|,\\
       \epsilon(p^1)-\epsilon(p^2)-\epsilon(p^3)+\epsilon(p^4)\equiv\pm s(p^1),
    \end{gather*}

\item \label{item:g6-7}
    \begin{gather*}
    \left|\begin{array}{ccc}
        C^1-2 & \Delta^2-1 & \Delta^3-1 \\
         O  & \Delta^4-1    & \Delta^5-1 \\
       \end{array}\right|,\\
       \epsilon(p^2)-\epsilon(p^3)-\epsilon(p^4)+\epsilon(p^5)\equiv\pm s(p^1),
    \end{gather*}

    \item \label{item:g6-8}
    \begin{gather*}
    \left|\begin{array}{ccc}
        \Delta^1-1 & \Delta^2-1 &    O     \\
        \Delta^3-1 &    O     & \Delta^4-1 \\
            O    & \Delta^5-1 & \Delta^6-1 \\
       \end{array}\right|,\\
       \epsilon(p^1)-\epsilon(p^2)-\epsilon(p^3)+\epsilon(p^4)+\epsilon(p^5)-\epsilon(p^6) \equiv 0
    \end{gather*}

\end{enumerate}

\begin{proof}
As for the previous theorem the proof is once more divided in smaller
lemmas each considering a particular case.
An $6$-cycle is formed by three  cycle columns called $y,z,v$ and three
cycle rows called $x,t,w$. 
In every $6$-cycle there are $6$ cycle points, they take the name of
the column and row where they lie:
$(x,y),(x,z),(t,y),(t,v),(w,v),(w,z)$. The notation is used to compute
the conditions.

\begin{lemma}\label{lem:g6-1} 
There is a $6$-cycle and there are no $4$-cycles in case~\ref{item:g6-1} if and only if
$$
   3|m  \mbox{ and } \quad
    \mbox{and} \quad s(p)=m/3 \,.
$$

\begin{proof}
There is a $6$-cycle and no $4$-cycle if and only if $g=6$. Applying Prop.
\ref{prop:girth-circulant} to the case $g=6$ and $M=C$, this is equivalent to
$$
   6= g =2 \frac{m}{\gcd(m,s)}
$$
i.e. $m= 3 \gcd(m,s)$. In particular, $3|m$ and $m/3 | s$, but $s\leq m/2$,
so that $s=m/3$. On the other hand, if $m$ is divisible by $3$ and $s=m/3$ then
$\gcd(m,s)=s=m/3$.
\end{proof}
\end{lemma}
\begin{lemma}\label{lem:g6-2} 
There is a $6$-cycle in case~\ref{item:g6-2}  if and only if
$$
s(p^2) \equiv \pm 2s(p^1) 
$$
\begin{proof}
It can be assumed that there is a $6$-cycle if and only if if both (cycle)
column $y$ and column $z$ lie in $C^1$ and column $v$ lies in $C^2$ (\figur~\ref{fig:circulant_6-1}).

\begin{figure}[h!tb]
\begin{center}
\psfrag{x}{x} \psfrag{t}{t} \psfrag{w}{w} \psfrag{y}{y} \psfrag{z}{z} \psfrag{v}{v}
\includegraphics[width=0.8\columnwidth]{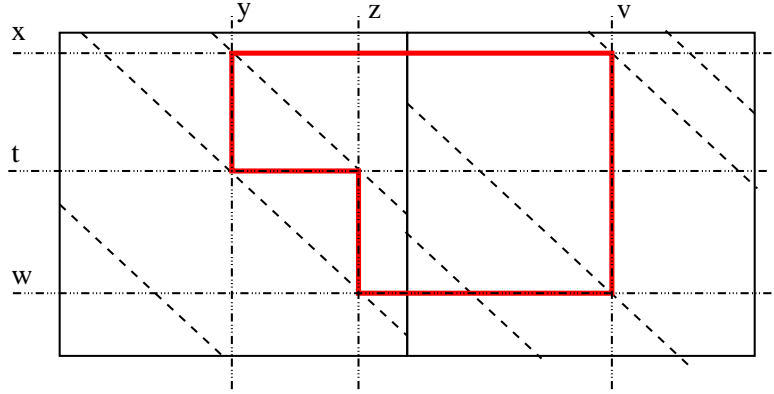}
\end{center}
\caption[$6$-cycle on two weight-$2$ circulants]{Example of $6$-cycle on two weight-$2$ circulants.}
\label{fig:circulant_6-1}
\end{figure}

Applying Proposition~\ref{prop:cas}-4 to cycle column $y$ and cycle column $z$ yields
\begin{equation} \label{eq:dim_g6-2-1}
x-t \equiv \pm s_j^1 \,,
\end{equation}
\begin{equation} \label{eq:dim_g6-2-2}
x-w \equiv \mp s_j^1 \,.
\end{equation}
Applying Proposition~\ref{prop:cas}-2 to cycle column $v$ yields
\begin{equation} \label{eq:dim_g6-2-3}
w-t \equiv \pm s_j^2 \,.
\end{equation}
Since $x-t = (x-w)+(w-t)$, from (\ref{eq:dim_g6-2-1}),
(\ref{eq:dim_g6-2-2}) and (\ref{eq:dim_g6-2-3}):
$$
    \pm s^1 \equiv \mp s^1 + \pm s^2 \,,
$$
from which the desired result is obtained:
$$
  s^2 \equiv 2 s^1 \,.
$$
\end{proof}
\end{lemma}
\begin{lemma}\label{lem:g6-3}
 There is a $6$-cycle in case~\ref{item:g6-3}  if and only if
  \begin{equation*}
   s^1  \pm s^2 \pm s^3 \equiv 0\,.
  \end{equation*}
\begin{proof}
It can be assumed that there is a $6$-cycle if and only if 
column $y$ lie in $C^1$, column $z$ lie in $C^2$ and column $v$ lies
in $C^3$ (\figur~\ref{fig:circulant_6-2}).

\begin{figure}[h!tb]
\begin{center}
\psfrag{x}{x} \psfrag{t}{t} \psfrag{w}{w} \psfrag{y}{y} \psfrag{z}{z} \psfrag{v}{v}
\includegraphics[width=0.8\columnwidth]{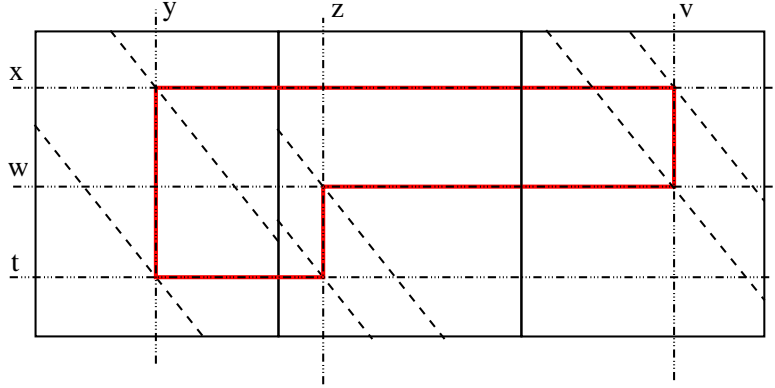}
\end{center}
\caption[$6$-cycle on three weight-$2$ circulants]{Example of $6$-cycle on three weight-$2$ circulants.}
\label{fig:circulant_6-2}
\end{figure}

Applying Proposition~\ref{prop:cas}-2 to cycle column $y$ yields
\begin{equation} \label{eq:dim_g6-3-1}
x-t \equiv \pm s^1 \,,
\end{equation}
Applying Proposition~\ref{prop:cas}-2 to cycle column $z$ yields
\begin{equation} \label{eq:dim_g6-3-2}
t-w \equiv \pm s^2 \,,
\end{equation}
Applying Proposition~\ref{prop:cas}-2 to cycle column $v$ yields
\begin{equation} \label{eq:dim_g6-3-3}
w-t \equiv \pm s^3 \,,
\end{equation}
Since $ (x-t)+(w-x)+(t-w)=0$, from (\ref{eq:dim_g6-3-1}), (\ref{eq:dim_g6-3-2}) and (\ref{eq:dim_g6-3-3}):
$$
    \pm s^1  \pm s^2  \mp s^3 \equiv 0\,,
$$
from which the desired result is obtained.
Note how the sign of one of the separations, $s^1$ in the main
theorem, can be fixed. In fact if  $- s^1 + s^2 -  s^3 \equiv 0$ then
also  $ s^1 - s^2 +  s^3 \equiv 0$ and this is part of the conditions.
 
\end{proof}
\end{lemma}

\begin{lemma}\label{lem:g6-5} 
There is a $6$-cycle in case~\ref{item:g6-5} if and only if
$$
\epsilon(p^2)-\epsilon(p^2)\equiv \pm s(p^1) \pm s(p^3)
$$
\begin{proof}
It can be assumed that there is a $6$-cycle if and only if simultaneously cycle column $y$
lies in $C^1$, cycle points $(x,z)$ and $(t,v)$ lie in $\Delta^2$ and cycle row $w$ lies
in $C^3$ (\figur~\ref{fig:circulant_6-3}).

\begin{figure}[h!tb]
\begin{center}
\psfrag{x}{x} \psfrag{t}{t} \psfrag{w}{w} \psfrag{y}{y} \psfrag{z}{z} \psfrag{v}{v}
\includegraphics[width=0.8\columnwidth]{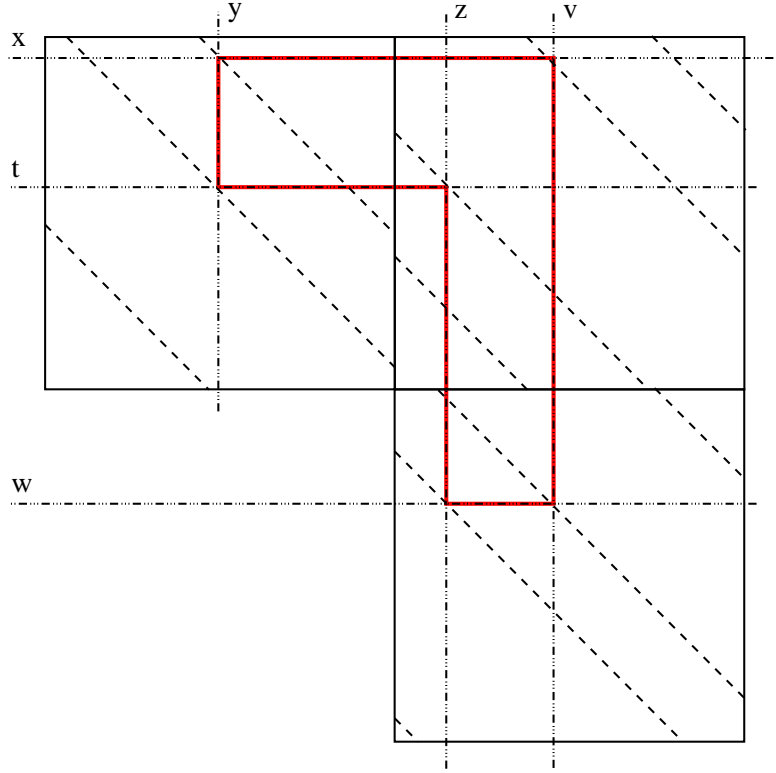}
\end{center}
\caption[$6$-cycle on three weight-$2$ circulants]{Example of $6$-cycle on three weight-$2$ circulants.}
\label{fig:circulant_6-3}
\end{figure}

Since cycle column $y$ lies in $C^1$, applying Prop.~\ref{prop:cas}-2:
\begin{equation} \label{eq:dim_g6-5-1}
x- t \equiv \pm s^1 \,.
 \end{equation}
Since cycle row $w$ lies in $C^3$, applying Proposition~\ref{prop:cas}-3:
\begin{equation} \label{eq:dim_g6-5-2}
v-z \equiv \pm s^3\,.
 \end{equation}
Since cycle points $(x,z)$ and $(t,v)$ lie in $\Delta^2$, applying Lemma~\ref{lem:id}:
\begin{equation} \label{eq:dim_g6-5-3}
   z \equiv x +\epsilon^2, \qquad v \equiv t + \epsilon^2 \,.
 \end{equation}
Substituting (\ref{eq:dim_g6-5-3}) and (\ref{eq:dim_g6-5-1}) into
 (\ref{eq:dim_g6-5-2}) 
the desired result is obtained.

\end{proof}
\end{lemma}

\begin{lemma}\label{lem:g6-6} 
There is a $6$-cycle in case~\ref{item:g6-6}  if and only if
$$
\ep(p^1)-\ep(p^2)-\ep(p^3)+\ep(p^4) \equiv \pm s(p^1)
$$
\begin{proof}
It may be assumed that there is a $6$-cycle if and only if, simultaneously, cycle column $y$
lies in $C^1$, cycle point $(x,z)$ lies in $C^1$, cycle point $(w,z)$ lies in
$\Delta^3$, cycle point $(w,v)$ lies in $\Delta^4$ and cycle point $(t,v)$ lies in
$\Delta^2$ (\figur~\ref{fig:circulant_6-4}).
\begin{figure}[h!tb]
\begin{center}
\psfrag{x}{x} \psfrag{t}{t} \psfrag{w}{w} \psfrag{y}{y} \psfrag{z}{z} \psfrag{v}{v}
\includegraphics[width=0.8\columnwidth]{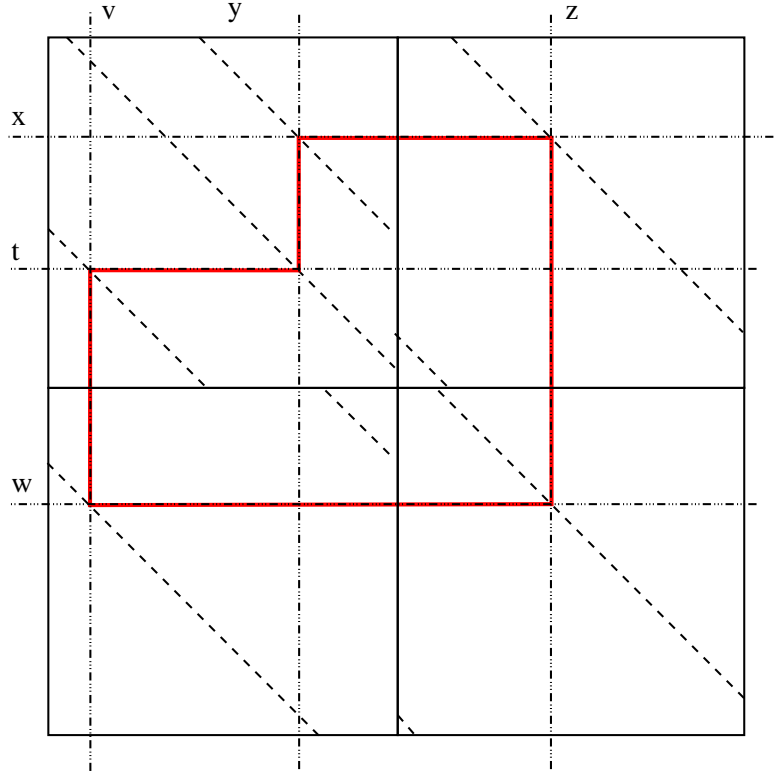}
\end{center}
\caption[$6$-cycle on one weight-$2$ and three weight-$3$  circulants]{Example of $6$-cycle on one weight-$2$ and three weight-$3$  circulants.}
\label{fig:circulant_6-4}
\end{figure}
Since cycle column $y$ lies in $C^1$, applying  Proposition~\ref{prop:cas}-2:
\begin{equation} \label{eq:dim_g6-6-1}
x- t \equiv \pm s^1 \,.
 \end{equation}
Since cycle point $(x,z)$ lies in $\C^1$,  applying Proposition~\ref{prop:cas}-1:
\begin{equation} \label{eq:dim_g6-6-2}
z \equiv x +\ep(p^1) \,.
 \end{equation}
Since cycle points $(w,v)$ lies in $\Delta^4$,  applying Proposition~\ref{prop:cas}-1:
\begin{equation} \label{eq:dim_g6-6-3}
v \equiv w +\ep(p^4) \,.
 \end{equation}
Since cycle points $(w,z)$ lies in $\Delta^3$,  applying Lemma~\ref{lem:id}:
\begin{equation} \label{eq:dim_g6-6-4}
z \equiv w +\ep(p^3)\,.
 \end{equation}
Since cycle points $(t,v)$ lies in $\Delta^2$,  applying Lemma~\ref{lem:id}:
\begin{equation} \label{eq:dim_g6-6-5}
v \equiv t +\ep(p^2) \,.
 \end{equation}
The desired result is obtained from (\ref{eq:dim_g6-6-1}), (\ref{eq:dim_g6-6-2}), (\ref{eq:dim_g6-6-3}), (\ref{eq:dim_g6-6-4})
 and (\ref{eq:dim_g6-6-5}).
\end{proof}
\end{lemma}
\begin{lemma}\label{lem:g6-7} 
There is a $6$-cycle in case~\ref{item:g6-7}  if and only if
$$
\ep(p^2)-\ep(p^3)-\ep(p^4)+\ep(p^5) \equiv \pm s(p^1)
$$
\begin{proof}
It may be assumed that there is a $6$-cycle if and only if, simultaneously, cycle column $y$
lies in $C^1$, cycle point $(x,z)$ lies in $\Delta^2$, cycle point $(w,z)$ lies in
$\Delta^4$, cycle point $(w,v)$ lies in $\Delta^5$ and cycle point $(t,v)$ lies in
$\Delta^3$ (\figur~\ref{fig:circulant_6-5}).
\begin{figure}[p]
\begin{center}
\psfrag{x}{x} \psfrag{t}{t} \psfrag{w}{w} \psfrag{y}{y} \psfrag{z}{z} \psfrag{v}{v}
\includegraphics[width=0.8\columnwidth]{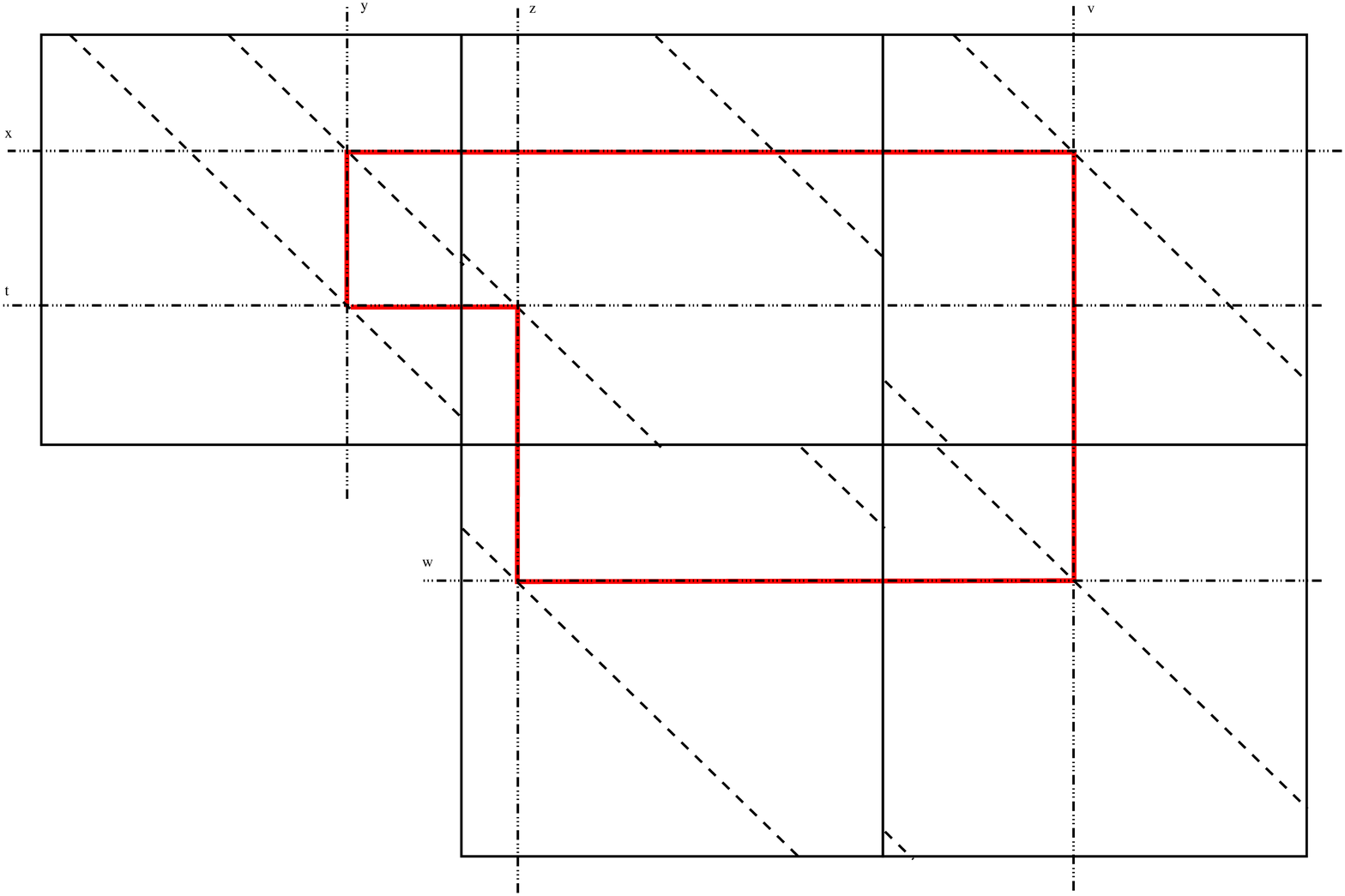}
\end{center}
\caption[$6$-cycle on one  weight-$2$ and four  weight-$1$ circulants]{Example of $6$-cycle on one  weight-$2$ and four  weight-$1$ circulants.}
\label{fig:circulant_6-5}
\end{figure}
Since cycle column $y$ lies in $C^1$, applying Proposition~\ref{prop:cas}-2:
\begin{equation} \label{eq:dim_g6-7-1}
x- t \equiv \pm s^1 \,.
 \end{equation}
Since cycle points $(x,z)$ lies in $\Delta^2$, applying Proposition~\ref{prop:cas}-1:
\begin{equation} \label{eq:dim_g6-7-2}
z \equiv x +\ep(p^2) \,.
 \end{equation}
Since cycle points $(w,v)$ lies in $\Delta^5$, applying Proposition~\ref{prop:cas}-1:
\begin{equation} \label{eq:dim_g6-7-3}
v \equiv w +\ep(p^5) \,.
 \end{equation}
Since cycle points $(w,z)$ lies in $\Delta^4$, applying  Lemma~\ref{lem:id}:
\begin{equation} \label{eq:dim_g6-7-4}
z \equiv w +\ep(p^4)\,.
 \end{equation}
Since cycle points $(t,v)$ lies in $\Delta^3$, applying Lemma~\ref{lem:id}:
\begin{equation} \label{eq:dim_g6-7-5}
v \equiv t +\ep(p^3) \,.
 \end{equation}
The desired result is obtained from  (\ref{eq:dim_g6-7-1}), (\ref{eq:dim_g6-7-2}), (\ref{eq:dim_g6-7-3}), (\ref{eq:dim_g6-7-4})
 and (\ref{eq:dim_g6-7-5}).
\end{proof}

\end{lemma}
\begin{lemma}\label{lem:g6-8} 
There is a $6$-cycle in case~\ref{item:g6-8}  if and only if 
$$
\epsilon(p^1)-\epsilon(p^2)-\epsilon(p^3)+\epsilon(p^4)+\epsilon(p^5)-\epsilon(p^6) \equiv 0
$$
\begin{proof}
It can be assumed that there is a $6$-cycle if and only if, simultaneously, 
cycle point $(x,y)$ lies in $\Delta^1$,
cycle point $(t,y)$ lies in $\Delta^3$,
cycle point $(x,z)$ lies in $\Delta^2$,
cycle point $(w,z)$ lies in $\Delta^5$,
cycle point $(w,v)$ lies in $\Delta^6$ and
cycle point $(t,v)$ lies in $\Delta^4$ (\figur~\ref{fig:circulant_6-6}).
\begin{figure}[p]
\begin{center}
\psfrag{x}{x} \psfrag{t}{t} \psfrag{w}{w} \psfrag{y}{y} \psfrag{z}{z} \psfrag{v}{v}
\includegraphics[width=0.8\columnwidth]{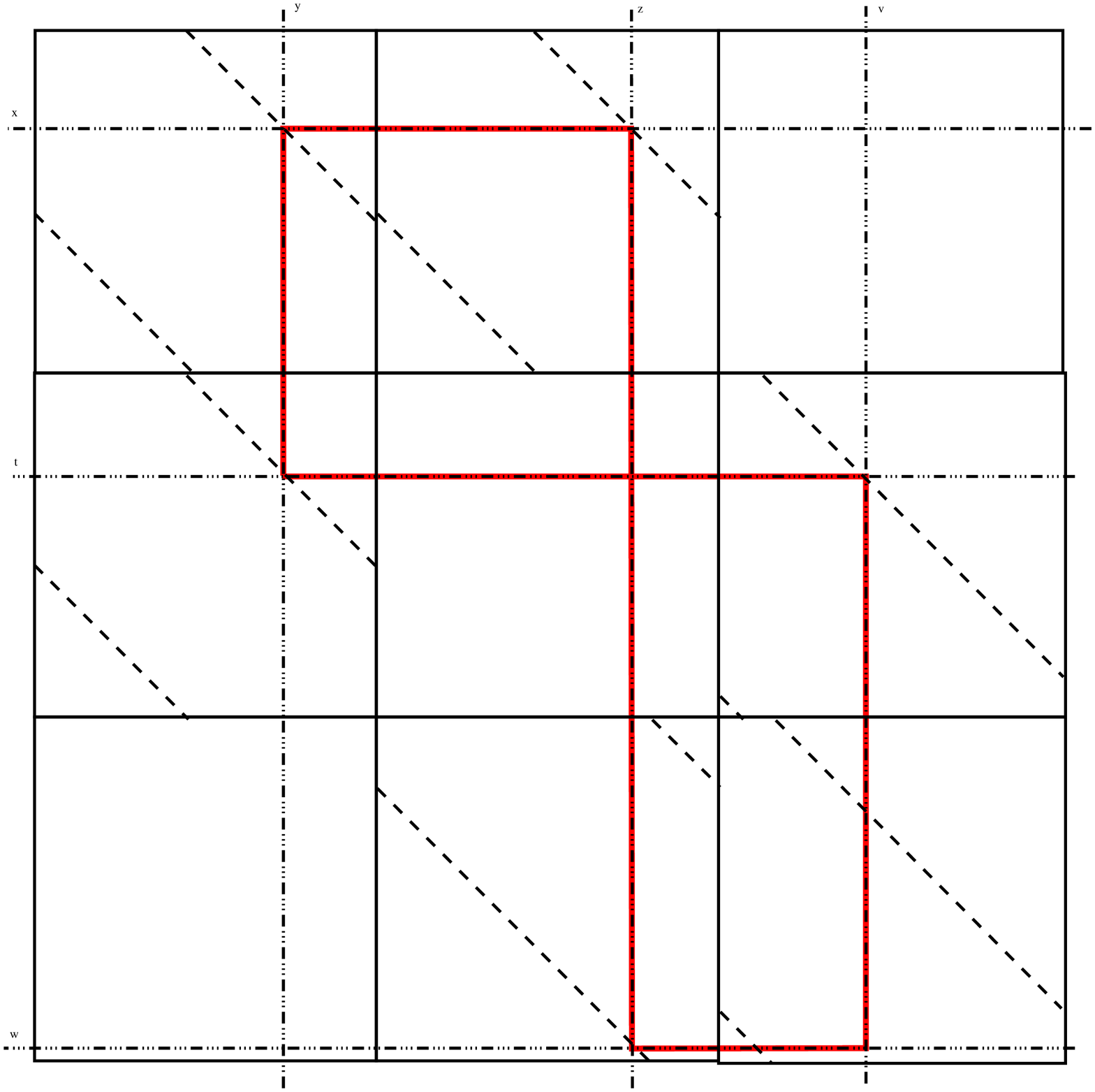}
\end{center}
\caption[$6$-cycle on two weight-$2$ circulants]{Example of $6$-cycle on six weight-$1$ circulants.}
\label{fig:circulant_6-6}
\end{figure}
Since cycle points $(x,y)$ lies in $\Delta^1$,  applying  Proposition~\ref{prop:cas}-1:
\begin{equation} \label{eq:dim_g6-8-1}
y \equiv x +\ep(p^1) \,.
 \end{equation}
 Since cycle points $(t,y)$ lies in $\Delta^3$, applying  Proposition~\ref{prop:cas}-1:
\begin{equation} \label{eq:dim_g6-8-2}
y \equiv t +\ep(p^3) \,.
 \end{equation}
Since cycle points $(x,z)$ lies in $\Delta^2$, applying  Proposition~\ref{prop:cas}-1:
\begin{equation} \label{eq:dim_g6-8-3}
z \equiv x +\ep(p^2) \,.
 \end{equation}
Since cycle points $(w,v)$ lies in $\Delta^6$,  applying Proposition~\ref{prop:cas}-1:
\begin{equation} \label{eq:dim_g6-8-4}
v \equiv w +\ep(p^6) \,.
 \end{equation}
Since cycle points $(w,z)$ lies in $\Delta^5$, applying  Lemma~\ref{lem:id}:
\begin{equation} \label{eq:dim_g6-8-5}
z \equiv w +\ep(p^5)\,.
 \end{equation}
Since cycle points $(t,v)$ lies in $\Delta^4$, applying   Lemma~\ref{lem:id}:
\begin{equation} \label{eq:dim_g6-8-6}
v \equiv t +\ep(p^4) \,.
 \end{equation}
The desired result is obtained from (\ref{eq:dim_g6-8-1}), (\ref{eq:dim_g6-8-2}), (\ref{eq:dim_g6-8-3}), (\ref{eq:dim_g6-8-4}), (\ref{eq:dim_g6-8-5})  and (\ref{eq:dim_g6-8-6}).
\end{proof}

\end{lemma}

It has hence been proved that the conditions listed on the statement
considers all the possible $6$-cycles that can exist 
on the studied \QC~matrices.

\end{proof}
\end{theorem}

\unitlength=1mm
\newcommand{\downLineOneOne}{
  \begin{picture}(0,0)(0,0)
    \put(0.5,-24){\tikz\draw[red,very thick] (0mm,0mm) -- (0mm,-26mm);}
\end{picture}}
\newcommand{\downLineTwoOne}{
  \begin{picture}(0,0)(0,0)
    \put(0.5,-12){\tikz\draw[red,very thick] (0mm,0mm) -- (0mm,-14mm);}
\end{picture}}
\newcommand{\downLineThreeOne}{
  \begin{picture}(0,0)(0,0)
    \put(0.5,-32){\tikz\draw[red,very thick] (0mm,0mm) -- (0mm,-34mm);}
\end{picture}}
\newcommand{\downLineFourOne}{
  \begin{picture}(0,0)(0,0)
    \put(0.5,-12){\tikz\draw[red,very thick] (0mm,0mm) -- (0mm,-14mm);}
\end{picture}}
\newcommand{\downLineFiveOne}{
  \begin{picture}(0,0)(0,0)
    \put(0.5,-20){\tikz\draw[red,very thick] (0mm,0mm) -- (0mm,-22mm);}
\end{picture}}

\newcommand{\orLineOneOne}{
  \begin{picture}(0,0)(0,0)
    \put(0.5,1.5){\tikz\draw[red,very thick] (0mm,0mm) -- (11mm,0mm);}
\end{picture}}
\newcommand{\orLineTwoOne}{
  \begin{picture}(0,0)(0,0)
    \put(0.5,1.5){\tikz\draw[red,very thick] (0mm,0mm) -- (21mm,0mm);}
\end{picture}}
\newcommand{\orLineThreeOne}{
  \begin{picture}(0,0)(0,0)
    \put(0.5,1.5){\tikz\draw[red,very thick] (0mm,0mm) -- (11mm,0mm);}
\end{picture}}
\newcommand{\orLineFourOne}{
  \begin{picture}(0,0)(0,0)
    \put(0.5,1.5){\tikz\draw[red,very thick] (0mm,0mm) -- (47mm,0mm);}
\end{picture}}
\newcommand{\orLineFiveOne}{
  \begin{picture}(0,0)(0,0)
    \put(0.5,1){\tikz\draw[red,very thick] (0mm,0mm) -- (58mm,0mm);}
\end{picture}}

Following two examples. 
In the first  the polynomial $p(x)=1+x^2$ with $m=6$ for such
 polynomial $s(p)=2$ hence $s(p)=m/3$ and for condition $1$  a
 $6$-cycle exist. 

  \begin{equation*}
          \left[
            \begin{array}{cccccc}
              \downLineOneOne \orLineOneOne 1 & 0 & \downLineTwoOne 1 & 0 & 0 &0\\
              0 & 1 & 0 & 1 & 0 &0\\
              0 & 0 & \orLineOneOne 1 & 0 & \downLineTwoOne 1 &0\\
              0 & 0 & 0 & 1 & 0 &1\\
              \orLineTwoOne 1 & 0 & 0 & 0 & 1 &0\\
              0 & 1 & 0 & 0 & 0 &1
            \end{array}\right]
      \end{equation*}
In the second example the first  polynomial is $p^1(x)=1+x^2$ with $m=7$ for such
 polynomial $s(p^1)=2$  and the second is $p^2(x)=x+x^5$ for which
  $s(p^2)=4$ hence $s^2(p)=2s^1(p)$ and for condition $2$  a
 $6$-cycle exist. 
\begin{equation*}
\begin{array}{cc}
          \left[
            \begin{array}{ccccccc}
              0 & \downLineThreeOne \orLineThreeOne 1 & 0 &  \downLineFourOne 1 & 0 &0 &0\\
              0 & 0 & 1 &  0 & 1 &0 &0\\
              0 & 0 & 0 &  \orLineFourOne 1 & 0 &1 &0\\
              0 & 0 & 0 &  0 & 1 &0 &1\\
              1 & 0 & 0 &  0 & 0 &1 &0\\
              0 & \orLineFiveOne  1 & 0 &  0 & 0 &0 &1\\
              1 & 0 & 1 &  0 & 0 &0 &0
            \end{array}\right]
&
          \left[
            \begin{array}{ccccccc}
              0 & 1 & 0 &  0 & 0 &1 &0\\
              0 & 0 & 1 &  0 & 0 &0 &1\\
              1 & 0 & 0 &  \downLineFiveOne 1 & 0 &1 &0\\
              0 & 1 & 0 &  0 & 1 &0 &1\\
              0 & 0 & 1 &  0 & 0 &1 &0\\
              0 & 0 & 0 &  1 & 0 &0 &1\\
              1 & 0 & 0 &  0 & 1 &0 &0
            \end{array}\right]
\end{array}
\end{equation*}

\subsection{Conditions for the 8-cycles}\label{subsec:polysTOcycles-8}
The following lemma reduces the number of cycle configurations to be considered.

\begin{lemma} \label{lemma:2C=g8}
For any matrix $M \in \C_{m,\alpha,\beta,\gamma}$ s.t. at least two
weight-$2$ circulants lie  in the same row or column then the girth is
less or equal to $8$.
\begin{proof}
The case of two weight-$2$ circulants laying in the same d.r is
considered, the other case is the transpose of such case.
when two weight-$2$  circulants lie in the same d.r. The following cycle configuration always arises
\begin{eqnarray*}
      & \left|\begin{array}{cc}
            C^1-4 & C^2-4\\
           \end{array}\right|.
         \end{eqnarray*}

\begin{figure}[h!tb]
\begin{center}
\psfrag{a}{y} \psfrag{b}{u} \psfrag{c}{z}  \psfrag{c}{v} 
\psfrag{x}{x} \psfrag{t}{w} \psfrag{w}{t} \psfrag{l}{l}
\includegraphics[width=0.85\columnwidth]{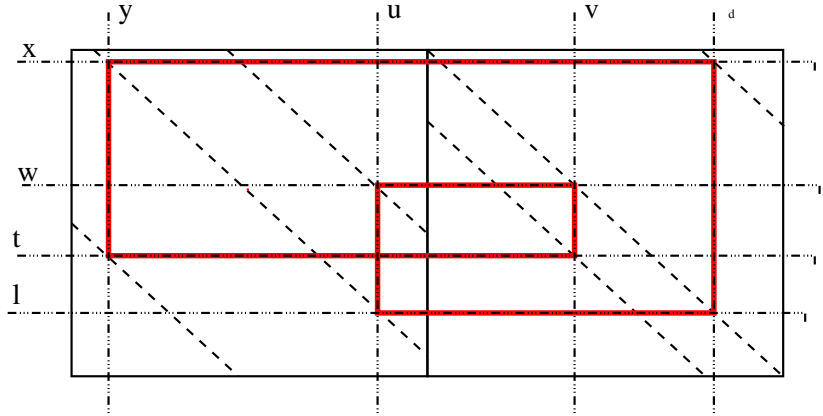}
\end{center}
\caption[$8$-cycle on two weight-$2$ circulants]{Example of $8$-cycle on two weight-$2$ circulants.}
\label{fig:circulant_8-1}
\end{figure}

It can be assumed that there is a  $8$-cycle if columns $y$ and $u$ lie in the first $C$ and columns  $v$ and $z$ lie in the second $C$.
Cycle column $y$ and cycle column $u$ satisfy
\begin{equation} \label{eq:lemma_g8-2-1}
x-t \equiv \pm s^1 \,,
\end{equation}
\begin{equation} \label{eq:lemma_g8-2-2}
w-l \equiv \pm s^1 \,.
\end{equation}

Cycle column $v$ and cycle column $z$ satisfy
\begin{equation} \label{eq:lemma_g8-2-3}
t-w \equiv \pm s^2 \,,
\end{equation}
\begin{equation} \label{eq:lemma_g8-2-4}
x-l \equiv \pm s^2 \,.
\end{equation}
Combining
equations~(\ref{eq:lemma_g8-2-1}),~(\ref{eq:lemma_g8-2-2}),(~\ref{eq:lemma_g8-2-3}),(~\ref{eq:lemma_g8-2-4}) 
an equation of the form $ \pm s^1\pm s^1 \pm s^2 \pm s^2\equiv 0$ is
obtained.
In particular consider the case  $s^1- s^1+ s^2+ s^2$
it is equal to $0$ independently to the values of the two separations.
\end{proof}
\end{lemma}
The previous lemma shows how any \HH matrix of a LDPC \QC~code may have girth bigger than 8
if and only if, 
or it contain only weight-$1$ circulants or the  weight-$2$ circulants
are not in the same 
row or column. Hence the lemma gives a practical guideline to the construction of \QC~codes with high girth.

The following theorem lists all the possible configurations that may
contain cycles of length $8$.
Beside each configuration the conditions on separations and exponents under which such cycles exist are given.
\newpage
\begin{theorem}\label{the:conditions8}
Let be $M \in \C_{m,\alpha,\beta,\gamma}$. The configurations in $M$
that may contain a cycles of length exactly $8$,  are the following 
\footnote{ Configurations with two or more weight-$2$ circulants in
the same row or column are  not listed since they always contain a cycle of at most $8$
(Lemma~\ref{lemma:2C=g8}), 
hence they do not add any new information.
}
\begin{enumerate}
    \item \label{item:g8-1}
          \begin{align*}
            &\left| C-8 \right|,& s(p)=m/4
          \end{align*}
    \item \label{item:g8-6}
    \begin{gather*}
            \left|\begin{array}{cc}
        C^1-5      & J^2-1 \\
        J^3-1 & \Delta^4-1 \\
       \end{array}\right|,\\
       \epsilon(p^1)-\epsilon(p^2)-\epsilon(p^3)+\epsilon(p^4)\equiv\pm 2s(p^1),
        \end{gather*}
    \item \label{item:g8-7}
    \begin{gather*}
            \left|\begin{array}{cc}
        C^1-4      & J^2-2 \\
        0        & \C^3-2 \\
       \end{array}\right|,\\
       \pm s(p^3) \equiv 2s(p^1) 
        \end{gather*}
    \item\label{item:g8-10}
    	\begin{gather*}
            \left|\begin{array}{cc}
        C^1-3      & J^2-1 \\
        J^3-1 & C^4-3 \\
       \end{array}\right|,\\
\epsilon(p^1)-\epsilon(p^2)-\epsilon(p^3)+\epsilon(p^4)\equiv \pm s(p^1) \pm s(p^3),
        \end{gather*}
    \item\label{item:g8-11}
     \begin{gather*}	
        \left|\begin{array}{cc}
        \Delta^1-2      & \Delta^2-2 \\
        \Delta^3-2 & \Delta^4-2 \\
       \end{array}\right|,\\
\epsilon(p^1)+\epsilon(p^1)-\epsilon(p^2)-\epsilon(p^2)-\epsilon(p^3)-\epsilon(p^3)+\epsilon(p^4)+\epsilon(p^4)\equiv 0\,,
\end{gather*}
\item \label{item:g8-13}
\begin{gather*}
            \left|\begin{array}{ccc}
        C^1-4 & J^2-1 & J^3-1 \\
         O  & \Delta^4-1    & \Delta^5-1 \\
       \end{array}\right|,\\
\epsilon(p^2)-\epsilon(p^3)-\epsilon(p^4)+\epsilon(p^5)\equiv \pm 2s(p^1),
    \end{gather*}
\item \label{item:g8-15}
\begin{gather*}
            \left|\begin{array}{ccc}
        C^1-3     & O   & J^2-1 \\
      J^3-1  &  C^4-2    & J^5-1 \\
       \end{array}\right|,\\
\epsilon(p^1)-\epsilon(p^2)-\epsilon(p^3)+\epsilon(p^5)\equiv \pm s(p^1) \pm s(p^4),
    \end{gather*}
\item \label{item:g8-17}
\begin{gather*}
            \left|\begin{array}{ccc}
       \Delta^1-2 & \Delta^2-1   & \Delta^3-1 \\
       \Delta^4-2 & \Delta^5-1   & \Delta^6-1 \\
       \end{array}\right|,\\
\epsilon(p^1)+\epsilon(p^1)-\epsilon(p^2)-\epsilon(p^3)-\epsilon(p^4)-\epsilon(p^4)+\epsilon(p^5)+\epsilon(p^6)=0\,,
\mbox{ or }\\
\epsilon(p^2)-\epsilon(p^3)+\ep(p^4)-\ep(p^4)-\epsilon(p^5)+\epsilon(p^6)=\pm s(p^1)\,,
\mbox{ or }\\
\epsilon(p^2)-\epsilon(p^3)+\ep(p^1)-\ep(p^1)-\epsilon(p^5)+\epsilon(p^6)=\pm s(p^4)\,.
    \end{gather*}
\item \label{item:g8-18}
\begin{gather*}
            \left|\begin{array}{ccc}
       C^1-2 & J^2-2   &  O\\
       O        & J^3-2   & C^4-2 \\
       \end{array}\right|,\\
  s(p^1)\equiv  s(p^4),
    \end{gather*}
\item \label{item:g8-20}
\begin{gather*}
            \left|\begin{array}{cccc}
       C^1-2 & O  & J^3-1 & J^4-1 \\
         O   &  C^2-2   & J^5-1 & J^6-1 \\
       \end{array}\right|,\\
\epsilon(p^3)-\epsilon(p^4)-\epsilon(p^5)+\epsilon(p^6)\equiv \pm s(p^1) \pm s(p^2),
    \end{gather*}
\item \label{item:g8-21}
\begin{gather*}
            \left|\begin{array}{cccc}
       \Delta^1-1 & \Delta^2-1  & \Delta^3-1 & \Delta^4-1 \\
       \Delta^5-1 & \Delta^6-1  & \Delta^7-1 & \Delta^8-1 \\
       \end{array}\right|,\\
\epsilon(p^1)+\epsilon(p^2)-\epsilon(p^3)-\epsilon(p^4)-\epsilon(p^5)-\epsilon(p^6)+\epsilon(p^7)+\epsilon(p^8)=0\,, \mbox{ or}\\
\epsilon(p^1)-\epsilon(p^2)+\epsilon(p^3)-\epsilon(p^4)-\epsilon(p^5)+\epsilon(p^6)-\epsilon(p^7)+\epsilon(p^8)=0\,, \mbox{ or}\\
\epsilon(p^1)-\epsilon(p^2)-\epsilon(p^3)+\epsilon(p^4)-\epsilon(p^5)+\epsilon(p^6)+\epsilon(p^7)-\epsilon(p^8)=0\,.
    \end{gather*}
    \item \label{item:g8-22}
    \begin{gather*}
            \left|\begin{array}{ccc}
           C^1-3 & J^2-1 &    O     \\
           J^3-1 &    O     & \Delta^4-1 \\
            O    & \Delta^5-1 & \Delta^6-1 \\
       \end{array}\right|,\\
\epsilon(p^1)-\epsilon(p^2)-\epsilon(p^3)+\epsilon(p^4)+\epsilon(p^5)-\epsilon(p^6)\equiv \pm s(p^1),
    \end{gather*}
    \item \label{item:g8-23}
    \begin{gather*}
            \left|\begin{array}{ccc}
          J^1-2 & \Delta^2-1 &    \Delta^3-1     \\
        C^4-2 &    O     & O \\
            O    & \Delta^5-1 & \Delta^6-1 \\
       \end{array}\right|,\\
\epsilon(p^2)-\epsilon(p^3)-\epsilon(p^5)+\epsilon(p^6)\equiv \pm s(p^4)
    \end{gather*}

    \item \label{item:g8-24}
    \begin{gather*}
            \left|\begin{array}{ccc}
        \Delta^1-2 & \Delta^2-1 & \Delta^3-1     \\
        \Delta^4-1 & \Delta^5-1 & O \\
        \Delta^6-1 &  O       & \Delta^7-1 \\
       \end{array}\right|,\\
\epsilon(p^1)+\epsilon(p^1)-\epsilon(p^2)-\epsilon(p^3)-\epsilon(p^4)+\epsilon(p^5)-\epsilon(p^6)+\epsilon(p^7)=0\,, \mbox{ or}\\
\epsilon(p^2)-\epsilon(p^3)+\epsilon(p^4)-\epsilon(p^5)-\epsilon(p^6)+\epsilon(p^7)=\pm s(p^1),
    \end{gather*}
 \item \label{item:g8-25}
 \begin{gather*}
            \left|\begin{array}{ccc}
            C^1-2 &    O     & O \\
            J^2-1 & \Delta^3-1 & O\\
            J^4-1 & J^5-1 &    \C^6-2     \\
       \end{array}\right|,\\
\epsilon(p^2)-\epsilon(p^3)-\epsilon(p^4)+\epsilon(p^5 )\equiv \pm s(p^1) \pm s(p^6),
    \end{gather*}
\item \label{item:g8-26-a}
    \begin{gather*}
            \left|\begin{array}{cccc}
        \Delta^1-2 & \Delta^2-1 & \Delta^3-1 & \Delta^4-1\\
        \Delta^5-1 & \Delta^6-1 &     O    & 0 \\
           O     &   O      & \Delta^7-1 & \Delta^8-1\\
       \end{array}\right|,\\
\epsilon(p^2)-\epsilon(p^3)-\epsilon(p^4)+\epsilon(p^5)+\epsilon(p^6)-\epsilon(p^7)=\pm s(p^1),
    \end{gather*}
\item \label{item:g8-26}
    \begin{gather*}
            \left|\begin{array}{cccc}
        C^1-2 & J^2-1 & J^3-1 & O\\
           O     & \Delta^4-1 &     O    & \Delta^5-1\\
           O     &   O      & \Delta^6-1 & \Delta^7-1\\
       \end{array}\right|,\\
\epsilon(p^2)-\epsilon(p^3)-\epsilon(p^4)+\epsilon(p^5)+\epsilon(p^6)-\epsilon(p^7)=\pm s(p^1),
    \end{gather*}
    \item \label{item:g8-27}
    \begin{gather*}
            \left|\begin{array}{ccccc}
        \Delta^1-1 & \Delta^2-1 &     O    & O\\
        \Delta^3-1 &    O     & \Delta^4-1 & O\\
           O     & \Delta^5-1 &     O    & \Delta^6-1\\
           O     &   O      & \Delta^7-1 & \Delta^8-1\\
       \end{array}\right|,\\
\epsilon(p^1)-\epsilon(p^2)-\epsilon(p^3)+\epsilon(p^4)+\epsilon(p^5)-\epsilon(p^6)-\epsilon(p^7)+\epsilon(p^8)=0,
    \end{gather*}
\end{enumerate}
\begin{proof}
Proving this theorem is particularly long, the arguments used are
similar to the ones used to prove the two previous theorems.
The detailed proof can be found in Appendix.

\end{proof}
\end{theorem}

The theorems above listed all the conditions (regarding the exponents and separations)
that must be satisfied for a cycle of length less than ten to exist
on a generic  \QC~\HH matrix.
Checking all the conditions listed in a \HH matrix can be a daunting task.
The complexity raises from the fact that  for every configuration all the possible
row and column permutations must be considered.
The following sections present several classes of \QC~codes that have
some regularity and  reduce the total number of conditions needed.



\section{A first subclass of codes}\label{sec:oneH_oneI}
In this section a first family of \QC~matrices is presented. 
The structure of this class has been developed to achieve high
sparsity and to ease implementation aspects.
This class of codes contain the matrices $M$ in which each d.c. can have at most one $H$ and/or one
$I$, where $I$ is the identity matrix.
Notation  $D_{m,\alpha ,\beta ,\gamma}$ denotes  such a class. 
Theorems~\ref{the:gencase4},~\ref{the:gencase6} and~\ref{the:gencase8}
can be applied  to $D_{m,\alpha,\beta ,\gamma}$ .
The use of $I$ sub-matrices reduces the number of variables present in
the conditions to be check.

Theorem~\ref{the:condition4},~\ref{the:condition6},~\ref{the:conditions8}
are applied to  $D_{m,\alpha ,\beta,\gamma}$.
\begin{proposition}\label{prop:oneH_oneI_cycles}
Let $ M \in D_{m,\alpha,\beta,\gamma}$. The d.m.'s in $M$ that contain cycles of length $<8$ and the associate conditions are
the following:
\begin{enumerate}
\item Containing $4$-cycles:
    \begin{doubleenum}
    \item \begin{align*}
            &\left| H-4 \right|, & s(p)=m/2
          \end{align*}
    \item \begin{align*}
            &\left| H^1-2 \quad H^2-2 \right|, & s(p^1)=s(p^2)
          \end{align*}
    \item \begin{align*}
        &\left|\begin{array}{cc}
            H^1-1 & I-1\\
            I-1 & H^2-1 \\
           \end{array}\right|, & \epsilon(p^1)+\epsilon(p^2) \equiv 0
         \end{align*}
    \item \begin{align*}
        &\left|\begin{array}{cc}
            H^1-1 & H^2-1\\
            I-1 & I-1 \\
           \end{array}\right|, & \epsilon(p^1) - \epsilon(p^2) \equiv 0
         \end{align*}

    \end{doubleenum}
\item Containing $6$-cycles:
    \begin{doubleenum}
    \item \begin{align*}
            &\left| H-6 \right|,& s(p)=m/3
          \end{align*}
    \item \begin{align*}
            &\left| H^1-4 \quad H^2-2 \right|, &  s(p^2)\equiv \pm2s(p^1)
           \end{align*}
    \item \begin{align*}
            &\left| H^1-2 \quad H^2-2 \quad H^3-2  \right|, &
    s(p^1)\pm s(p^2)\pm  s(p^3) &\equiv 0
          \end{align*}
    \item\begin{align*}
        &\left|\begin{array}{cc}
        H^1-2 & I-2 \\
         O  & H^2-2 \\
       \end{array}\right|,  & s(p^1)=s(p^2)
    \end{align*}
    \item\begin{align*}
    &\left|\begin{array}{cc}
       H^1-3 & H^2-1\\
        I-1 & I-1 \\
       \end{array}\right|,& \epsilon(p^1)-\epsilon(p^2) \equiv \pm s(p^1)
    \end{align*}
    \item\begin{align*}
    &\left|\begin{array}{cc}
       H^1-3 & I-1\\
        I-1 & H^2-1 \\
       \end{array}\right|, & \epsilon(p^1)+\epsilon(p^2) \equiv \pm s(p^1)
    \end{align*}
    \item \begin{align*}
    &\left|\begin{array}{ccc}
        H^1-2 & H^2-1 & H^3-1 \\
         O  &   I-1    & I-1 \\
       \end{array}\right|&  \epsilon(p^2)-\epsilon(p^3) \equiv \pm s(p^1)
    \end{align*}
    \item\begin{align*}
    &\left|\begin{array}{ccc}
        H^1-2 &    H^2-1   & I-1 \\
         O  & I-1 & H^3-1 \\
       \end{array}\right|, & \epsilon(p^2)+\epsilon(p^3) \equiv \pm s(p^1)
        \end{align*}
    \item\begin{align*}
    &\left|\begin{array}{ccc}
        H^1-2 &    I-1   & I-1 \\
         O  & H^2-1 & H^3-1 \\
       \end{array}\right|&  \epsilon(p^2)-\epsilon(p^3) \equiv \pm s(p^1)
      \end{align*}
    \item \begin{align*}
    &\left|\begin{array}{ccc}
        H^1-1 & H^2-1 &    O     \\
        I-1 &    O  & H^3-1 \\
        O   & I-1   & I-1 \\
       \end{array}\right|&  \epsilon(p^1)-\epsilon(p^2)+\epsilon(p^3) \equiv 0
    \end{align*}

    \item \begin{align*}
    &\left|\begin{array}{ccc}
        H^1-1 & I-1 &    O     \\
        I-1 &    O  & H^3-1 \\
        O   & H^2-1   & I-1 \\
       \end{array}\right|&  \epsilon(p^1)+\epsilon(p^2)+\epsilon(p^3) \equiv 0
    \end{align*}

\end{doubleenum}
\end{enumerate}

\begin{proof}
 The theorem can be demonstrated using a case by case analysis of the
 configurations presented in previous sessions.
\end{proof}
\end{proposition}

\section{Bresnan Codes}\label{sec:Bresnan_code}
A variation of the previous class of codes that 
first appeared in~\cite{BresnanThesis:04} and later studied by
mean of \GR~ bases in~\cite{Marta_tesi:04,Sala:05}, is investigated.
This class is  being used to reduce hardware complexity of 
the decoder in~\cite{Christian_ECCTD:05}.
Such  class of codes is here referred to with the name ''Bresnan
codes'', they consist of two $[ \alpha \times \alpha ]$ square blocks of circulants. Each square
block has weight-$2$ circulants in the main diagonal and identities in
another diagonal.
The structure of the parity check matrix allows the codes to be  $(3,6)$-regular LDPC codes and reduces the number of conditions that need to be
satisfied to have girth $8$.

\begin{definition}[Bresnan codes]\label{def:ourH}
Let $\alpha,m$ be positive integers such that $\alpha,m \geq 4$.
Then $\mathcal H_{m,\alpha}$ denotes the class of the
$( m \alpha \times 2m\alpha )$
matrices of the form
$$
H=\left[\begin{array}{ccccc|ccccc}
H^1_1 & 0 & \ldots & 0 & I & H^2_1 & I & 0 & \ldots & 0  \\
I & H^1_2 & 0 & \ldots & 0 & 0 & H^2_2 & I &  & \vdots  \\
0 & I & \ddots  &\ddots  & \vdots  & \vdots & 0 & \ddots & \ddots & 0 \\
\vdots &  & \ddots & \ddots & 0 & 0 &  &\ddots & \ddots & I  \\
0  & \ldots & 0 & I & H^1_\alpha & I & 0 & \ldots  & 0 & H^2_\alpha \\
\end{array} \right].
$$
where every $H^c_i$, with $c \in \{1,2\}$ and $i \in \{1,2,..,\alpha\}$,
is an $m\times m$ binary weight-$2$ circulant matrix and $I$ is
the $m\times m$ identity matrix.
Given any matrix $H\in \mathcal H_{m,\alpha}$,
$p_i^c$ denotes the polynomial of $H_i^c$
\end{definition}
%
The following fact is obvious.
\begin{fact}
Let $\alpha ,m$ be positive integers such that $\alpha,m \geq 4$,
then
$$
           \mathcal H_{m,\alpha}=\C_{m,\alpha,2,1}
$$
\end{fact}

The following theorem, presented in \cite{Marta_preprint:05} and later
generalized in~\cite{martinaLDPC}, gives a condition to ensure
that \HH has full rank.
\begin{proposition} \label{prop:gcd}
Let $\alpha ,m$ be positive integers such that $\alpha,m \geq 4$.
For any matrix $H$ in ${\mathcal H}_{m,\alpha}$,
let $C$ be the $[N,K,d]$ quasi-cyclic code with parity-check matrix $H$.
Suppose
$$\gcd(1+\PPP,x^m+1)=1$$
Then $N=2K$.
\end{proposition}

To further simplify the notation, the sentence ``d.s. $A$
is under d.s. $B$'' means that both $A$ and $B$ belong to the same
d.c. but that the d.r. of $A$ is d.r. $i_A$ and the d.r. of $B$ is
d.r. $i_B$, with $i_A > i_B$ is used. Similarly for ``$A$ is at the right of $B$''
and other intuitive positional expressions.

The Bresnan Codes are a subclass of the class presented in
section~\ref{sec:oneH_oneI} and it is possible to apply theorem~\ref{prop:oneH_oneI_cycles} to it.
It is clear that any configuration
presented in the theorems is possible except for 
2.8, 2.10-2.14. Moreover for configurations  2.9
only these special cases can arise:
\begin{enumerate}
    \item \begin{align*}
      &\left|\begin{array}{ccc} 
             H^1_j-1     &      0      &   I^2_j-1    \\
         I^1_{j+1}-1 & H^1_{j+1}-2 & H^2_{j+1}-1
             \end{array}\right|,\,
             j  \in \{1,..,\alpha-1\}, & \epsilon(p^1_j)+\epsilon(p^2_{j+1})\equiv \pm s(p^1_{j+1})
           \end{align*}
    \item \begin{align*}
      &\left|\begin{array}{ccc} 
             H^1_j-1     & H^2_j-2 & I^2_j-1    \\
         I^1_{j+1}-1 &   0     & H^2_{j+1}-1
             \end{array}\right|,\,
             j  \in \{1,..,\alpha-1\}, & \epsilon(p^1_j)+\epsilon(p^2_{j+1})\equiv \pm s(p^2_j)
    \end{align*}
    \item \begin{align*}
      &\left|\begin{array}{ccc}  
             H_1^1-2 & I_1^1 -1     & H_1^2-1      \\
            0    & H_\alpha^1-1 & I_\alpha^2-1
       \end{array}\right|, & \epsilon(p^1_\alpha)+\epsilon(p^2_1)\equiv \pm s(p^1_1)
        \end{align*}
    \item \begin{align*}
      &\left|\begin{array}{ccc}  
             I_1^1 -1     & H_1^2-1      & 0             \\
         H_\alpha^1-1 & I_\alpha^2-1 & H^2_\alpha-2  \\
       \end{array}\right|, & \epsilon(p^1_\alpha)+\epsilon(p^2_1)\equiv \pm s(p^2_\alpha).
       \end{align*}

\end{enumerate}

The following important theorem provides a {\em complete}
characterization for cycles of length $<8$ in $\mathcal
H_{m,\alpha}$. 
A definition is needed in order to present compact statements.

\begin{definition}
Let $\alpha\geq 4$ then  $J_\alpha$ denotes the sub-set of $\NNin^2$
such that $(i,j)\in J_\alpha$ if and only if
$i,j\in \{1,\ldots,\alpha\}$, for $j \equiv i+1 \mod(\alpha)$.
\end{definition}
%
%
%
\begin{proposition}
\label{prop:QC_Bresnan_conditions}
Let $\alpha,m \geq 4$ and $ M \in \mathcal H_{m,\alpha}$.
\begin{itemize}
\item
There is a cycle of length 4 in $M$ {\bf if and only if} at least
one of the next conditions holds:
 \begin{enumerate}

    \item $m$ is even and there is $1\leq i\leq \alpha$ and $c\in\{1,2\}$
      such that
      $$s(p_i^{c}) = \frac{m}{2} \,, $$
    \item there is $1\leq i\leq \alpha$ s.t.
      $$s(p_i^1)=s(p_i^2) \,.$$
    \item there is $(i,j)\in J_\alpha$ s. t.
      $$\ep(p_i^1) + \ep(p_j^2) \equiv 0  \,.$$
    \end{enumerate}

\item There is a cycle of length 6 in $M$ {\bf if and only if} at least
one of the next conditions holds:
 \begin{enumerate}

\item $m$ is divisible by $3$ and there is
      $1\leq i\leq \alpha$ and $c\in\{1,2\}$
      such that
      $$s(p_i^{c}) = \frac{m}{3} \,,$$
\item there is $1\leq i\leq \alpha$ s.t.
      $$
      s(p_i^2)\equiv \pm 2 s(p_i^1), \quad \mbox{or} \quad
      s(p_i^1)\equiv \pm 2 s(p_i^2) \,,
      $$
\item there is $(i,j)\in J_\alpha$ s. t.
      $$
      s(p^1_i) = s(p^1_j)  \,,
      $$
\item there is $(i,j)\in J_\alpha$ s. t.
      $$
      s(p^2_i) = s(p^2_j)  \,,
      $$
\item there is $(i,j)\in J_\alpha$ s. t.
      $$
      s(p^1_i) = s(p^2_j)  \,,
      $$
\item there is $(i,j)\in J_\alpha$ s. t.
      $$
      \ep(p^1_i)+\ep(p^2_j) \equiv \pm s(p^1_i)  \,,
      $$
\item there is $(i,j)\in J_\alpha$ s. t.
      $$
      \ep(p^1_i)+\ep(p^2_j) \equiv \pm s(p^2_i)  \,,
      $$
\item there is $(i,j)\in J_\alpha$ s. t.
      $$
      \ep(p^1_i)+\ep(p^2_j) \equiv \pm s(p^1_j)  \,,
      $$
\item there is $(i,j)\in J_\alpha$ s. t.
      $$
      \ep(p^1_i)+\ep(p^2_j) \equiv \pm s(p^2_j)    \,.
      $$
\end{enumerate}
\end{itemize}
\end{proposition}
At this stage, it is finally possible to write down the theorem that
allows the construct of Bresnan Codes with girth at least $8$.
\begin{theorem}
\label{the:bresnan_final}
Let $\alpha,m \geq 4$ and $ M \in \mathcal H_{m,\alpha}$.
Let $g$ be the girth of the Tanner graph of $M$.
Then $g\geq 8$ {\bf if and only if} all the following conditions hold:
\begin{enumerate}
\item for any
      $1\leq i\leq \alpha$ and $c\in\{1,2\}$
\begin{align*}
     &s(p_i^{c}) \not= \frac{m}{3},
     &s(p_i^{c}) \not= \frac{m}{2},
\end{align*}

\item for any $1\leq i\leq \alpha$,
$$
\begin{array}{ccc}
    s(p_i^2) \not= s(p_i^1),
    &s(p_i^2) \not \equiv \pm 2 s(p_i^1),
    &s(p_i^1) \not \equiv \pm 2 s(p_i^2),
\end{array}
$$
\item for any $(i,j)\in J_\alpha$,
$$
\begin{array}{cccc}
       s(p^1_i) \not = s(p^1_j), &s(p^2_i) \not = s(p^2_j), &s(p^1_i) \not = s(p^2_j),
 \end{array}
$$
$$
      \ep(p_i^1) + \ep(p_j^2) \not\equiv 0 ,
$$
$$
\begin{array}{cc}
      \ep(p^1_i)+\ep(p^2_j) \not\equiv \pm s(p^1_i), &\ep(p^1_i)+\ep(p^2_j) \not\equiv \pm s(p^2_i),\\
      \ep(p^1_i)+\ep(p^2_j) \not\equiv \pm s(p^1_j), &\ep(p^1_i)+\ep(p^2_j) \not\equiv \pm s(p^2_j).
 \end{array}
$$

\end{enumerate}
\end{theorem}

\subsection{Existence of Bresnan Codes}\label{subsec:search}
An exhaustive search has been carried out to show that collections of
circulants that satisfy Theorem~\ref{the:bresnan_final} exist.
First some value of $m$ can be excluded because they cannot have solutions.

\begin{proposition}
Let $\alpha\geq 4$ and $ M \in \mathcal H_{m,\alpha}$.
There does not exist any $ M \in \mathcal H_{m,\alpha}$ with $m \leq
10$ such that $g\geq 8$.
\begin{proof}
The aim of the following proof is to show that  with $m \leq 10$ the
number of conditions is too big to allow any solution.
The conditions related to the sum of exponent in~\ref{the:bresnan_final} will be considered.
For sake of simplicity the various separations are called 
$s(p_i^1)=a$, $s(p_i^2)=b$, $s(p_j^1)=c$, $s(p_j^2)=d$,
note  also that the conditions of Theorem~\ref{the:bresnan_final} imply
 that $\{a,b,c\}$ are different and that  $\{a,d,c\}$ are different but they allow $b=d$.
The second exponent is written as the sum between the first
and the separation 
(\emph{i.e. $e_i^1(2)=e_i^1(1)+a$}),
with this notation it is possible to write the condition 
$e^1_i(2)+e^2_j(1) \neq b,$ as $e^1_i(1)+e^2_j(1) \neq b-a$,
The list of all conditions related to the sum of exponents in theorem~\ref{the:bresnan_final}, using
the new notation, is the follow:
\begin{equation}\label{eq:list_condition}  
\begin{split}
	e^1_i(1)+e^2_j(1) &\neq 0,a,b,c,d,\\
	e^1_i(1)+e^2_j(1) &\neq -a,-c,-b,-d,\\
	e^1_i(1)+e^2_j(1) &\neq b-a,c-a,d-a,-2a,-b-a,-c-a,-d-a\\
	e^1_i(1)+e^2_j(1) &\neq a-c,b-c,d-c,-a-c,-b-c,-2c,-d-c\\
	e^1_i(1)+e^2_j(1) &\neq b-a-c,c-a-c,d-a-c,\\
	e^1_i(1)+e^2_j(1) &\neq -2a-c,-b-a-c,-2c-a,-d-a-c
\end{split}
\end{equation}
 
The conditions are symmetrical for $a$ and $c$ so it can be supposed,
 without loss of generality, that $a<c$.

The case with $b \neq d$ is considered first.
Under the condition that $\{a,b,c,d\}$ are different and considering
that the separations are less than $m/2$, the first two rows
 of the previous list form nine independent conditions that must be
 satisfied.
 They can be satisfied if and only if $m > 9$.
 To prove that there are no solution for $ m = 10$ 
it is necessary to  find another condition that is independent from
 the previous nine.
Condition $e^1_i(1)+e^2_j(1) \neq b-a$ is chosen, this new condition is
 independent from the previous nine if $b-a$ 
is different from $\{0,a,b,c,d,-a,-b,-c,-d\}$.
 The following investigate what happen if $b-a$ is equal to any of these.
$$
\begin{array}{ll}
      b-a=0 => b=a \mbox{ Not possible} \\
      b-a=a => b=2a \mbox{ Not possible} &   b-a=-a => b=0 \mbox{ Not possible}   \\
      b-a=b => a=0 \mbox{ Not possible} &  b-a=-b => a=2b \mbox{ Not possible}  \\
      b-a=c => b=a+c &  b-a=-c => a=c+b => a<c \mbox{ Not possible}\\
      b-a=d => b=d+a & b-a=-d => a=b+d\\  
 \end{array}
$$
Some cases have been marked impossible because, if they occur, they will
break conditions related to the separation, in
Theorem~\ref{the:bresnan_final}.
Such cases cannot give codes with girth higher than $8$ so there is no need to study if the
new condition is independent.
Next it is  analyzed what happen in the cases that cannot be
discharged, considering that with $m\leq10$ $\{a,b,c,d\} \leq 4$.
\begin{description}
\item[b=a+c] The conditions $a<c$ and $a\neq c$ imply that $b\geq3$.
	\begin{itemize}
	\item If $b=3$ then $a=1$ and $c=2$ but in this case
	$e^1_i(1)+e^2_j(1) \neq -2c-a $ is an independent condition,
	in fact the set $\{0,a,b,c,d,-a,-b,-c,-d\}$ became
	$\{0,1,3,2,4,9,7,8,6\}$ and $-2c-a= 5$. Note that $d$  must be
	$4$ to be different from $\{a,b,c\}$ and $d \leq m/2$.
	\item If $b=4$ then $a=1$ and $c=3$ (since $a\neq 2$) but in
	this case 
        $e^1_i(1)+e^2_j(1) \neq -2a-c $ is an independent condition,
	\end{itemize}
\item[b=d+a] The condition  $a\neq d$ implies that $b\geq3$.
	\begin{itemize}
	\item If $b=3$ then or $a=1$ and $d=2$ or $a=1$ and $d=2$
		\begin{itemize}
		\item if $a=1$ and $d=2$ then does not exist $c$ such that $c > a$, $c\neq d$, $c\neq b$ and $c\neq 2a$,
		\item if $a=2$ and $d=1$  then $c=4$ (since $c\neq b$)
	and $e^1_i(1)+e^2_j(1) \neq m-d-c $ is an independent
	condition.
        In fact with this values the conditions became $
	e^1_i(1)+e^2_j(1) \&\neq \{0,1,2,3,4,9,8,7,6\}$ but  $m-d-c =
	10-1-4=5 $
		\end{itemize}
	\item If $b=4$ then or $a=1$ and $d=3$ or $a=3$ and $d=1$ (since $a\neq d$)
		\begin{itemize}
		\item if $a=1$ and $d=3$  then $c=2$ (since $c\neq b$) in this case $e^1_i(1)+e^2_j(1) \neq m-d-c $ is still an independent condition,
		\item if $a=3$ and $d=1$  then does not exist $c$ such that $c > a$ and  $c\neq b$.
		\end{itemize}
	\end{itemize}
\item[a=b+d] Considering that $a<c<m/2$ then or $a=2$ or $a=3$.
	\begin{itemize}
	\item if $a=2$ then $b=d=1$ that is not possible since $b\neq 2a$.
	\item if $a=3$ then $d\neq 2$ since $c\neq 2d$ so $d=1$ and $b=2$, in this case $e^1_i(1)+e^2_j(1) \neq m-d-c $ is still an independent condition,
	\end{itemize}
\end{description}

It has been shown how if $ d\neq b$  it always exist a new independent
 condition.
Hence a group of ten conditions that must be satisfied exist, this implied that for  $ M \in \mathcal H_{m,\alpha}$ and
$b \neq d$, it is possible to have $g\geq 8$ if and only if $m>10$.

Next step is to find a group of ten independent condition in the case $b=d$.
If $b=d$ the first two rows of~\ref{eq:list_condition} reduce to seven
independent 
conditions so that they cannot be satisfied for $m<8$. 
To prove the proposition it is necessary to find three more independent conditions.

The condition $e^1_i(1)+e^2_j(1) \neq b-a$  is reconsidered when $b=d$.
Following the previous reasoning this is an independent condition if
$b-a$ is different from $\{0,a,b,c,-a,-b,-c\}$.

 The following studies what happen if $b-a$ is equal to any of these.
$$
\begin{array}{ll}\label{eq:list_condition2}  
      b-a=0 => b=a \mbox{ Not possible} \\
      b-a=a => b=2a \mbox{ Not possible} &   b-a=-a => b=0 \mbox{ Not possible}   \\
      b-a=b => a=0 \mbox{ Not possible} &  b-a=-b => a=2b \mbox{ Not possible}  \\
      b-a=c => b=a+c &  b-a=c => a=c+b => a<c \mbox{ Not possible}\\    
 \end{array}
$$

For the case $b=a+c$, equivalent to the cases ($a=1$,$b=3$,$c=2$) and
($a=1$,$b=4$,$c=3$), it is easy to verify that condition $-2a-c$ is
an independent condition for $m \leq 10$.
Note how for $m=8$ the case $a=1$,$b=4$,$c=3$ is not allowed since
$b=m/2$.

%

Hence $e^1_i(1)+e^2_j(1) \neq b-a$ (or $e^1_i(1)+e^2_j(1) \neq m-2a-c$
in the case $b=a+c$)  is a new independent condition for $m\leq10$.
A set of eight independent conditions for $m\leq 10$ has been
obtained, another two are still needed.
The two conditions needed must be independent  for $m=\{9,10\}$ but not
necessary for $m=8$, since it has been proved that for $m\leq8$ there
cannot be Bresnan codes with girth more than $8$.

Condition $e^1_i(1)+e^2_j(1) \neq b-c$ is considered next,
 this new condition is independent from the previous eight if $d-c$ is different from $\{0,a,b,c,-a,-b,-c,b-a\}$.
Note the case $b=a+c$ for which the eight conditions are
 $\{0,a,b,c,-a,-b,-c,-2a-c\}$  ( and not $\{0,a,b,c,-a,-b,-c,b-a\}$) will be considered,
 together with other particular cases, at the end.

 The following considers what happen if $b-c$ is equal to any of these, considering that $d=b$
$$
\begin{array}{ll}
      b-c=0 => d=c \mbox{ Not possible}\\
      b-c=a => b=a+c &   b-c=-a => c=a+b \\
      b-c=b => c=0 \mbox{ Not possible} &  b-c=-b => c=2d \mbox{ Not possible}  \\
      b-c=c => d=2c \mbox{ Not possible} &  b-c=-c => b=0 \mbox{ Not possible} \\
      b-c=b-a => a=c \mbox{ Not possible} 
 \end{array}
$$
As mention above, the case $b=a+c$ will be treated apart, the only case to consider is when $c=a+b$ .
\begin{itemize}
\item $c=a+b$;  $m\leq10$ implies $c\leq4$
	\begin{itemize}
	\item if $c=2$ then $a=b$ that is not allow,
	\item if $c=3$ then $a=2b$ or $b=2a$ that is not allow,
	\item if $c=4$ then the only two possible situations are
	$a=1, b=3$ or $a=3, b=1$. These are not possible if $m=9$
	because one separation is equal to $m/3$.
        In the case $m=10$ $e^1_i(1)+e^2_j(1) \neq m-a-b-c$ is a new
	independent condition.
        \end{itemize}
\end{itemize}

This proves how $e^1_i(1)+e^2_j(1) \neq d-c$ is an independent
condition for $m\leq10$ apart 
for the cases ($a=1$,$b=3$,$c=4$) and  ($a=3$,$b=1$,$c=4$) in which case $e^1_i(1)+e^2_j(1) \neq -a-b-c$ is.

%

At this point it is possible to build a set of nine independent
conditions for $m\leq10$ .
Hence it is proved that for $m\leq 9$ there cannot be a code with girth$\geq8$. 
It is now necessary to prove the existence 
of at least another independent condition for $m=10$.
Condition  $e^1_i(1)+e^2_j(1) \neq b-a-c $ is now considered.
The situations when this new condition is dependent to
the condition $e^1_i(1)+e^2_j(1) \neq 0,a,b,c,-a,-b,-c,b-a,b-c$ are
studied next.
In other word situations when  $b-a-c$ is equal to any of
${0,a,b,c,-a,-b,-c,b-a,b-c}$ are considered .
The special cases ($a=1$,$b=3$,$c=4$) and ($a=3$,$b=1$,$c=4$) will be considered at the end.
$$
\begin{array}{cc}
      b-a-c=0 => b=a+c  & \\
      b-a-c=a => b=2a+c  &   b-a-c=-a => b=c  \mbox{ Not possible}   \\
      b-a-c=b => a=-c \mbox{ Not possible} &  b-a-c=-b => 2b=a+c  \\
      b-a-c=c => b=a+2c &  b-a-c=-c => b=a<c \mbox{ Not possible}\\        
 \end{array}
$$
The case $b=a+c$ will be treated a part and it is of no concern at the moment.
What happen in the cases that cannot be discharged is studied;
 remembering that with $m\leq10$.
\begin{itemize}
\item[b=2a+c] since $a<c$ and $b<5$  $b$ is $4$. This implies $a=1$ and $c=2$. This is not possible since $c=2b=2d$
\item[2b=a+c] since $a<c$ and $b<5$ it is :
	\begin{itemize}
	\item if $b=2$ then $a=1$ and $c=3$.  This is not possible since $b=2a$
	\item if $b=3$ then $a=2$ and $c=4$.  In this case $-b-a$ is a new independent condition
	\item if $b=4$ this is not possible since does not exist $a,c$ s.t. $a+c=8$ and $a<c\leq4$.
	\end{itemize}
		
\item[b=2c+a] this is impossible since $a<c$ implies $b=2c+a>4$
\end{itemize}
This proves how $e^1_i(1)+e^2_j(1) \neq b-a-c$ is an independent condition for $m\leq10$ apart for the case ($a=2$,$b=3$,$c=4$) in which case $e^1_i(1)+e^2_j(1) \neq -b-a$ is.

It has so been proved that the set of conditions\\
 $e^1_i(1)+e^2_j(1) \neq 0,a,b,c,-a,-b,-c,b-a,b-c,b-a-c$ \\
 (or in the case ($a=2$,$b=3$,$c=4$) the set $e^1_i(1)+e^2_j(1) \neq 0,a,b,c,-a,-b,-c,b-a,b-c,-b-a$) consist on a set of independent inequalities that must be satisfied, hence this required that $m>10$.

To prove the proposition it is necessary to consider the four special
cases that has been neglected : ($a=1$,$b=3$,$c=2$),
($a=1$,$b=4$,$c=3$), ($a=1$,$b=3$,$c=4$) and ($a=3$,$b=1$,$c=4$) and
demonstrate that for any of these cases 
exist three independent conditions to add to the basic set  $e^1_i(1)+e^2_j(1) \neq 0,a,b,c,-a,-b,-c$, when $m\leq10$.

These special cases do not need to be considered for $m=8$ since it has
 already been proved that there is no Bresnan codes with $g\geq 8$ in
 such case.
Moreover, in all four cases one of the separation is equal to three
 but this is not allow if $m=9$, 
hence any of this cases cannot happen if $m=9$.
 
 It is only necessary to consider these cases for $m=10$; every case is considered singularly.
 \begin{description}
\item[($a=1$, $b=3$, $c=2$)]It is straight forward to verify that the
following set of conditions are independent 
$\{0,a,b,c,-a,-b,-c,-2a-c,-a-b-c,-b-c\}$
\item[($a=1$, $b=4$, $c=3$)]  It is straight forward to verify that the
following set of conditions are independent
 $\{0,a,b,c,-a,-b,-c,-2a-c,-a-b-c,-2a\}$
\item[($a=1$, $b=3$, $c=4$)]It is straight forward to verify that the
 following set of conditions are independent
 $\{0,a,b,c,-a,-b,-c,b-a,-a-c,-2a\}$
\item[($a=3$, $b=1$, $c=4$)]It is straight forward to verify that the
 following set of conditions are independent
 $\{0,a,b,c,-a,-b,-c,b-a,-a-b-c,-b-c\}$
\end{description}
It has so been proved that for every $m\leq 10$  it is always possible to
find a set of $m$ independent values from whom the sum of exponents
must differ.
This implies that for  $m\leq 10$  it is never possible to meet such
conditions hence to find a Bresnan code with girth  $8$.
 \end{proof}
\end{proposition}
Table~\ref{tab:bresnan_num_sol} shows how many solutions exist for any
value of $\alpha$ and $m$.
The percentage given beside some values indicates that due to memory overflow the complete
search could not be completed. The given number is an estimate  based
on that percentage.
Table~\ref{tab:bresnan_timing} shows the time taken to find all the
possible codes with given $\alpha$ and $m$.
It can be seen that the number of Bresnan codes increases with the
dimension of the circulants ($m$) and with the dimension of the block ($\alpha$).

\begin{sidewaystable}[]
\renewcommand{\arraystretch}{1.3}
\caption[Bresnan codes search results]{Number Bresnan codes for given $\alpha$ and $m$} \label{tab:bresnan_num_sol}
\centering
\begin{tabular}{|*{8}{|c}||}
\hline
\hline
$m\setminus\alpha$&4 & 5 & 6 & 7 & 8 & 9 & 10 \\
\hline
\hline
11&0& 0& 0& 0& 0& 0& 0\\
\hline
12&0& 0& 0& 0& 0& 0& 0\\
\hline
13&6.1e3 &0& 1.1e6 & 5.7e6 &9.4e6& 7.5e8& 3.4e9(20\%)\\
\hline
14&3.7e3 &2.0e5& 1.1e7 &1.6e7& 8.5e8 &2.4e10(27\%) &1.1e11(3.8\%)\\
\hline
15&5.6e5 &6.9e6 &2.8e8 &1.3e10(82\%) &1.0e11(5.9\%)&&\\
\hline
16&2.2e6 &2.1e8 &1.2e10 &2.2e11(14\%)&&&\\
\hline
17&1.5e8 &2.2e10 &2.8e12&2.8e14(0.4\%)&&&\\
\hline
18&2.4e8 &4.2e10 &5.5e12&&&&\\
\hline
19&9.0e9 &3.2e12&1.4e15(0.3\%)&&&&\\
\hline
20&2.3e10&7.3e12(27\%)&&&&&\\
\hline
21&1.1e11&5.2e13(17\%)&&&&&\\
\hline
22&4.8e11&4e14(4.3\%)&&&&&\\
\hline
23&4.0e12&&&&&&\\
\hline
\end{tabular}
\end{sidewaystable}
\clearpage
\begin{table}[htb]
\renewcommand{\arraystretch}{1.3}
\caption[Bresnan codes search timing]{Timing (in seconds) for find all Bresnan codes for given $\alpha$ and $m$}\label{tab:bresnan_timing}
\centering
\begin{tabular}{|*{8}{|c}||}
\hline
\hline
$m\setminus\alpha$&4 & 5 & 6 & 7 & 8 & 9 & 10 \\
\hline
\hline
11& 0&  0&  0&  1&  1&  1&  1\\
\hline
12& 0&  0&  0&  1&  1&  1&  1\\
\hline
13& 0&  0&  1&  7&  70& 8.6e2&4.8e3\\
\hline
14 &0&  1&  5&  58& 1.1e3&1.0e4&6.4e4\\
\hline
15& 0&  5&  1.2e2&3.4e3 &7.1e4&&\\
\hline
16& 1&  61& 3.1e3&1.5e5&&&\\
\hline
17& 22& 2.9e3&3.7e5&4.5e7&&&\\
\hline
18& 15& 2.3e3&3.3e5&&&&\\
\hline
19& 4.5e2&1.6e5&8.3e7&&&&\\
\hline
20& 8.8e2&4.7e5&&&&&\\
\hline
21&2.1e3&1.2e6&&&&&\\
\hline
22&8.1e3&7.3e6
&&&&&\\
\hline
23&4.7e4&&&&&&\\
\hline

\hline
\end{tabular}
\end{table}
%
\subsection{Performance} \label{subsec:performance}
A study of the performances of the Bresnan codes when compared with
random like code is presented here.

All simulations presented in the remainder of this
 thesis  have been obtained using a maximum limit of $20$
 iterations. For every SNRdB point a minimum of 10 block errors has
 been found. Error bars are plotted on plots in this thesis. 
The error associated with the Frame Error rate is defined as (\cite{MacKayComparison:98}):
\begin{equation}
FER\pm=FER\times\exp\left(\sqrt{\frac{N-N_{error}}{N\times N_{error}}}\right),
\end{equation}
where $N$ is the number simulation trials and $N_{error}$ is the
number of frame errors recorded. 
This is to be considered an important aspect of the performance graphs
often overlooked.

\begin{figure}[p]
\begin{center}
\includegraphics[angle=-90,width=0.8\columnwidth]{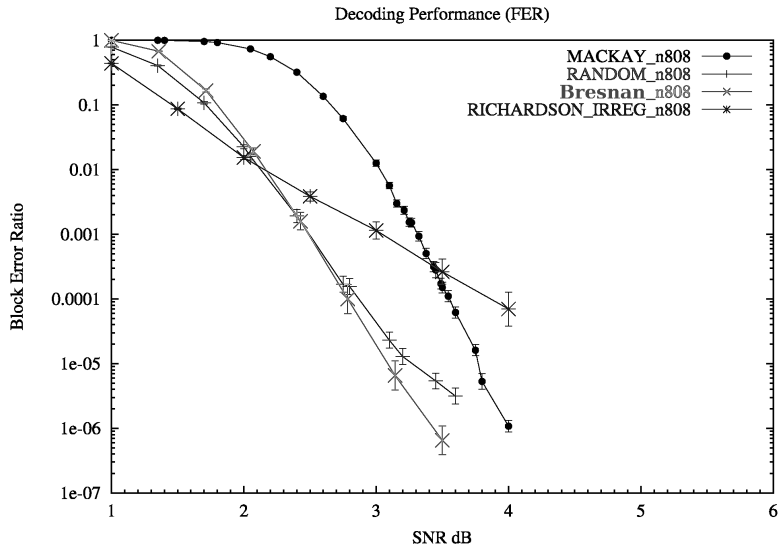}
\end{center}
\caption[Performance of a $N=808$ Bresnan codes]{Block error rate
performance for $N=808$ Bresnan code compared with three random like
codes of the same length obtained from MacKay, Richardson and random
permutations methods.}
\label{fig:comp_QC808_FER}
\end{figure}

The simulations show that the codes of the family here developed behave very well,
even for medium lengths.
First the result for a $N=808$ code.
In \figur~\ref{fig:comp_QC808_FER} a comparison
between one of the new codes and three codes taken from known good
families is provided.
All codes are LDPC codes, with rate $1/2$ and length $N=808$.
The new code is a regular code with a structure, while the others are
randomly constructed using some optimization method.
To be more precise, these three literature codes are:
\begin{itemize}
\item  A MacKay code with $N=808,\, K=404,\, d_v=3,\, d_c=6$  obtained with the
method presented in~\cite{MacKayComparison:98}; 

\item  A randomly constructed code with $N=808,\, K=404,\, d_v=3,\, d_c=6$
       this code has been obtained by running the optimized edge-placement
       algorithm for five passes (see for example~\cite{BresnanThesis:04});
\item  A Richardson irregular code with  $N=808,\, K=404,$ this
       code has been created using the optimized weight distributions  proposed by Richardson~\etal~\cite{Richardson&Shokrollahi&Urbanke01density,Richardson&Urbanke01capacity};
\end{itemize}


%

The performance of the new code is close to that of the random-edge
one, with the Random code hitting an error floor around $3 db$m $1e-05$.
The MACKAY code performs worse than the new codes losing $0.5 dB$ at $1e-06$.
%
The Richardson code has particularly poor performance, it is possible
that the code constructed is a bad representative  of the ensemble of
which it is part. It is known that for medium and shorter length codes
the performance of the single codes varies widely from the average
performance of the ensemble and from the performance of long codes for
the same construction.
 On the other hand the situation highlights the problem of finding good random
codes: creation and simulation of many codes are necessary to find a
good one. 

\begin{figure}[p]
\begin{center}
\includegraphics[angle=-90,width=0.8\columnwidth]{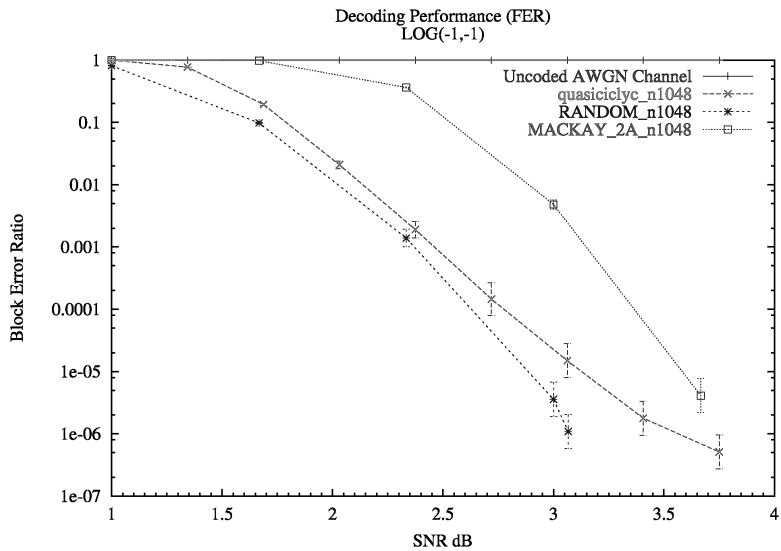}
\end{center}
\caption[Performance of a $N=1048$ Bresnan codes
 performance]{ Block error rate
performance for $N=1048$ Bresnan code compared with the a code
obtained with the  random permutation method  and one with  MacKay methods.} \label{f:comp_small_QC}
\end{figure}

Figure~\ref{f:comp_small_QC} shows
the performances comparison between $N=1048$ $R=1/2$ codes constructed with MacKay method
\cite{MacKayComparison:98}, Random code~\cite{Mao:AHSFGLDPCCASBL} and a \QC~code of this class.
It can be seen that the performance of the \QC~code is close to the
 one of the  random one,
 it is slightly worse when the code meets a noise floor. The performance of the MacKay code in the $1dB-4dB$ range is worse than the \QC~code.
For low values of SNR the \QC~code performance compares well
with the random code, and both are superior to the MacKay one. 

\begin{figure}[htbp]
\begin{center}
\includegraphics[angle=-90,width=0.8\columnwidth]{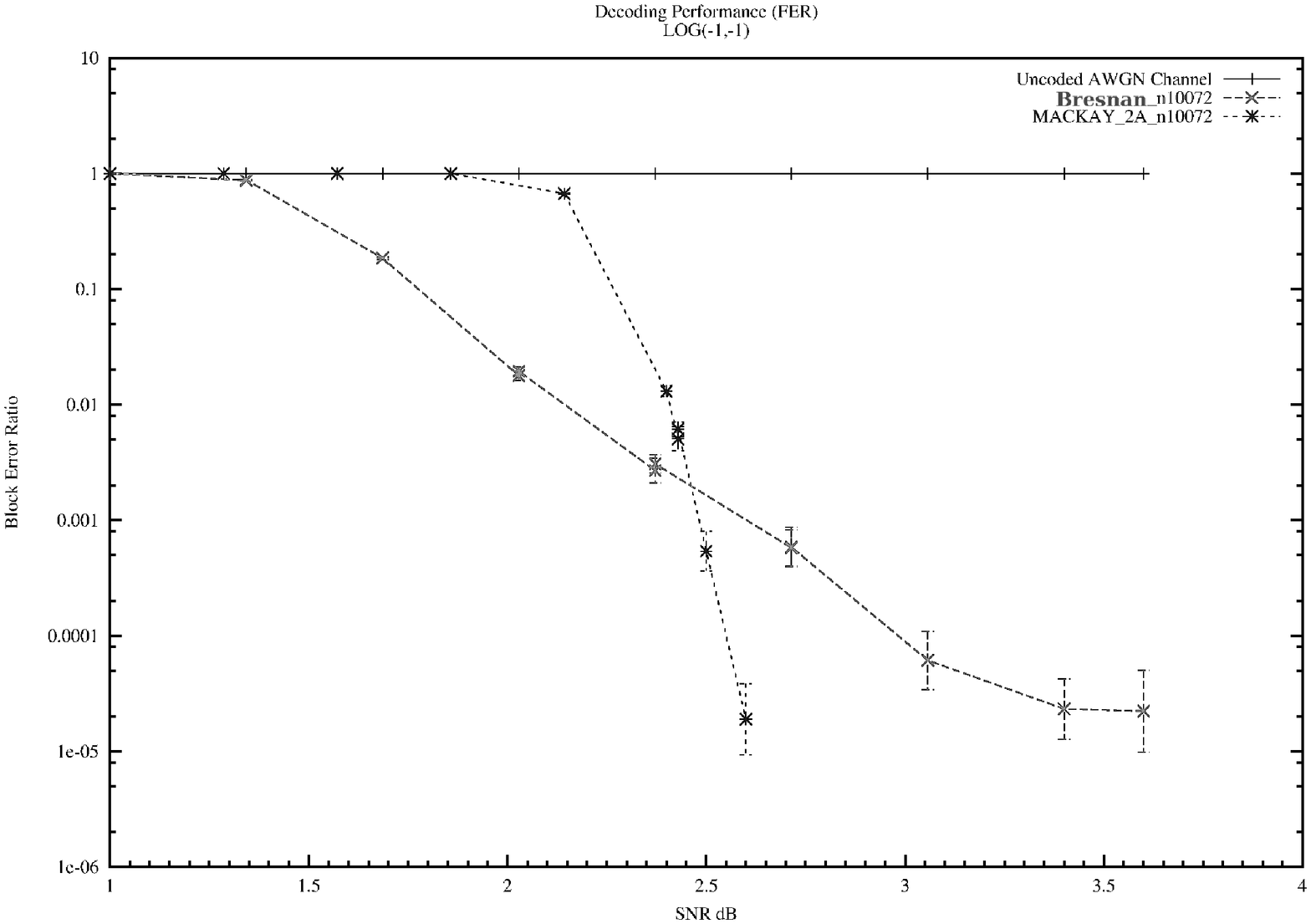}
\end{center}
\caption{Block error rate performance for $N\approx10,000$. A
Bresnan \QC~code and a MacKay code are compared.}
\label{fig:comp_QC10k_FER}
\end{figure}

Even if the main interest of this thesis is on medium length codes it
is interesting to evaluate the construction presented for longer codes
to have an idea of how such family performs in such case, for this
reason two code with  $ N\approx 10,000$, are presented.
In \figur~\ref{fig:comp_QC10k_FER} a Bresnan code is compared with a
MacKay code, it can be seen how the MacKay code has  a much better defined
waterfall region that make the code outperform the Bresnan code for high
SNR, this could be due to low minimum distance or to the presence of
trapping sets.
The result is in line with what is a know fact: analytic codes do not
perform as well as random codes for high length.
Moreover it proves that the MacKay codes for this
length start to behave as  it is theoretically proved for infinite
length.

Finally it is important to consider that  for the
concentration theorem~\cite{Richardson&Shokrollahi&Urbanke01density} it is more likely that a code taken at
random from the ensemble performs as well as the average performance
of the ensemble for this value of $N$. 
%
%
%
%
It is important to observe that the construction of good random codes,
such as the three confronted here, requires a consistent amount of computation,
while the new codes can be obtained immediately, thanks to their
simple structure.
\section{Extension of the Bresnan Codes}\label{sec:extendedBresnan}
In this section an approach to construct Bresnan-like codes
with rate higher than $1/2$ is presented that maintain girth $8$.

The Bresnan codes are formed by two square blocks of circulants, the
codes developed in this section are formed by three blocks. A rate of $2/3$ is
achievable in such way.
The method is expandable to more blocks  and  
any rate of the type $(l-1)/l$ can be achieved.

\begin{definition}[Rate $2/3$ Bresnan codes]\label{def:3_4ourH}
Let $\alpha ,m$ be positive integers such that $\alpha > 4, m\geq 3$.
$\mathbb H_{m,\alpha}$ denotes the class of the
$( m \alpha \times 3m\alpha )$
matrices of the form
$$
H=\left[\begin{array}{ccccc|ccccc|ccccc}
H^1_1 & 0 & \ldots & 0 & I & H^2_1 & I & 0 & \ldots & 0  & H^3_1 & 0 & I & \ldots & 0 \\
I & H^1_2 & 0 & \ldots & 0 & 0 & H^2_2 & I &  & \vdots   & 0  & H^3_2 & 0 & I & \ldots \\
0 & I & \ddots  &\ddots  & \vdots  & \vdots & 0 & \ddots & \ddots & 0 & \vdots  & 0 & \ddots & \ddots & \ddots \\
\vdots &  & \ddots & \ddots & 0 & 0 &  &\ddots & \ddots & I & \vdots &  & \ddots & \ddots & 0\\
0  & \ldots & 0 & I & H^1_\alpha & I & 0 & \ldots  & 0 & H^2_\alpha & 0 & I & \ldots  & 0 & H^3_\alpha\\
\end{array} \right].
$$
where every $H^c_i$, with $c \in \{1,2\}$ and $i \in \{1,2,..,\alpha\}$,
is an $m\times m$ binary weight-$2$ circulant matrix and $I$ is
the $m\times m$ identity matrix.
\end{definition}

\begin{figure}[htbp]
    \centering
    \includegraphics[angle=-90,width=0.8\columnwidth]{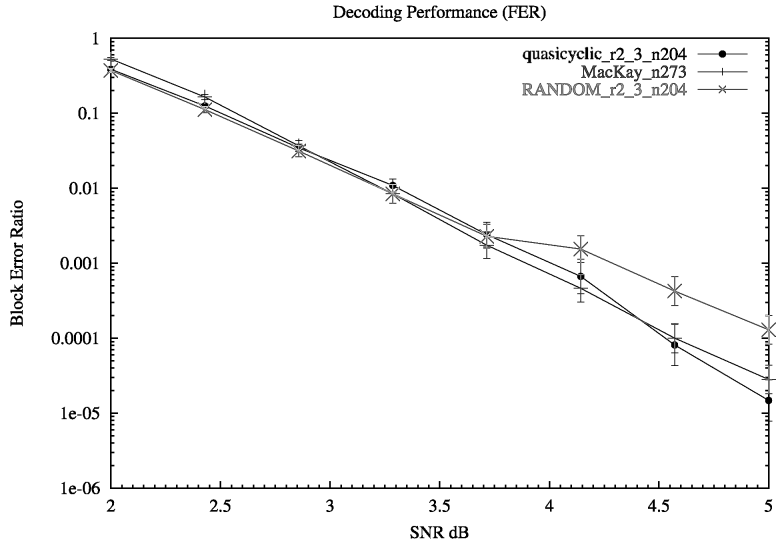}
    \caption[Performance of $R=2/3$, $N \approx 200$ girth eight \QC~code]{Block error rate performances comparison of a
    regular \QC~codes with $R=2/3$ versus a MacKay codes and a random
    permutation codes of similar   rate and length  $N \approx 200$}
    \label{fig:24_n200}
\end{figure}

To simplify the discussion the three blocks that form the \HH matrix
are called $B_1$, $B_2$ and $B_3$ and $\HHin =[B_1| B_2 | B_3]$.
The following theorem lists all the conditions that must
hold for an \HH~matrix of this type to have girth exactly eight.
\begin{theorem} \label{the:2_3_final}
Let $m\geq 3$, $\alpha > 4$ and $ M \in \mathbb H_{m,\alpha}$.
Let $g$ be the girth of the Tanner graph of $M$.
Then $g=8$ {\bf if and only if} all the following conditions hold:
\begin{enumerate}
\item for any
      $1\leq i\leq \alpha$ and $c\in\{1,2,3\}$

\begin{align*}
     &s(p_i^{c}) \not= \frac{m}{3},
     &s(p_i^{c}) \not= \frac{m}{2},
\end{align*}

\item for any $1\leq i\leq \alpha$ and $c,d \in\{1,2,3\}$ and $ c\neq d $,
$$
\begin{array}{ccc}
    s(p_i^c) \not= s(p_i^d),
    &s(p_i^c) \not \equiv \pm 2 s(p_i^d),
\end{array}
$$
\item for any $1\leq i\leq \alpha$,\\
$
\begin{aligned}
  \qquad  s(p_i^1) \pm s(p_i^2) \pm s(p_i^3) \not\equiv 0,
\end{aligned}
$
\item for any $(i,j)\in J_\alpha$,
 \begin{doubleenum}
\item $ \ep(p_i^1) + \ep(p_j^2) \not\equiv 0 ,$
\item $ \ep(p^1_i)+\ep(p^2_j) \not\equiv \{\pm s(p^1_i),\pm s(p^2_j)\},$
\item $\ep(p^1_i)+\ep(p^2_j) \not\equiv \{\pm s(p^1_j),\pm s(p^2_i),\pm s(p^3_i),\pm s(p^3_j)\}.$
\end{doubleenum}

\item for any $1\leq i\leq \alpha$,\\
$
\begin{aligned}
       \quad s(p^1_i) &\not \equiv \{s(p^1_{i+1}), s(p^2_{i+1}), s(p^3_{i+1}), s(p^3_{i+2})\},\\
       \quad s(p^2_i) &\not \equiv \{s(p^2_{i+1}), s(p^3_{i+2})\},\\       
       \qquad s(p^3_i) &\not \equiv \{s(p^3_{i+2}), s(p^2_{i+1})\},
 \end{aligned}
$

\item for any $1\leq i\leq \alpha$,
\begin{center}
 \begin{doubleenum}
\item $ \ep(p^1_i)+\ep(p^1_{i+1}) +\ep(p^3_{i+2}) \not\equiv 0$
\item $ \ep(p^2_i)+\ep(p^2_{i+1}) -\ep(p^3_{i+1}) \not\equiv 0$
\item $ \ep(p^1_i)-\ep(p^2_i)    +\ep(p^3_{i+1}) \not\equiv 0$
\item $ \ep(p^1_i)-\ep(p^2_{i+2}) +\ep(p^3_{i+2}) \not\equiv 0$
\end{doubleenum}
\end{center}
\end{enumerate}
\begin{proof}
It is necessary to prove that the conditions listed in the statement cover all the
possible cycle configurations existing for the construction presented.
 To prove
this it is necessary to show that  the configurations, regarding cycles of
length $4$ and $6$ in  theorem~\ref{prop:oneH_oneI_cycles}, or are not possible for this
construction, or are associated with conditions listed in the
statement.
\begin{itemize}
\item Configurations 1.1 and 2.1 in theorem~\ref{prop:oneH_oneI_cycles}
are evidently associated with the conditions in point 1.
\item Configurations 1.2 and 2.2 are covered by the conditions in point 2.
\item Point 3 considers configuration 2.3  in
theorem~\ref{prop:oneH_oneI_cycles}.
\item Configuration 1.3 is considered by condition in point 4.1. 
Note that due to the position of the diagonal of identities in $B_3$
such configuration can exist only between $B_1$ and $B_2$.
\item Configuration 1.4 cannot exist in the proposed class thanks to
the position of the identities diagonals. There cannot be any  couple of
columns that have $H$ and $I$ sub-matrices in the same two rows.
\item Configuration 2.4 in considered by the conditions in point 5. In fact
every $H$ sub-matrix  has three $I$ sub-matrices lying  in the same
row and one in the same column.  Hence for every $H$ sub-matrix six possible cycles of the type 2.4 can exist, three formed by the
 three identities in the same row and three formed by the identity in
the same column and the three $H$ sub-matrices lying in the row of such $I$.\\
The conditions for cycles to existing in such configuration are:
\begin{equation*}
\begin{aligned}
       s(p^1_i) & \equiv \{s(p^1_{i-1}), s(p^2_{i+1}), s(p^3_{i+2})\},\\
       s(p^1_i) & \equiv \{s(p^1_{i+1}), s(p^2_{i+1}), s(p^3_{i+1})\},\\
       s(p^2_i) & \equiv \{s(p^2_{i-1}), s(p^2_{i+1}), s(p^3_{i+2})\},\\   
       s(p^2_i) & \equiv \{s(p^2_{i-1}), s(p^2_{i-1}), s(p^3_{i-2})\},\\       
       s(p^3_i) & \equiv \{s(p^3_{i+2}), s(p^1_{i-1}), s(p^2_{i+1})\},\\       
       s(p^3_i) & \equiv \{s(p^3_{i-2}), s(p^1_{i-2}), s(p^2_{i-2})\},
 \end{aligned}
\end{equation*}
Eliminating from the list above the conditions that are equivalent the set of conditions
in  point 5 is obtained. 
\item Configurations 2.5 cannot exist for the same reason discussed for
configuration 1.4, the same is true for configurations 2.7 and 2.9 of theorem~\ref{prop:oneH_oneI_cycles}.
\item Cycles from configuration 2.6 can exist only between
sub-matrices in $B_1$  and sub-matrices in $B_2$ . In fact the position of the diagonal of identities in
$B_3$ does not allow any of the columns in the third block to
overlap in two positions with any of the other columns.
Conditions in point 4.2 evidently cover all the possible
cycles of this type.
\item For the same reason of the previous point, configuration 2.8 can exist only when the 
$\left|\begin{smallmatrix}
        H^2-2 & I\\
         I   & H^3-2\\
\end{smallmatrix}\right|$
columns are in $B_1$ and $B_2$. The remaining column can be in any of
the three blocks, the relative cycles  are
avoided by the conditions in point 4.3.
\item Cycles from configurations 2.10 and 2.11 are associated to conditions in
point 6.
To prove this it is necessary to show that the conditions listed in
point 6 consider \textbf{all} possible cycles that can arise from the
two configurations.

Both configurations require that the distance from an $H$ sub-matrix
and the identity in the same column is the sum of the distances from the
other two  $H$ sub-matrices and relative identities.
It is evident from the position of the identities diagonals
that such case is possible only if one column lies in $B_3$
and the other two columns do not. 
If the other columns both lie in $B_1$ it is only a matter of
connecting the sub-matrices to see that the resulting cycle is part of
configuration 2.11 and it is not allowed by the condition~6.1.
If the  other columns both lie in $B_2$ it is only a matter of
connecting the sub-matrices to see that the resulting cycle is part of
configuration 2.10 and it is not allowed by the condition~6.2.
If one of the remaining column lies in $B_1$ and the other
in $B_2$  there are two possibilities to form a cycle, both
corresponding to configuration 2.10. 
The fist case has the $H$ sub-matrices in
$B_1$  and $B_2$ lying in the same d.r., the second case  has  the $H$ sub-matrices in
$B_2$  and $B_3$ lying in the same d.r. The resulting cycles
are avoided by conditions 6.3 and 6.4 respectively.

Note that if the two  $H$ sub-matrices in
$B_1$  and $B_3$  lie in the same d.r. is not possible to form a
cycle due to the position of the identities diagonals
\end{itemize}
It has been proved that all the possible cycles that can exist in the \HH
matrix for this class of codes is avoided by one of the condition listed.
\end{proof}
\end{theorem}
The performance of such class of codes is presented next.
Figure~\ref{fig:24_n200}  shows the comparison between a code from the class
of \QC~codes presented versus a random code and a MacKay code obtained
from~\cite{MacKay:EOSGC}, the codes have length $N \approx 200$. The codes have similar length
and rate, the MacKay code is slightly longer but has higher rate
$R\approx 0.7$. It can be seen how the
\QC~code and MacKay code perform closely and slightly better than the
random code.

\begin{figure}[htbp]
    \centering
    \includegraphics[angle=-90,width=0.8\columnwidth]{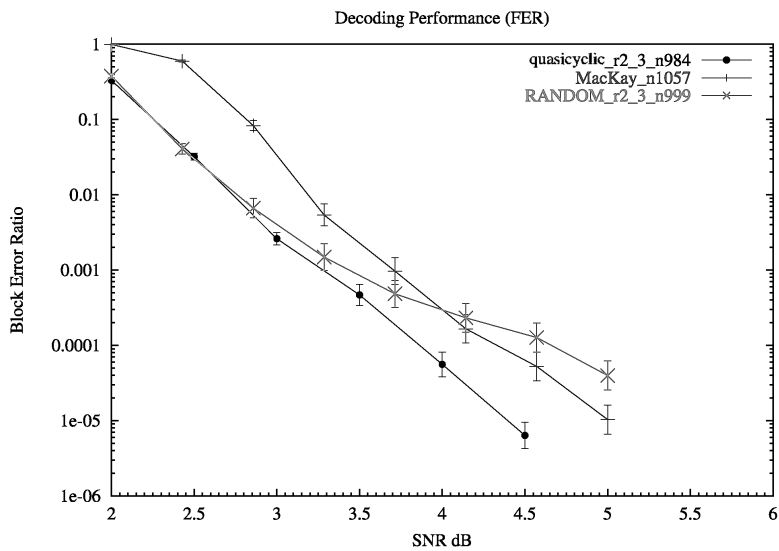}
    \caption[Performance of $R=2/3$, $N \approx 1,000$ girth eight \QC~code]{Block error rate performances comparison of a
    regular \QC~codes  versus a random constructed code and a MacKay
    code, $N \approx 1,000$. All codes have similar rate.}
    \label{fig:24_n1000}
\end{figure}

The comparison  of the same families of codes for $N\approx 1,000$ is
presented in \figur~\ref{fig:24_n1000}. The codes have similar length
and rate, the MacKay code is slightly longer but has higher rate
$R\approx 0.85$. For such length the proposed \QC~code outperforms both
the random code and the MacKay code. In particular the random code
seems to hit an early error floor and the MacKay code perform few tens
of dB worse for all the SNR range considered.

\begin{figure}[htb]
    \centering
    \includegraphics[angle=-90,width=0.8\columnwidth]{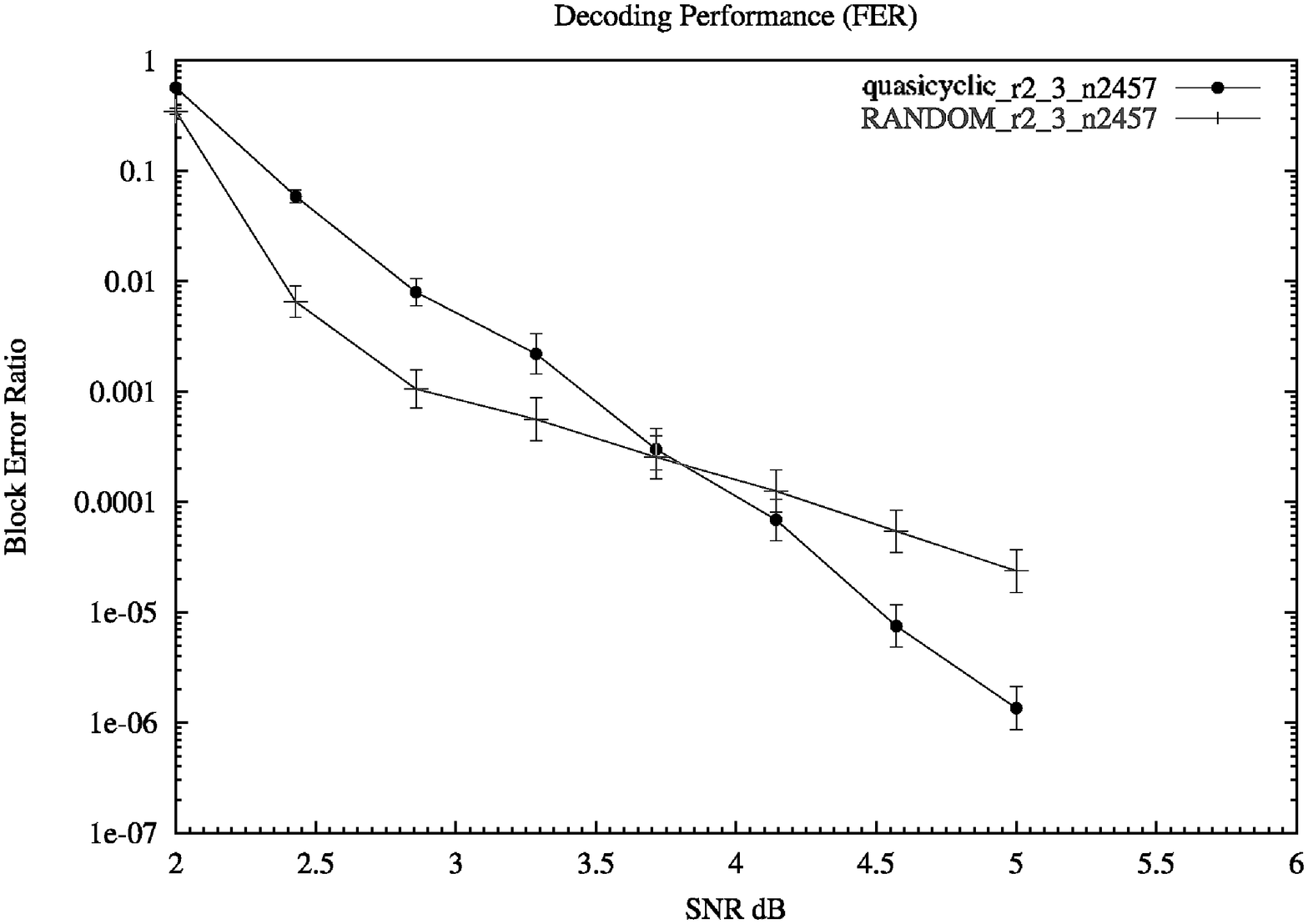}
    \caption[Performance of $R=2/3$, $N \approx 2,000$ girth eight \QC~code]{Block error rate performances comparison of a
    regular \QC~codes  versus a random constructed code with same rate
    and length $N \approx 2,000$}
    \label{fig:24_n2k}
\end{figure}

Finally \figur~\ref{fig:24_n2k} presents the comparison of a \QC~code
with a random code for $N=2457$. Unfortunately it has not been
possible to obtain a MacKay code with such rate and length due to
failure of the search carried out.
The random code performs slightly better than the\QC~code for low SNR
but it is affected by early error floor that makes its performance
poor at higher SNR. In contrast the \QC~code does not show the
presence of any early error floor.

It can be concluded that the construction proposed is a valid
alternative to random like codes for high rate. The \QC~codes, for the
dimensions considered,  perform at least as well as other
constructions but offer the advantage 
of reduced complexity and minimize memory requirement.
Those are important aspects for many
applications where high rate codes are used.

\section{Two families of \QC~codes with girth 10}\label{sec:qc_g10}
In this section two families of \QC~LDPC codes with girth at least 10
and rate $1/2$ are presented.
The first construction is a $(2,4)$-regular codes, the second a
$(3,6)$-regular codes. The constructions are based on the idea of
using blocks of circulants to compose the \HH~matrix. The circulants
lie in the diagonals of the block, this help to detect  the
possible configurations arising and can be exploited to reduce
hardware complexity of the decoding algorithm (see~\cite{Christian_ECCTD:05}).

\subsection{Regular $(2,4)$}\label{sec:qc_g10-2_4}
This family of codes has been studied first because its structure
drastically reduce the number of conditions necessary to ensure high girth.

\begin{definition}[ $(2,4)$-regular \QC~codes girth 10]\label{def:g10-2_4ourH}
Let $\alpha ,m$ be positive integers such that $\alpha > 4, m\geq 3$.
Notation $\mathbb C_{m,\alpha}$ denotes the class of the
$( m \alpha \times 2m\alpha )$
matrices of the form
\begin{equation*}
\mathbf{H}=\left[B_1\, |\, B_2\right]
\end{equation*}

where
\begin{equation*}
B_1=\left[\begin{array}{ccccc}
H_0 & 0 & \ldots & 0 & 0  \\
0 & H_1 & 0 & \ldots & 0  \\
0 & 0 & \ddots  &\ddots  & \vdots \\
\vdots &  & \ddots & \ddots & 0  \\
0  & \ldots & 0 & 0 & H_{\alpha-1}  \\
\end{array} \right]
\end{equation*}
and
\begin{equation*}
B_2=\left[\begin{array}{ccccc}
I & 0 & \ldots & J_0 & 0  \\
0 & I & 0 & \ldots & J_1  \\
J_2 & 0 & \ddots  &\ddots  & \vdots \\
\vdots &  & \ddots & \ddots & 0  \\
0  & \ldots & J_{\alpha-1} & 0 & I  \\
\end{array} \right].
\end{equation*}
Where every $H_i$, with  $i \in \{0,1,..,\alpha-1\}$,
is an $m\times m$ binary weight-$2$ circulant matrix, $J_i$ are binary weight-$1$ circulant matrices  and $I$ is
the $m\times m$ identity matrix.
\end{definition}
Reducing the theorems studied in section~\ref{sec:polysTOcycles}
for this class of matrices is a long task. This can be eased by
considering that some of the cycle configurations listed in
theorems~\ref{the:condition4},~\ref{the:condition6},~\ref{the:conditions8}
can exist only if the distance from the main diagonal and the diagonal
of $J$'s in $B_2$  is a particular value.
Such distance is called $\delta$.

The following theorem lists all the conditions that must
hold for an \HH~matrix of this class to have girth at least $10$.

\begin{theorem}\label{the:Condition_QC_g10_2-4}
Let $m\geq 3$, $\alpha > 4$ and \HH $\in\mathbb C_{m,\alpha}$.
Let $g$ be the girth of the Tanner graph of \HH.
Then $g\geq 10$ {\bf if and only if} all the following conditions hold
for any    $0\leq i\leq \alpha-1$.
\begin{enumerate}
\item

$$
\begin{array}{ccc}
     s(p_i) \not= \frac{m}{2}, &s(p_i) \not= \frac{m}{3},  &s(p_i) \not= \frac{m}{4}, 
\end{array}
$$

\item \begin{align*}
     s(p_i) \not= s(p_{i+\delta}),
\end{align*}

\item

\begin{equation*}
   \mbox{or} \quad \delta \not= \alpha/2,\quad \mbox{or} \quad
\left\{
\begin{aligned}
  &3.1 \quad \ep(p_i)^J + \ep(p_{i+\delta})^J    &\not\equiv &0 ,               &\mbox{and}\\
  &3.2 \quad \ep(p_i)^J + \ep(p_{i+\delta})^J    &\not\equiv &\pm s(p_i)            &\mbox{and}\\
  &3.3 \quad \ep(p_i)^J + \ep(p_{i+\delta})^J    &\not\equiv &\pm s(p_{i+\delta})      &\mbox{and}\\
  &3.4 \quad  2\ep(p_i)^J + 2\ep(p_{i+\delta})^J &\not\equiv &0,   &\mbox{and}\\
  &3.5 \quad \ep(p_i)^J + \ep(p_{i+\delta})^J    &\not\equiv &\pm 2s(p_i)           &\mbox{and}\\
  &3.6 \quad \ep(p_i)^J + \ep(p_{i+\delta})^J    &\not\equiv &\pm 2s(p_{i+\delta})     &\mbox{and}\\
  &3.7 \quad \ep(p_i)^J + \ep(p_{i+\delta})^J    &\not\equiv &\pm s(p_i)\pm s(p_{i+\delta})\\
\end{aligned}
\right.
\end{equation*}

\item 

\begin{equation*}
   \mbox{or}\quad \delta \not= \pm \alpha/3,\quad \mbox{or} \quad
\left\{
\begin{aligned}
  &4.1  \quad \ep(p_i) + \ep(p_{i+\delta})+ \ep(p_{i+2\delta}) \not\equiv 0 &\mbox{and}\\
  &4.2  \quad \ep(p_i) + \ep(p_{i+\delta})+ \ep(p_{i+2\delta}) \not\equiv \pm s(p_i)\\
\end{aligned}
\right.
\end{equation*}

\item 

\begin{flalign*}
   \mbox{or}\quad \delta \not= \pm \alpha/4,\quad \mbox{or}
   \quad \ep(p_i) + \ep(p_{i+\delta})+ \ep(p_{i+2\delta}) + \ep(p_{i+3\delta}) + \ep(p_{i+4\delta})\not\equiv 0\\
\end{flalign*}

\end{enumerate}

Where the index operations are modulo $alpha$.

\begin{proof}
It is necessary to prove that the conditions listed cover all the
possible cycle configurations existing in such class of codes.
 To prove
this it is necessary to show that  the configurations  listed in
theorems~\ref{the:condition4},~\ref{the:condition6} and~\ref{the:conditions8}  are not possible for this
construction or are associated with a condition listed in the
statement.
First the configurations that can have $4$-cycles, listed in
theorem~\ref{the:condition4}, are considered.
\begin{itemize}
\item Configuration~\ref{item:g4-1} is associated with the first condition in
point 1.
\item Configuration~\ref{item:g4-2} cannot exist because there cannot
be two $H$ sub-matrices in the same row or column.
\item Configuration~\ref{item:g4-3} can exist in this construction
only if $\delta=\alpha /2$. In any other case there cannot be four
sub-matrices lying in two columns and two rows. If  $\delta=\alpha /2$
condition 3.1 assures that the cycle cannot exist.
\end{itemize}
The configurations that can have $6$-cycles, listed in
theorem~\ref{the:condition6}, are considered next.
\begin{itemize}
\item Configuration~\ref{item:g6-1}  is associated with the second condition in 
point 1.
\item Configurations~\ref{item:g6-2} and~\ref{item:g6-3} cannot exist
because there cannot be two $H$ sub-matrices in the same row or column.

\item Configurations~\ref{item:g6-5} and~\ref{item:g6-6} cannot exist because for this
construction every $H$ sub-matrix lies alone in a column.

\item For the same reason discusses for configuration 3 of
theorem~\ref{the:condition4}, configuration~\ref{item:g6-7} can exist in this construction
only if $\delta=\alpha /2$.
If  $\delta=\alpha /2$ conditions 3.2 and 3.3  assure that the cycle
cannot exist.

\item Configurations~\ref{item:g6-8} requires that  the distance from
two sub-matrices in the same column is the sum of the 
two other such distances for the remaining d.c.'s in the configuration. 
It is evident that all the columns must lie in $B_2$ since in $B_1$
there is only one sub-matrix in each column. Moreover such
situation exists only if $\delta=\pm \alpha /3$. 
If  $\delta=\pm \alpha /3$ conditions 4.1 assures that the cycle
cannot exist.
To determine the sign of the terms in the condition it is sufficient
to consider that there cannot be two $J$'s in the same column. Hence
the $J$'s take alternate positions in the cycle configuration hence
they all have the same sign.
\end{itemize}
The configurations that can have $8$-cycles, listed in
theorem~\ref{the:conditions8} are considered next.
\begin{itemize}
\item Configuration~\ref{item:g8-1} is associated with the third condition in
point 1.
\item
Configurations~\ref{item:g8-6},~\ref{item:g8-7},~\ref{item:g8-10},~\ref{item:g8-15},~\ref{item:g8-22} 
and~\ref{item:g8-25}
cannot exist because for this
construction every $H$ sub-matrix lies alone in a column.

\item For the same reason discussed for configuration 3 of
theorem~\ref{the:condition4}, configuration~\ref{item:g8-11} can exist in this construction
only if $\delta=\alpha /2$.
If  $\delta=\alpha /2$ conditions 3.4  assures that the cycles
cannot exist.

\item For the same reason of previous point, configuration~\ref{item:g8-13} can exist in this construction
only if $\delta=\alpha /2$.
If  $\delta=\alpha /2$ conditions 3.5 and 3.6  assure that the cycles
cannot exist.
\item Configuration~\ref{item:g8-17} cannot exist because the structure
of the \HH does not allow to  have $6$ sub-matrices lying in two rows and
three columns.
\item Configuration~\ref{item:g8-18} is evidently associated with
condition 2.
\item  Configuration~\ref{item:g8-20}  can exist only with
$\delta=\alpha /2$ for the same reason discussed in previous cases. 
If  $\delta=\alpha /2$ conditions 3.7  assures that the cycles
cannot exist.
\item  Configuration~\ref{item:g8-21} cannot exist because the structure
of the \HH does not allow to  have $8$ sub-matrices lying in two rows and
four columns.

\item  Configuration~\ref{item:g8-23}  cannot exist because it implies
 the existence of three non zero sub-matrices in the same column or
 the existence of a column with
an $H$ and one identity, both cases are not possible due to the  particular structure of the \HH matrix.
\item  Configuration~\ref{item:g8-24}  cannot exist because it implies
to have three non zero sub-matrices in the same column that is not
possible due to the particular structure of the \HH matrix.
\item  Configuration~\ref{item:g8-26-a}  cannot exist because it implies
to have four  non zero sub-matrices in the same column or row that is
not possible  due to the  particular structure of the \HH matrix.
\item Configuration~\ref{item:g8-26} requires that  the distance from
an two sub-matrices in the same column is the sum of the 
two other such distances for the remaining d.c.'s in the configuration.
It is evident that the columns not containing $H$ must lie in $B_2$.
 Moreover such situation exist only if $\delta=\pm \alpha /3$. 
If  $\delta=\pm \alpha /3$ conditions 4.2 assures that the cycle
cannot exist. The same reasoning applied to
configuration~\ref{item:g6-8}  of theorem~\ref{the:condition6} can be
used to explain the signs of the terms of the condition.
\item Configurations~\ref{item:g8-27} requires that  the distance from
the  two sub-matrices in the same column is the sum of the 
three other such distances for the remaining d.c.'s in the
configuration. 
It is evident that all the columns must lie in $B_2$. Moreover such
situation exist only if $\delta=\pm \alpha /4$. 
If  $\delta=\pm \alpha /4$ conditions 5 assures that the cycles
cannot exist.

\end{itemize}
It has been proved that all the possible cycles that can exist in the \HH
matrix for this class of codes is avoided by one of the condition listed.
\end{proof}

\end{theorem}

Aside from the  degenerative cases where $\delta=\alpha/2$, $\delta=\alpha/3$ and
$ \delta=\alpha/4 $ the conditions to check are extremely simple.
Even in the degenerative cases the list of conditions is quite short.
Thanks  to this, finding exponents and conditions that allows girth $10$ is
an easy task. 
Following the performances of three of such codes are presented. The
codes have been obtained avoiding the degenerative cases of $\delta$
and imposing the conditions on the separations.

\begin{figure}[htbp]
    \centering
    \includegraphics[angle=-90,width=0.8\columnwidth]{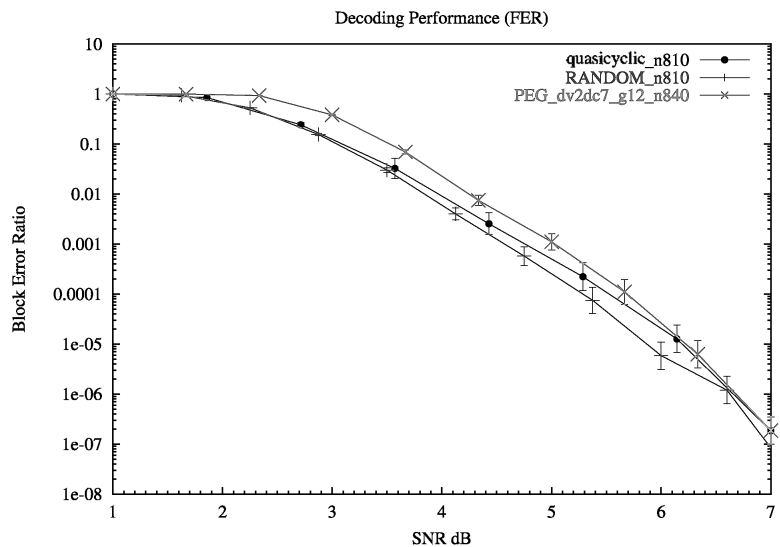}
    \caption[Performance of $(2,4)$-regular, $N \approx 800$, girth ten \QC~code]{Block error rate performances comparison of a \QC~code
    compared with a random code and a PEG code with    $N\approx 800$}
    \label{fig:24n810g16}
\end{figure}

In \figur~\ref{fig:24n810g16} the block error ratio versus noise
performance  graph of a small code ($N=810$) from this family  can be seen.
The code is compared with  a random code.
Both codes have same length and are $(2,4)$-regular.
The random codes has been obtained with a optimal permutations method
and has a girth of $14$ and an average girth of $15.7$, the \QC~codes
has girth $10$.
In the same figure the \QC~code is compared also with a more
advanced  Progressive Edge Grow (PEG) code
from~\cite{Venkiah08_improvedGirthPEG,Hu_Eleftheriou:05}. The code is  $(2,7)$-regular.
It can be seen how this code performs as well as the random code and  compares evenly with a
carefully constructed code, this
is a good result for such class of codes.

\begin{figure}[htbp]
    \centering
    \includegraphics[angle=-90,width=0.8\columnwidth]{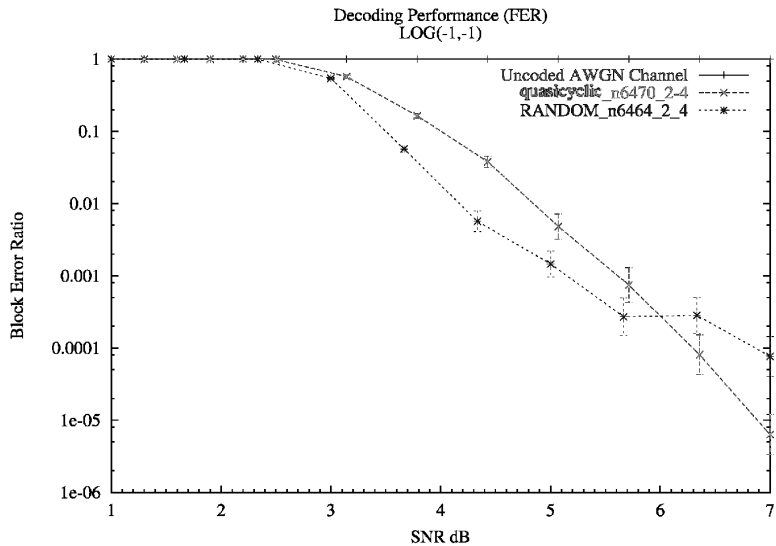}
    \caption[Performance of $(2,4)$-regular, $N \approx 6,500$, girth ten \QC~code]{Block error rate performances comparison of a \QC~code compared with a random code. Both codes are $(2,4)$-regular with $N \approx 6,500$}
    \label{fig:24n6464g10}
\end{figure}

Figure~\ref{fig:24n6464g10} presents a comparison of
a \QC~code from this family versus a random code in the case of longer
codes. Both the codes have $N \approx 6,500$.
It can be seen how the performance of the \QC~code with girth 10 do not hit an error floor as soon
as the random code. The random code has average girth of $20.37$ but
girth of only $6$.

\begin{figure}[htbp]
    \centering
    \includegraphics[angle=-90,width=0.8\columnwidth]{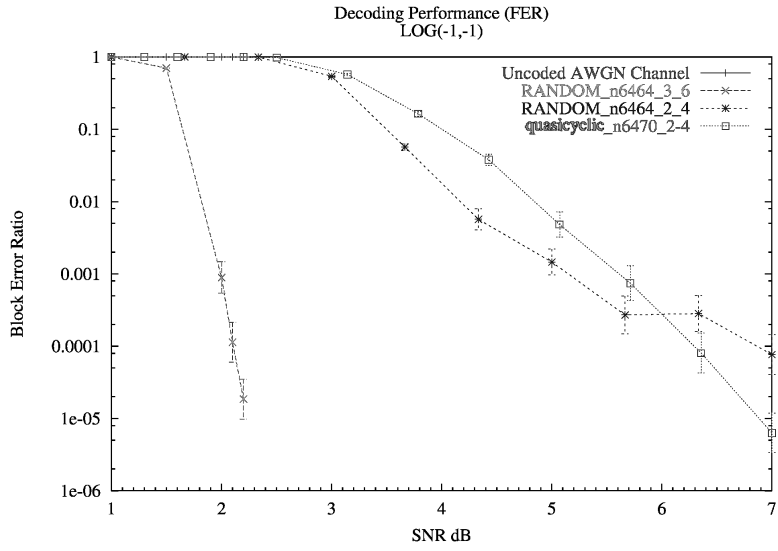}
    \caption[Performance comparison of a $(2,4)$-regular versus a $(3,6)$-regular \QC~codes]{Block error rate performances comparison of a $(2,4)$-regular \QC~codes versus a $(3,6)$-regular code $N \approx 800$}
    \label{fig:24VS36_RND}
\end{figure}

Regular $(2,4)$-LDPC codes are not often used in practice due to their poor
performances.
As an example of this a comparison between a $(3,6)$-regular Gallagher
code and the PEG code is presented in \figur~\ref{fig:24VS36_RND}
It can be seen how the $(3,6)$- regular code has up to 5db gain when
compared with $(2,4)$-regular codes both random and \QC.
For this reason the interest is on build $(3,6)$-regular \QC~LDPC codes with girth ten.

\subsection{Regular $(3,6)$}\label{sec:qc_g10-3_6}
The construction presented here is the result of few considerations.
Lemma~\ref{lemma:2C=g8} shows how any \QC~LDPC code with girth at
least $10$ cannot exist if two weight-$2$ circulants lie in the same row or column.
For this reason all weight-2 circulants are placed in the main diagonal of
$B_1$. 
To keep the column weight equal to three it is necessary to insert
another weight-1 circulant for each column and this is done inserting another diagonal of identities.
The second block $B_2$ cannot contain weight two circulants hence  any row and column
must contain three weight-$1$ circulants to form a $(3,6)$-regular code.
Aiming to maintain some regularity to allow reduced hardware complexity
of the decoding algorithm, the circulant matrices have been disposed in three diagonals.
To simplify  the number and complexity of the checks necessary to
guarantee girth $10$ identity matrices, as 
special case of weight-$1$ circulant matrices, have been used where
possible.The resulting family of codes is the following.
\begin{definition}[$(3,6)$-regular \QC~code]\label{def:g10-(3,6)ourH2}
Let $\alpha ,m$ be positive integers such that $\alpha >  4, m\geq 3$.
Notation  $\mathbb D_{m,\alpha}$ denotes the class of the
$( m \alpha \times m\alpha )$
matrices of the form
\begin{equation*}
\mathbf{H}=\left[B_1\, |\, B_2\right]
\end{equation*}
where
\begin{equation*}
B_1=\left[\begin{array}{ccccc}
H_0 & 0 & \ldots & 0 & I  \\
I & H_1 & 0 & \ldots & 0  \\
0 & I & \ddots  &\ddots  & \vdots \\
\vdots &  & \ddots & \ddots & 0  \\
0  & \ldots & 0 & I & H_{\alpha-1}  \\
\end{array} \right]
\end{equation*}

and
$$
B_2=\left[\begin{array}{cccccccccccc}
J_0     & 0    &\ldots&   0  &    I & 0    &\ldots&     0&    I & 0    &\ldots& 0  \\
0       & J_1  & 0    &\ldots&   0  & I    &     0&\ldots&   0  &  I   &\ldots&\vdots  \\
 \vdots &      &\ddots&      &      &      &\ddots&      &      &      &\ddots&  0\\
 \vdots &      &      &\ddots&      &      &      &\ddots&      &      &      &  I\\
I       &    0 &      &      &\ddots&      &      &      &\ddots&      &      &  0\\
0       &\ddots&      &      &      &\ddots&      &      &      &\ddots&      &  0\\
\vdots  &      &\ddots&      &      &      &\ddots&      &      &      &\ddots&  0\\
        &      &      &\ddots&      &      &      &\ddots&      &      &      &  I\\
I       &      &      &      &\ddots&      &      &      &\ddots&      &      &  0 \\
0       &\ddots&      &      &      &\ddots&      &      &      &\ddots&      &\vdots\\
\vdots  &      &\ddots&      &      &      &\ddots&      &      &      &\ddots&  \\
0       &      &      & I    &\ldots&      &      & I    &\ldots&      &      &J_{\alpha-1} \\
\end{array} \right].
$$

where every $H_i$, with  $i \in \{0,1,..,\alpha-1\}$,
is an $m\times m$ binary weight-$2$ circulant matrix and $I$ is
the $m\times m$ identity matrix, and $J_i$  are weight-$1$ circulant matrices.
\end{definition}

\begin{figure}[htb]
\psfrag{d2}{$\delta_2$}\psfrag{d3}{$\delta_3$}\psfrag{d32}{$\delta_{32}$}
    \centering
    \includegraphics[angle=0,width=0.8\columnwidth]{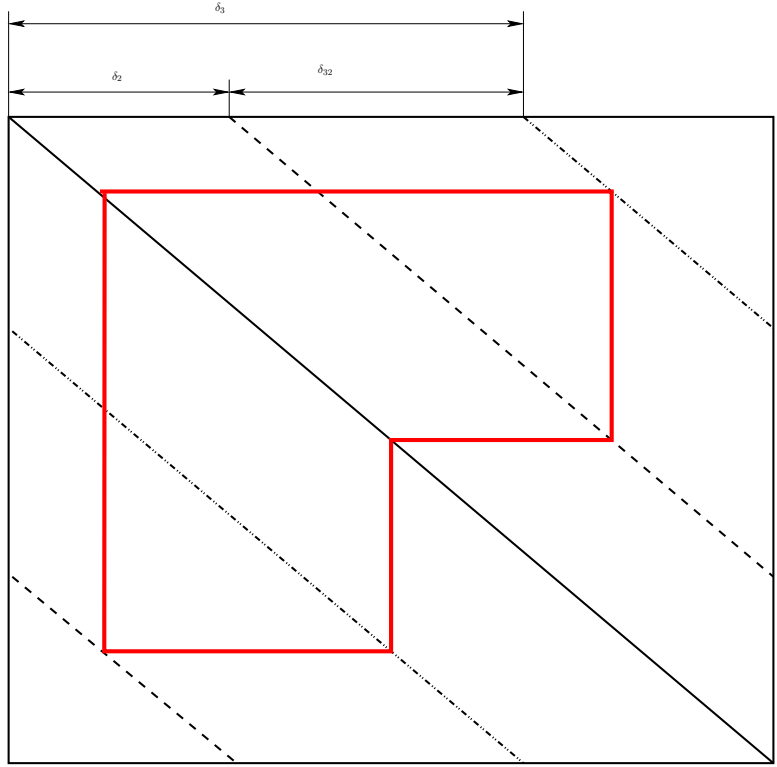}
    \caption[Unavoidable $6$-cycle configuration]{Example of configuration existing independently from the
    values of the deltas}
    \label{fig:g6-configuration}
\end{figure}

Some of the cycle configurations listed in
theorems~\ref{the:condition4},~\ref{the:condition6} and~\ref{the:conditions8}
exist only if the distance from the main diagonal and the diagonals
of identities in $B_2$ take particular values.
Such distances are called $\delta_2,\delta_3$, and for
simplicity the notation $\delta_3-\delta_2$ is called  $\delta_{32}$.
Such values can be seen in \figur~\ref{fig:g6-configuration}.
Moreover  the position of the diagonal of identities in $B_1$ is
called  $\delta_1$. For the construction presented here
$\delta_1=\alpha-1$.

Note that it is not possible to substitute the diagonal of $J$'s with
identities. 
In fact in such case $6$-cycle configurations will always arise independently 
from the positions of the diagonals.
The existence of this configurations carry a parallel with the
existence of $6$-cycles on a weight-$3$ circulant
matrix~\cite{BondHuiSch01}.
A graphical representation of such case is given in \figur~\ref{fig:g6-configuration}.
It can be seen that the particular $6$-cycle exists  if
$\delta_2+\delta_{32}=\delta_3$ but
$\delta_{2}+\delta_3-\delta_2=\delta_3$ hence the condition is always
satisfied and the cycle exists
independently from the values of the deltas. If all the matrix
involved in the cycle are identities it is not possible to avoid such cycle.
Setting the main diagonal to contain $J$'s, and not identities, it is
possible to 
avoid such cycle by imposing proper conditions on the exponents.

A new class of matrices is presented next. The matrices are part of
the previous class but some limitations on the position of the
diagonals of identities are added.
Such limitations on the values of deltas are not strictly necessary
but help to reduce the number of possible configurations, and so the
complexity of the conditions to check.
\begin{definition}[$(3,6)$-regular \QC~code girth 10]\label{def:g10-(3,6)ourH2-g10}
Let  $\mathbb E_{m,\alpha}$ denotes the class of the
$( m \alpha \times m\alpha )$
matrices in $\mathbb D_{m,\alpha}$ that satisfy the following
conditions on the distances of the diagonals.
\begin{equation}\label{eq:36_reg_delat_conditions}
\begin{array}{lll}
 1&\quad
\delta_x \not= \alpha/2,\; \delta_x \not=\pm \alpha/3, \; \delta_x \not=\pm \alpha/4
&\mbox{ where } x \in\{2,3,32\}\\

 2&\quad
\delta_x \not=\pm \delta_y,\;
&\mbox{ where } x,y \in\{1,2,3,32\}\; \mbox{and}\;x\neq y \\
 3&\quad
\delta_x \pm \delta_y \pm \delta_z \not= 0,
&\mbox{ where } x,y,z \in\{2,3,32\}\\

 4&\quad
\delta_x \pm \delta_y \pm \delta_z \pm \delta_w\not= 0,
&\mbox{ where } x,y,z,w \in\{1,2,3,32\}\\
\end{array}
\end{equation}
\end{definition}

Once those conditions (on the deltas) are verified it is possible to
apply the theorems studied in section~\ref{sec:polysTOcycles} to 
obtain the  set of conditions (on the circulants) that assure girth at
least 10.

\begin{theorem}\label{the:Condition_QC_g10_3-6}
Let $m\geq 3$, $\alpha\geq 4$ and \HH $\in {\mathbb E_{m,\alpha}}$.
Let $g$ be the girth of the Tanner graph of \HH.
Then $g\geq 10$ {\bf if and only if} all the following conditions hold
for any    $0\leq i\leq \alpha-1$ :
\begin{enumerate}

\item
$$
\begin{array}{ccc}
     s(p_i) \not= \frac{m}{2}, & s(p_i) \not= \frac{m}{3},  &s(p_i) \not= \frac{m}{4}, 
\end{array}
$$

\item $\qquad$          
    \begin{doubleenum}
    \item  $s(p_i) \not\equiv s(p_{i+\delta_1}),$
    \item  $s(p_i) \not\equiv \pm  2s(p_{i+\delta_1}),$
    \item  $s(p_{i+\delta_1}) \not  \equiv \pm 2s(p_i),$
    \end{doubleenum}

\item 
$$
\begin{array}{cc}
    s(p_i) \not  \equiv  s(p_{i+\delta_x}),&\mbox{ where } x\in\{2,3,32\}\\\end{array}
$$

\item  $\qquad\qquad\qquad$          
    \begin{doubleenum}
\item $ \ep(p_i)^J - \ep(p_{i\pm\delta_{32}})^J \not\equiv 0, $
\item $ \ep(p_i)^J - \ep(p_{i+\pm\delta_{32}})^J \not\equiv \pm s(p_i),$
\end{doubleenum}

\item $\qquad\qquad\qquad$          
    \begin{doubleenum}
\item  $\ep(p_i)^H - \ep(p_{i\delta_{32}})^H \not\equiv 0,$ 
\item  $\ep(p_i)^H
- \ep(p_{i-\delta_x})^H+ \ep(p_{i+1})^J- \ep(p_{i})^J \not\equiv
0\qquad  \mbox{ where } x\in\{2,3\}$
\end{doubleenum}

\end{enumerate}
\begin{proof}
It is necessary to prove that the conditions listed cover all the
possible cycle configurations existing in such class of codes.
 To prove
this it is necessary to show that  the configurations  listed in
theorems~\ref{the:condition4},~\ref{the:condition6} and~\ref{the:conditions8}, or are not possible for this
construction, or are associated with a condition listed in the
statement.
First the configurations that can have $4$-cycles, listed in
theorem~\ref{the:condition4} are considered.
\begin{itemize}
\item Configuration~\ref{item:g4-1} is associated with the first condition in
point 1.
\item Configuration~\ref{item:g4-2} cannot exist because there cannot be
two $H$ sub-matrices in the same row or column.
\item Configuration~\ref{item:g4-3} cannot exist in this construction
and with the conditions imposed on the deltas.
In fact such configuration can exist only if there are four
sub-matrices lying in two column and two rows. This can happen only if 
two deltas are equals  modulo alpha, or if one of the delta is
equal to $\alpha/2$. These cases are not
allowed by conditions 1 and 2 in definition~\ref{def:g10-(3,6)ourH2-g10}
hence the cycle configuration cannot exist.
\end{itemize}
The configurations that can have $6$-cycles, listed in
theorem~\ref{the:condition6} are considered next.
\begin{itemize}
\item Configuration~\ref{item:g6-1}  is associated with the second condition in
point 1.
\item Configurations~\ref{item:g6-2} and~\ref{item:g6-3} cannot exist
because there cannot be two $H$ sub-matrices in the same row or column.
\item Configuration~\ref{item:g6-5} contains two $C$ sub-matrices in the
two d.c. hence both the columns must be in the $B_1$.
 Every  $H$ sub-matrix can form a cycle only with the row above
and below itself. All this cases are considered by condition 2.1.
The remaining sub-matrix in the configuration is an identity hence the
two exponent terms are not present in the final condition.
\item Configurations~\ref{item:g6-6} and~\ref{item:g6-7} cannot
exist for the same reason discussed for configuration~\ref{item:g4-3} of theorem~\ref{the:condition4}.
\item Configurations~\ref{item:g6-8} requires that  the distance from
two sub-matrices in the same column is the sum of the 
two other such distances for the remaining d.c.'s in the
configuration. 
 This is not allowed by condition 3 in
 definition~\ref{def:g10-(3,6)ourH2-g10} with the exception of the case depicted in
\figur~\ref{fig:g6-configuration}.
Such cycles are avoided by condition 4.1.
\end{itemize}
The configurations that can have $8$-cycles, listed in
theorem~\ref{the:conditions8} are considered next.
\begin{itemize}
\item Configuration~\ref{item:g8-1} is associated with the third condition in
point 1.
\item
Configurations~\ref{item:g8-6},~\ref{item:g8-10},~\ref{item:g8-11},~\ref{item:g8-13},~\ref{item:g8-15},~\ref{item:g8-17},~\ref{item:g8-20},~\ref{item:g8-21},~\ref{item:g8-23},~\ref{item:g8-24},~\ref{item:g8-25}
and~\ref{item:g8-26-a}
cannot exist  for the same reason discussed for configuration~\ref{item:g4-3} of theorem~\ref{the:condition4} .
\item The same reasoning done for configuration~\ref{item:g6-5} of 
theorem~\ref{the:condition6} applies
to configuration~\ref{item:g8-7} and all the possible resulting
cycles are avoided  by conditions 2.2 and 2.3.
\item Configuration~\ref{item:g8-18} contains two $C$ sub-matrices in the
two d.c. hence both the columns must be in $B_1$.
 The remaining column contains two $J$ sub-matrices hence it
must lie $B_2$. The sub-matrices in the column in $B_2$
are spaced at distance $\delta_x$ with $x \in \{2,3,32\}$ hence all the possible
cycles are considered by condition 3.
\item Configuration~\ref{item:g8-22} requires that  the distance from
two sub-matrices in the same column is the sum of the 
two other such distances for the remaining d.c.'s in the
configuration. 
 This is not
allowed by condition 3 in
definition~\ref{def:g10-(3,6)ourH2-g10}. Note that the presence of
a weight-$2$ circulant forces one d.c.  to be in $B_1$ hence the case  depicted in
\figur~\ref{fig:g6-configuration} does not exist here.
\item Following the same reasoning done for configuration
~\ref{item:g6-8}  of theorem~\ref{the:condition6}. Configuration
~\ref{item:g8-26} exists only if the three columns lying
in B2 have the configuration depicted in \figur~\ref{fig:g6-configuration} and the last
column lies in B1. 
Such cycles are avoided by condition 4.2.
\item Configuration~\ref{item:g8-27} is associated to conditions in
point 5. To prove this it is necessary to show that the conditions listed in
point 5 consider \textbf{all} possible cycles that can arise from the
configuration.

The configuration requires that the distance of  two sub-matrices in the same column is the sum of the
three other such distances for the remaining d.c.'s in the
configuration. 
Such configurations are not possible thanks to condition 4 in
definition~\ref{def:g10-(3,6)ourH2-g10}.
A particular case remains to be considered. If two columns of the
cycle configuration lie in $B_1$ and two in $B_2$ then the existence
of this 
configurations carry a parallel with the
existence of $8$-cycles on two  weight-$2$ circulants in the same row
or column (\figur~\ref{fig:circulant_8-1}). It has been proved in lemma~\ref{lemma:2C=g8} that such cycle exists
independently from the value of the separations. In this case the
delta's take the part of the separation, hence the cycle configuration
exists independently from the value of delta's.
But $B_2$ has three diagonals so three different cycles can exist, and
such cycles are avoided by the three conditions in point~5.

To better understand this passage consider that the two cycle columns in
$B_1$ contain an identity and an $H$ sub-matrix each, hence exponents
of two $H$ sub-matrices must be part of the condition.
From \figur~\ref{fig:circulant_8-1} it can be seen that the cycle exists
only if both the cycle columns in $B_2$ contain two sub-matrices
appertaining to the same two diagonals. There are three possibilities:
one sub-matrix is in the second diagonal and one in the main diagonal, 
one sub-matrix is in the third  diagonal and one in the main diagonal
and finally 
one sub-matrices in the second diagonal and one in the third diagonal.
The first two cases lead to condition 5.2 the third to condition 5.1.
\end{itemize}
It has been proved that all the possible cycles that can exist in the \HH
matrix for this class of codes is avoided by one of the condition listed.
\end{proof}
\end{theorem}

Following the performances of codes of this family for different
lengths are presented.

\begin{figure}[htbp]
    \centering
    \includegraphics[angle=-90,width=0.85\columnwidth]{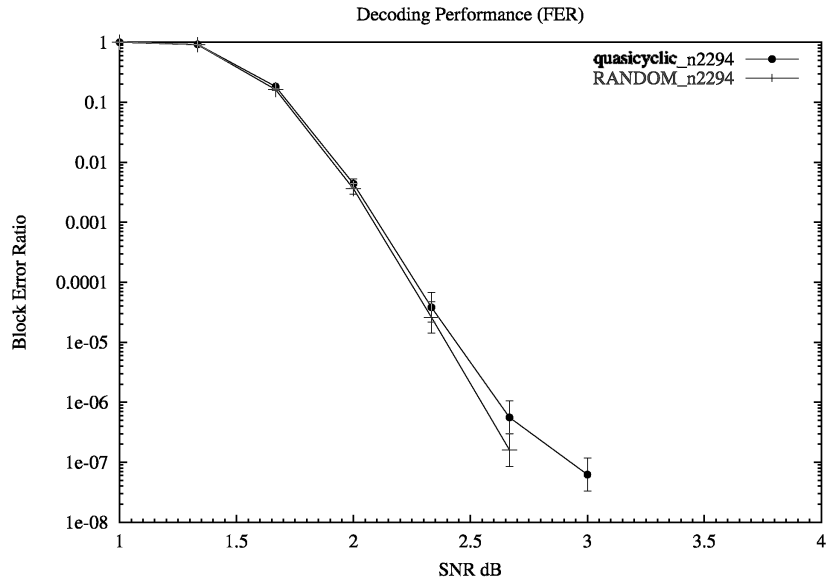}
    \caption[Performance of $(3,6)$-regular, $N \approx 2,200$, girth
    ten \QC~code]
{Block error rate performances of a girth 10 \QC~code compared with a random generated codes.  Both codes are $(3,6)$-regular with $N=2,294$ }
    \label{fig:36n2kg10}
\end{figure}

Figure~\ref{fig:36n2kg10} shows the performance of a medium
length \QC~code when compared with a random 
generated  code.
Both codes are $(3,6)$-regular codes of dimension $N=2,294$.
It can be seen how the \QC~codes from this family performs really close to the random code,
but it seems that the \QC~codes hit an error floor at $1e-09$ BER.

\begin{figure}[htbp]
    \centering
    \includegraphics[angle=-90,width=0.85\columnwidth]{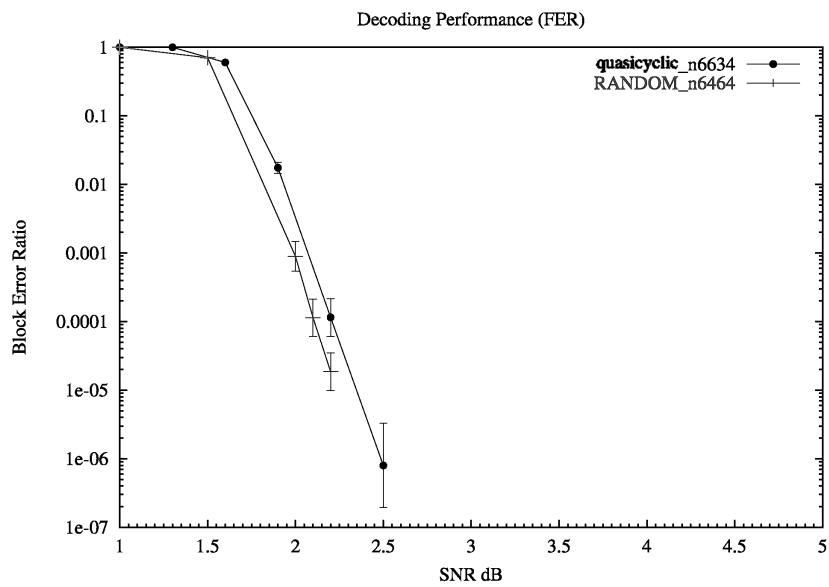}
    \caption[Performance of $(3,6)$-regular, $N \approx 6,500$, girth
    ten \QC~code]{Block error rate performances comparison of
    a \QC~code compared with a random code. 
Both codes are $(3,6)$-regular with  $N\approx 6,500$}
    \label{fig:36n6kg10}
\end{figure}

\begin{figure}[h!tb]
    \centering
    \includegraphics[angle=-90,width=0.8\columnwidth]{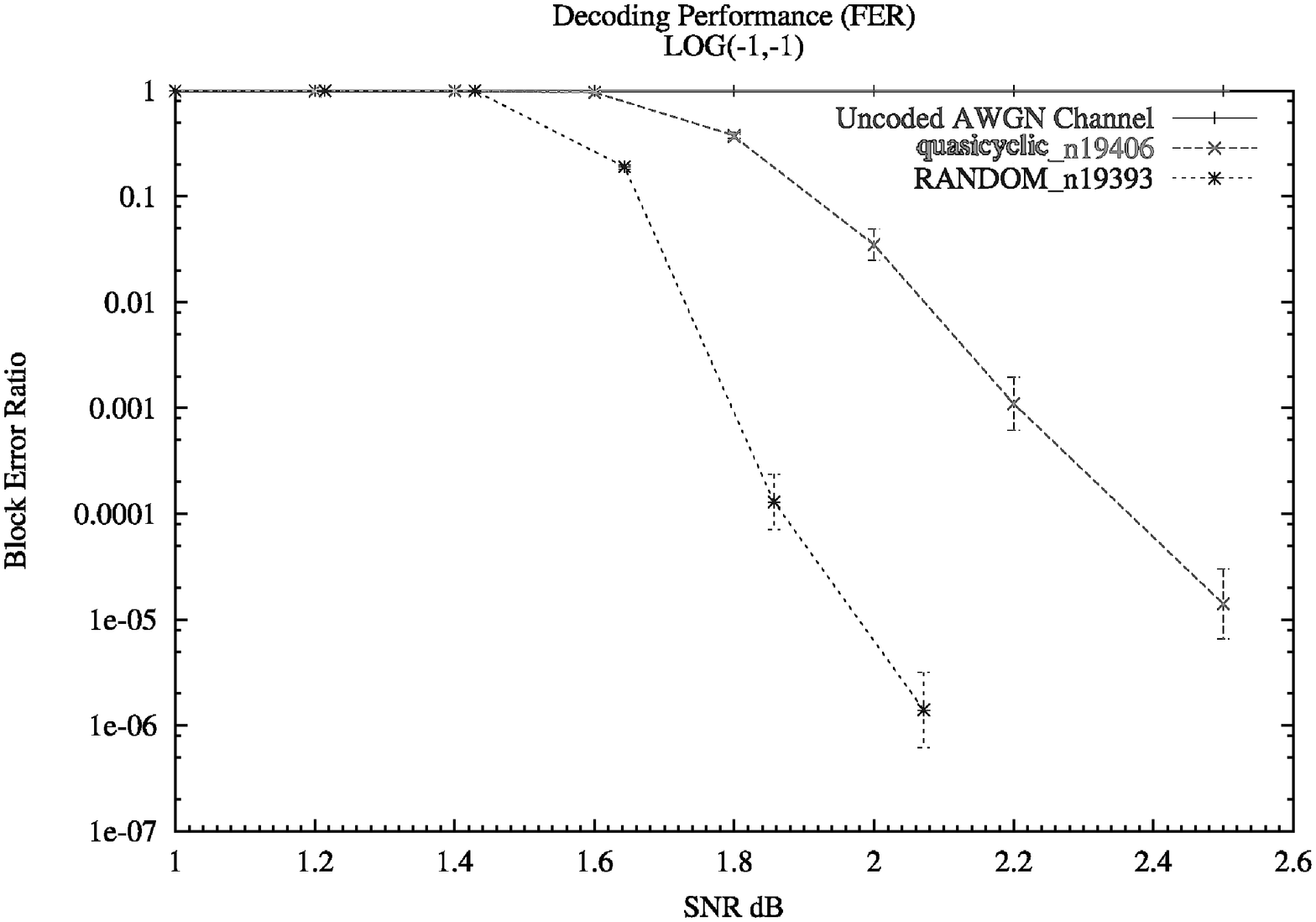}
    \caption[Performance of $(3,6)$-regular, $N \approx 19,500$, girth
    ten \QC~code]{Block error rate performances comparison of
    a \QC~code compared with a random code. 
Both codes are $(3,6)$-regular with  $N\approx 19,500$}
    \label{fig:36n19kg10}
\end{figure}

Even if out of the boundaries of the range of codes that are the main
focus of this thesis the comparison of \QC~codes with random codes
using longer codes is presented next for completeness.
Figure~\ref{fig:36n6kg10} and~\ref{fig:36n19kg10} present such a comparison.
Figure~\ref{fig:36n6kg10} shows the performances of two $(3,6)$-regular codes with $N \approx 6,500$ and 
\figur~\ref{fig:36n19kg10} shows the performances of two $(3,6)$-regular codes with $N \approx 19,500$ 
It can be seen the \QC~codes are quite close to the random codes.
With $N \approx 6,500$ only $0.1 $db, at  $1e-09$, divides the two codes, with  $N \approx 19,500$ the difference increased but
remain under the $0.5$db, at $1e-05$, that is a considerable achievement for
algebraic codes of this length.

\section{Conclusions anf future work}\label{sec:concl}
In this contribution a study on the presence of cycles on the \QC~LDPC codes
has been presented.
The aim of the study is to identify the configurations of circulant
matrices that allow cycles of a given length to exist. Once all such
configurations are known it is possible to associate to each of them a condition, on
the exponents and separations of the circulants involved in the cycle,
that must be satisfied for the cycle to exist.
The collection of these conditions gives a way to
construct \QC~codes without undesired cycle by ensuring that all the
conditions are not satisfied.
The chapter started with a novel definition of cycles in a matrix,
opposed to the normal definition that is given on graph. Such a
definition is used to prove  the ``isolation''
(lemma~\ref{lem:isolation}), the lemma is extensively used in the
search of which configurations are valid and can contain cycles.
The configurations that allows cycle of length less then $10$ are
found first for the generic case of a matrix that can be partitioned
in sub-matrices, then in the case of the matrices for \QC~codes.
Our approach is not restricted to circulant matrices of weight one as
is commonly done in literature but considers also circulant matrix of
weight-$2$ (weight-$3$ circulant matrices are of no interest for this
search because they have internal $6$-cycles).
The set of conditions for the generic case is found and then the
conditions are applied for various particular constructions.
The first class of \QC~codes studied are the Bresnan Codes~\cite{BresnanThesis:04},
several aspects of these codes are studied and their performances
presented.
Such class is then extended for higher rate codes.
Finally two new structures and the relative set of conditions that
guarantee girth at least $10$ are developed and their performances
presented.
The contribution of this work is not only  the four classes
of \QC~codes with high girth but mainly  the creation of  the list of 
conditions that every \QC~codes (of any possible construction method)
must satisfy to have a given girth. This not only covers all the
existing cases but can be used when constructing new structures
for \QC~codes to easily identify where the cycles can exist and how
to avoid them.

The methodology  presented in this thesis could be used
 to expand the understanding of \QC~codes. 
It would be possible to continue the search of codes with high girth
and look for the conditions that
guarantee girth twelve  but, in the authors opinion, such
work would be long, tedious and would carry  little benefit.
In fact, other parameters of the code, such as the minimum distance, the diameter, the
minimum stopping set and others, will became dominant and
prevent improvement in the performance of the code.
Of more interest would be to study if the methodology presented
could be applied to the problem of finding codes with good diameter
value.
It could be possible to develop a set of conditions that ensures the
diameter to be a certain value. The study of how these rules intersect with the
girth conditions could lead to classes of codes that can achieve given
girth and diameter and offer a good balance between the two.


\addcontentsline{toc}{chapter}{Bibliography}
\bibliographystyle{IEEEtran}
\bibliography{IEEEabrv,bibliography}
\section*{Appendix}
\subsection*{Proof of theorem \ref{the:gencase8}}\
\label{sec:AP:gencase8}

\begin{theorem} [\ref{the:gencase8}]
Let $B\in \M_{m,\alpha,\beta,\gamma}$.

The only possible $8$-cycle configurations are as follows
(1)-presentations):
%

\begin{enumerate}
\item \label{item:g8-1,1a} 
      $(1,1)$, $$\left| 8 \right|,$$
\item  \label{item:g8-1,2a}    $(1,2)$,\\
 $$
 \begin{array}{cc}
        \setcounter{newname}{1}
        \theenumi.\arabic{newname} \label{item:g8-1.2.1a}
      	\left|\begin{array}{cc}
        6 & 2 \\
       \end{array}\right|,&
        \addtocounter{newname}{1}
        \theenumi.\arabic{newname}
      \label{item:g8-1.2.2a}
      	\left|\begin{array}{cc}
        4 & 4 \\
       \end{array}\right|,
    \end{array}
$$
\item\label{item:g8-1,3a}
      $(1,3)$,
      $$\left|\begin{array}{ccc}
        4 & 2 & 2 \\
       \end{array}\right|,$$

\item\label{item:g8-1,4a}
      $(1,4)$,
      $$\left|\begin{array}{cccc}
        2 & 2 & 2 & 2\\
       \end{array}\right|,$$

\item \label{item:g8-2,2a}
      $(2,2)$,\\
   $$ \begin{array}{ccc}
        \setcounter{newname}{1}
        \theenumi.\arabic{newname}

	\label{item:g8-2,2.1a}
      	\left|\begin{array}{cc}
        5 & 1 \\
        1 & 1 \\
        \end{array}\right|,&
        \addtocounter{newname}{1}
        \theenumi.\arabic{newname}

	\label{item:g8-2,2.2a}
      	\left|\begin{array}{cc}
        4 & 2 \\
        2 & 0 \\
        \end{array}\right|,&
        \addtocounter{newname}{1}
        \theenumi.\arabic{newname}

	\label{item:g8-2,2.3a}
      	\left|\begin{array}{cc}
        4 & 2 \\
        0 & 2 \\
        \end{array}\right|,\\ \\
        \addtocounter{newname}{1}
        \theenumi.\arabic{newname}
	\label{item:g8-2,2.4a}
      	\left|\begin{array}{cc}
        3 & 1 \\
        3 & 1 \\
        \end{array}\right|,&
        \addtocounter{newname}{1}
        \theenumi.\arabic{newname}
	\label{item:g8-2,2.5a}
      	\left|\begin{array}{cc}
        3 & 1 \\
        1 & 3 \\
        \end{array}\right|,&
        \addtocounter{newname}{1}
        \theenumi.\arabic{newname}
	\label{item:g8-2,2.6a}
      	\left|\begin{array}{cc}
        2 & 2 \\
        2 & 2 \\
        \end{array}\right|,

	\end{array}
$$

\item \label{item:g8-2,3a}
      $(2,3)$,\\
   $$ \begin{array}{ccc}
        \setcounter{newname}{1}
        \theenumi.\arabic{newname}
	\label{item:g8-2,3.1a}
      	\left|\begin{array}{ccc}
        4 & 1 & 1 \\
        0 & 1 & 1 \\
        \end{array}\right|,&
        \addtocounter{newname}{1}
        \theenumi.\arabic{newname}
	\label{item:g8-2,3.2a}
      	\left|\begin{array}{ccc}
        3 & 2 & 1 \\
        1 & 0 & 1 \\
        \end{array}\right|,&
        \addtocounter{newname}{1}
        \theenumi.\arabic{newname}
	\label{item:g8-2,3.3a}
      	\left|\begin{array}{ccc}
        3 & 1 & 0 \\
        1 & 1 & 2 \\
        \end{array}\right|,\\ \\
        \addtocounter{newname}{1}
        \theenumi.\arabic{newname}
	\label{item:g8-2,3.4a}
      	\left|\begin{array}{ccc}
        2 & 2 & 2 \\
        2 & 0 & 0 \\
        \end{array}\right|,&
        \addtocounter{newname}{1}
        \theenumi.\arabic{newname}
	\label{item:g8-2,3.5a}
      	\left|\begin{array}{ccc}
        2 & 1 & 1 \\
        2 & 1 & 1 \\
        \end{array}\right|,&
        \addtocounter{newname}{1}
        \theenumi.\arabic{newname}
	\label{item:g8-2,3.6a}
      	\left|\begin{array}{ccc}
        2 & 2 & 0 \\
        2 & 0 & 2 \\
        \end{array}\right|,
	\end{array}
$$
\item\label{item:g8-2,4a}
      $(2,4)$,\\
   $$ 
   \begin{array}{ccc}
        \setcounter{newname}{1}
        \theenumi.\arabic{newname}
	\label{item:g8-2,4.1a}
      	\left|\begin{array}{cccc}
        2 & 2 & 1 & 1\\
        0 & 0 & 1 & 1\\
        \end{array}\right|,&
        \addtocounter{newname}{1}
        \theenumi.\arabic{newname}
	\label{item:g8-2,4.2a}
      	\left|\begin{array}{cccc}
        2 & 1 & 1 & 0\\
        0 & 1 & 1 & 2\\
        \end{array}\right|,&
        \addtocounter{newname}{1}
        \theenumi.\arabic{newname}

	\label{item:g8-2,4.3a}
      	\left|\begin{array}{cccc}
        1 & 1 & 1 & 1\\
        1 & 1 & 1 & 1\\
        \end{array}\right|,
	\end{array}
$$

\item\label{item:g8-3,3a}
      $(3,3)$,\\
   $$ \begin{array}{cccc}
        \setcounter{newname}{1}
        \theenumi.\arabic{newname}
	\label{item:g8-3,3.1a}
      	\left|\begin{array}{ccc}
        3 & 1 & 0 \\
        1 & 0 & 1 \\
        0 & 1 & 1 \\
        \end{array}\right|,&
        \addtocounter{newname}{1}
        \theenumi.\arabic{newname}
	\label{item:g8-3,3.2a}
      	\left|\begin{array}{ccc}
        2 & 1 & 1 \\
        2 & 0 & 0 \\
        0 & 1 & 1 \\
        \end{array}\right|,&
        \addtocounter{newname}{1}
        \theenumi.\arabic{newname}
	\label{item:g8-3,3.3a}
      	\left|\begin{array}{ccc}
        2 & 1 & 1 \\
        1 & 1 & 0 \\
        1 & 0 & 1 \\
        \end{array}\right|,&
        \addtocounter{newname}{1}
        \theenumi.\arabic{newname}
	\label{item:g8-3,3.4a}
      	\left|\begin{array}{ccc}
        2 & 0 & 0 \\
        1 & 1 & 0 \\
        1 & 1 & 2 \\
        \end{array}\right|.
	\end{array}
$$
\item\label{item:g8-3,4a}
$(3,4)$,\\
   $$ \begin{array}{cc}
        \setcounter{newname}{1}
        \theenumi.\arabic{newname}
	\label{item:g8-3,4.1a}
        \left|\begin{array}{cccc}
        2 & 1 & 1 & 0\\
        0 & 1 & 0 & 1\\
        0 & 0 & 1 & 1\\
        \end{array}\right|,&
        \addtocounter{newname}{1}
        \theenumi.\arabic{newname},
        \label{item:g8-3,4.2a}
        \left|\begin{array}{cccc}
        1 & 1 & 1 & 1\\
        1 & 1 & 0 & 0\\
        0 & 0 & 1 & 1\\
        \end{array}\right|,
        \end{array}
$$
\item\label{item:g8-4,4a}
      $(4,4)$,
     $$\left|\begin{array}{cccc}
        1 & 1 & 0 & 0\\
        1 & 0 & 1 & 0\\
        0 & 1 & 0 & 1\\
        0 & 0 & 1 & 1\\
        \end{array}\right|.$$

\end{enumerate}
\begin{proof}

Lemma~\ref{lem:class-2s-cycle} proves that these are all the possible dimensions that 8-cycle configurations can have. It is now necessary to prove that they are also all and only valid configurations.

Following from Theorem~\ref{th:possible_WV} the only possible \WV s for a 8-cycles are:
$$
\begin{array}{l}
\left[8\right],\\
\aout{\left[7,1\right]},\\
\left[6,2\right],\aout{\left[6,1,1\right]},\\
\aout{\left[5,3\right]},\aout{\left[5,2,1\right]},\left[5,1,1,1\right],\\
\left[4,4\right],\aout{\left[4,3,1\right]},\left[4,2,2\right],\aout{\left[4,2,1,1\right]},\left[4,1,1,1,1\right],\\
\aout{\left[3,3,2\right]},\left[3,3,1,1\right],\aout{\left[3,2,2,1\right]},\left[3,2,1,1,1\right],\left[3,1,1,1,1,1\right],\\
\left[2,2,2,2\right],\aout{\left[2,2,2,1,1\right]},\left[2,2,1,1,1,1\right],\left[2,1,1,1,1,1,1\right],\\
\left[1,1,1,1,1,1,1,1\right],\\
\end{array}
$$
\hrule
\vskip 2mm
\noindent
{\bf Configuration \ref{item:g8-1,1a}}
A type $(1,1)$ configuration can be generated only by \WV~$\left[8\right]$ and  it corresponds to case \ref{item:g8-1,1a}.\\
\hrule
\vskip 2mm
\noindent
{\bf Configuration \ref{item:g8-1,2a}}
Type $(1,2)$ configurations can be generated only by \WV s
$\left[6,2\right]$ and $\left[4,4\right]$. They correspond to the cases \ref{item:g8-1,2a}.1 and \ref{item:g8-1,2a}.2.\\
\hrule
\vskip 2mm
\noindent
{\bf Configuration \ref{item:g8-1,3a}}
A type $(1,3)$ configuration can be generated only by  \WV~ $\left[4,2,2\right]$ and it corresponds to case \ref{item:g8-1,3a}.
\hrule
\vskip 2mm
\noindent
{\bf Configuration \ref{item:g8-1,4a}}
A type $(1,4)$ configuration can be generated only by  \WV~
$\left[2,2,2,2\right]$ and it corresponds to case 
\ref{item:g8-1,4a}.
\hrule
\vskip 2mm
\noindent
{\bf Configuration \ref{item:g8-2,2a}}

Applying Theorem~\ref{th:r,c_VW}, configurations of type $(2,2)$ must have one of the following \WV s:
$$
\begin{array}{l}
\left[5,1,1,1\right],\left[4,2,2\right],\left[3,3,1,1\right],\left[2,2,2,2\right].
\end{array}
$$

\begin{itemize}
\item $\left[5,1,1,1\right]$  clearly can only generate configuration \ref{item:g8-2,2a}.1

\item $\left[4,2,2 \right]$ can only results in configurations
  \ref{item:g8-2,2a}.2 and \ref{item:g8-2,2a}.3. 
It is necessary to prove that they are the only two configurations
arising from this \WV.\\ $t_{1,1}$ 
is equal to 4 because it is the max.
There are two d.c.'s and for any of these d.c.'s  the column
weight  must be even and not zero 
(if there is a zero row then it falls in a smaller configuration). 
The sum of the column weights must be $8$ hence, or c-1 and c-2 column
weights are $4$, or c-1 weights $6$ and c-2 weights $2$ 
(the case c-1 column weight equal to  $2$ and c-2 column weight equal to $6$ is not considered because  $t_{1,1}=4$).
In the first case the only possibility is c-1=
	\small{$\left|\begin{array}{cc}
        4  &     0 \\
        \end{array}\right|^T$}, and c-2=
	\small{$\left|\begin{array}{cc}
        2 &      2 \\
        \end{array}\right|^T$}, this is configuration \ref{item:g8-2,2a}.3.
In the second case c-1 can only be 
	\small{$\left|\begin{array}{cc}
        4 &     2 \\
        \end{array}\right|^T$},  c-2 can be or 
	\small{$\left|\begin{array}{cc}
        2  &      0 \\
        \end{array}\right|^T$} and that is configuration \ref{item:g8-2,2a}.2, or
	\small{$\left|\begin{array}{cc}
        0 &       2 
        \end{array}\right|^T$ } that is  the transpose of  configuration \ref{item:g8-2,2a}.3.

\item $\left[3,3,1,1 \right]$ can only results in configurations
  \ref{item:g8-2,2a}.4 and \ref{item:g8-2,2a}.5. $t_{1,1}$ is equal to 3 because
  it is the max. 
As in the previous case or c-1 and c-2 have weight 4, or c-1 has weight 6 and c-2 has weight 2. \\
In the first case, or  c-1 is 
	\small{$\left|\begin{array}{cc}
        3  &        1 \\
        \end{array}\right|^T$} and c-2 is
	\small{$\left|\begin{array}{cc}
        1 &        3 \\
        \end{array}\right|^T$}, or  c-1 is 
	\small{$\left|\begin{array}{cc}
        3  &        1 \\
        \end{array}\right|^T$} and c-2 is
	\small{$\left|\begin{array}{cc}
        3 &        3 \\
        \end{array}\right|^T$},
the first correspond to configuration~\ref{item:g8-2,2a}.5  the second
to the transpose of  Configuration~\ref{item:g8-2,2a}.4.\\
In the second case  c-1 must be  
	\small{$\left|\begin{array}{cc}
        3  &        3 \\
        \end{array}\right|$}, and c-2 must be
	\small{$\left|\begin{array}{cc}
        1 &        1 \\
        \end{array}\right|^T$},   this is configuration \ref{item:g8-2,2a}.4.
\item $\left[2,2,2,2 \right]$ evidently can generate only configuration \ref{item:g8-2,2a}.6.\\
\end{itemize}

\hrule
\vskip 2mm
\noindent
{\bf Configuration \ref{item:g8-2,3a}}

Applying Theorem~\ref{th:r,c_VW}, configurations of type $(2,3)$ must have one of the following \WV s:
$$
\begin{array}{c}
\left[4,1,1,1,1\right],\left[3,3,1,1\right],\left[3,2,1,1,1\right],\left[3,1,1,1,1,1\right],\\
\left[2,2,2,2\right],\left[2,2,1,1,1,1\right].\\
\end{array}
$$

\begin{itemize}
\item $\left[4,1,1,1,1\right]$  is the \WV~ for configuration \ref{item:g8-2,3a}.1. This is the only possible configurations rising from such \WV~because otherwise there would be a column with odd weight.

\item $\left[3,3,1,1 \right]$ does not result in any valid configurations. In fact all entries are odd, hence every d.c. must have two elements to make its column weight even, but there are not sufficient entries to fill three columns.

\item $\left[3,2,1,1,1\right]$ can results only in configurations
  \ref{item:g8-2,3a}.2 and \ref{item:g8-2,3a}.3. It is necessary to prove that
  they are the only two configurations erasing from this \WV.\\
  $t_{1,1}$ is equal to 3 because it is the max.
As discussed previously the column weights of each column must be
 at least $2$. The sum of the columns weight must be $8$
 hence there must be one d.c. with weight $4$ and  two d.c.'s with weight $2$.
c-1 must have  weight $4$ since $t_{1,1}=3$. 
Considering also that there are two d.r.'s and the sum of their row
 weights must be $8$ then it must be either that  both the row weights are $4$ or one is $6$ and the other $2$.
If r-1 has weight $6$ then $t_{1,2}+t_{1,3}=3$ hence $t_{1,2}=2$ and
 $t_{1,3}=1$ 
( remember that $t_{1,2}\geq t_{1,3}$), 
applying Lemma \ref{lem:ganzo} c-2 c-3 configuration \ref{item:g8-2,3a}.2 is
 obtained.
 If r-1 has weight $4$ then $t_{1,2}+t_{1,3}=1$ that implies  $t_{1,2}=1$, $t_{1,3}=0$ ($t_{1,2}\geq t_{1,3}$), applying Lemma \ref{lem:ganzo}  configuration \ref{item:g8-2,3a}.3 is found. 

\item $\left[3,1,1,1,1,1\right]$ cannot generate valid configurations. In fact there are 6 entries for six positions so the configurations must be full, but this cause both d.r.'s to have odd row weight.
\item $\left[2,2,2,2\right]$ can only results in configurations
  \ref{item:g8-2,3a}.4 and \ref{item:g8-2,3a}.6. 
$t_{1,1}$ is equal to 2 because it is the max.  As for the previous case or r-1 has weight 6 and r-2 has weight 2 or both have weight 4. It is clear that one case generate \ref{item:g8-2,3a}.4 and the other \ref{item:g8-2,3a}.6 or equivalent.
\item $\left[2,2,1,1,1,1\right]$ can only generate configurations
  \ref{item:g8-2,3a}.5, in fact it is the only possible way to arrange 
the entries such that Lemma~\ref{lem:ganzo} is valid for every column.
\end{itemize}

\hrule
\vskip 2mm
\noindent
{\bf Configuration \ref{item:g8-2,4a}}

Applying Theorem~\ref{th:r,c_VW}, configurations of type $(2,4)$ must have one of the following \WV s:
$$
\begin{array}{c}
\left[2,2,1,1,1,1\right],\left[2,1,1,1,1,1,1\right],\left[1,1,1,1,1,1,1,1\right].\\
\end{array}
$$

\begin{itemize}

\item $\left[2,2,1,1,1,1\right]$ can only generate configurations
  \ref{item:g8-2,4a}.1 and \ref{item:g8-2,4a}.2. $t_{1,1}$ is equal to 2, and each d.c. must have column weight two because there are four non zero columns that must sum to 8.  
Considering also that there are two d.r.'s and the sum of their row
weights must be $8$  then or one has row weight $6$ and the other $2$
or both the row weights are $4$. In the first case the d.r. of weight 6
must be $\left[2,2,1,1\right]$ (the only four entries that sum to six),
applying Lemma~\ref{lem:ganzo} 
on all the d.c.'s configuration \ref{item:g8-2,4a}.1 or equivalent is
obtained.
 In the second case both rows must contain the entries $2,1,1,0$ (in
 the proper order), in fact if a d.r. contains $\left[2,2\right]$ then the other
 elements of that row must be zeros 
(otherwise the row weight would be  greater than 4) but also the
 elements of the two columns must be zero (column weight must be
 two). This situation cannot be for Lemma~\ref{lem:isolation}. Once
 fixed the first row the second row can be obtained  applying
 Lemma~\ref{lem:ganzo} on all the d.c.'s and configuration
 \ref{item:g8-2,4a}.1 or equivalent is obtained.
\item $\left[2,1,1,1,1,1,1\right]$ cannot generate valid configurations. In fact $t_{1,1}=2$ and Lemma~\ref{lem:ganzo} c-1 implies that  $t_{2,1}=0$. The other six positions must be filled with 1-elements but this cause both d.r.'s to have odd row weight.
\item $\left[1,1,1,1,1,1,1,1\right]$ clearly can only results in configuration \ref{item:g8-2,4a}.3.
\end{itemize}

\hrule
\vskip 2mm
\noindent
{\bf Configuration \ref{item:g8-3,3a}}

Applying Theorem~\ref{th:r,c_VW}, configurations of type $(3,3)$ must have one of the following \WV s:
$$
\begin{array}{c}
\left[4,1,1,1,1\right],\left[3,2,1,1,1\right],\left[3,1,1,1,1,1\right],\left[2,2,1,1,1,1\right],\left[2,1,1,1,1,1,1\right],\left[1,1,1,1,1,1,1,1\right].\\
\end{array}
$$
\begin{itemize}

\item $\left[4,1,1,1,1\right]$ cannot generate valid
  configurations. In fact  there are three d.c.'s and for any of this
  column the weight must be even and not zero, the only possibility to
  have a total sum of eight and column weight even is to have one
  column with weight $4$ and two with weight $2$; the same is true for
  the d.r.'s. This forces  the 4-element to be isolated and this cannot be.

\item $\left[3,2,1,1,1\right]$  cannot generate valid
  configurations. As proved in the previous case one d.c. and one d.r.
  have column weight four and the others weight two. $t_{1,1}$ is
  equal to  3, hence the row weight of r-1 and column weight of c-1 must be 4,
  this implies that the d.m. with the 2-element is isolated, since it
  can only  be in the row and column with weights 2:
\small{$\left|\begin{array}{ccc}
        3 & 1 & 0 \\
        1 & 1 & 0 \\
	0 & 0 & 2 \\
        \end{array}\right|$} and this cannot be.

\item $\left[3,1,1,1,1,1\right]$ can only generate
configuration \ref{item:g8-3,3a}.1. As proved in the previous case one
d.c. and one d.r.  have column weight and row weight four and the
others weight two, moreover  $t_{1,1}=3$. This implies that  in r-1 and c-1 there must be only
another $1$-element to complete r-1 and c-1 to row and column weight
four. Without loss of generality 
it is supposed them to be in $t_{1,2}$ and $t_{2,1}$.  In r-2 and c-2
now there must be only another $1$-element (column and row weights of
the column/rows different from the first must have weight $2$). If the 1-element falls in $t_{2,2}$ 
then there a  configuration of the type 
\small{$\left|\begin{array}{ccc}
        3 & 1 & 0 \\
        1 & 1 & 0 \\
	0 & 0 & x \\
        \end{array}\right|$} is obtained.\\
 This configuration is not possible for Lemma \ref{lem:isolation}. \\
If the 1-element does  not fall in $t_{2,2}$ then there must be a
1-element in position $t_{3,2}$  to complete c-2 and one in  $t_{2,3}$
to complete r-2, hence $t_{3,3}=1$ Lemma \ref{lem:ganzo} r-3 c-3. 
Configuration \ref{item:g8-3,3a}.1 is obtained.

\item $\left[2,2,1,1,1,1\right]$ can generate only  configurations
  \ref{item:g8-3,3a}.2 and  \ref{item:g8-3,3a}.4. c-1 has weight four,
  since $t_{1,1}=2$ and it is not possible to obtain row weight four
  without using a 2-element.
 Note that it is not possible to fix also r-1 to have weight four because the d.r. with weight four could be composed with the other 2-element. 
Two cases are considered. In the first case  c-1 is completed with a $2$-element, without loss of generality it is possible to suppose to be $t_{2,1}= 2$. In this case the first column would be:
\small{$\left| 2\ 2\ 0\right|^T$}.
In the second case  c-1 is completed with two $1$-elements, hence the first column would be:
\small{$\left| 2\ 1\ 1\ \right|^T$}.
\begin{itemize}
\item c-1 is \small{$\left|2\ 2\ 0\right|^T$}.
 One of r-1, r-2 must have row weight four and it must be completed
 with two  $1$-elements. Supposing this to be r-1, then for Lemma
 \ref{lem:ganzo} c-2 c-3 and the fact that r-2 has already weight 2,
 $t_{3,2}=t_{3,3}=1$.
Configuration \ref{item:g8-3,3a}.2 is obtained. It the d.r. with   row
weight four is  r-2 a configuration that is a row permutation of this
is obtained.
 \item c-1 is \small{$\left|2\ 1\ 1\right|^T$}.     
	If  $t_{1,2}=2$ and $t_{1,3}=0$, or $t_{1,2}=0$
        and $t_{1,3}=2$ the transpose of the previous case is obtained
        and the resulting configuration is the transpose of the
        previous configuration.
	Due to Lemma \ref{lem:ganzo} r-1  and the fact that the row sum
        cannot be $>4$ 
        there are only two possibility for the remaining elements of r-1:
        $t_{1,2}=t_{1,3}=1$ or $t_{1,2}=t_{1,3}=0$. The first case
        cannot be because there is no possibility to place the
        2-element satisfying Lemma \ref{lem:ganzo} r-2 r-3 c-2 c-3. In the
        second case for Lemma \ref{lem:ganzo} applied to r-2 (or r-3) 
        it must be $t_{2(3),2}=1$, $t_{2(3),3}=2$ or $t_{2(3),2}=1$,
        $t_{2(3),3}=0$. It is evident how choosing between one of
        these two possibilities fixes (Lemma \ref{lem:ganzo} c-2 c-3)
        the values in r-3 in a configuration that is (apart for row  or column permutation) configuration \ref{item:g8-3,3a}.4.
\end{itemize}

\item $\left[2,1,1,1,1,1,1\right]$ can generate only configuration
  \ref{item:g8-3,3a}.3. c-1 has weight four,
  since $t_{1,1}=2$ and it is the only 2-element. For the same reason
  r-1 has row weight four, hence $t_{1,2}=t_{1,3}=t_{2,1}=t_{3,1}=1$.
Applying Lemma~\ref{lem:ganzo} r-2, r-3, c-2, c-3 a configuration that is
(apart for row  or column permutation) Configuration \ref{item:g8-3,3a}.3 is
obtained.
\item $\left[1,1,1,1,1,1,1,1\right]$ cannot generate valid configurations. In fact it would be impossible to have column weight four.
\end{itemize}

\hrule
\vskip 2mm
\noindent
{\bf Configuration \ref{item:g8-3,4a}}

Applying Theorem~\ref{th:r,c_VW}, configurations of type $(3,4)$ must have one of the following \WV s:
$$
\begin{array}{c}
\left[2,2,1,1,1,1\right],\left[2,1,1,1,1,1,1\right],\left[1,1,1,1,1,1,1,1\right].\\
\end{array}
$$
\begin{itemize}
\item $\left[2,2,1,1,1,1\right]$ cannot generate valid
  configurations. In fact  there are four d.c.'s  and for any of this
  column the weight must be at least $2$.  This implies that for each
  column the weight is exactly $2$ otherwise the total sum would exceed  $8$.
 Also, there are three rows hence, as discussed previously for a
 similar case, the only possibility is to have one row with row weight
 $4$ and the other two with row weights $2$.
 If a 2-element lies on a d.r. with row weight two that row is
 complete ans so it is  column,  hence the 2-element is isolated and this is not
possible (Lemma~\ref{lem:isolation}).
 It must be that both the 2-elements lie in the row with row weight
 four,  nothing else lies in the such row, but this cannot be because
 they would be isolated since all d.c.'s have column weight two.
\item $\left[2,1,1,1,1,1,1\right]$ can only generate our
configuration \ref{item:g8-3,4a}.1. $t_{1,1}$ is equal to 2, hence  r-1 must
have row weight four, otherwise $T_{1,1}$ would be isolated. Hence
there are  two 1-elements in r-1. Applying Lemma~\ref{lem:ganzo} r-2, r-3,
c-2, c-3 
a configuration that is (apart for row  or column permutation)
Configuration \ref{item:g8-3,4a}.1 is obtained.
\item $\left[1,1,1,1,1,1,1,1\right]$ can generate only  configuration
  \ref{item:g8-3,4a}.2. The row with weight four must have all
$1$-elements,  suppose  this is  the first. 
Considering  Lemma \ref{lem:ganzo} applied to all columns and rows and
the fact that row weight of r-2 and r-3 are equal to $2$,  configuration \ref{item:g8-3,4a}.2 or equivalent is obtained.
\end{itemize}

\hrule
\vskip 2mm
\noindent
{\bf Configuration \ref{item:g8-4,4a}}

Applying Theorem~\ref{th:r,c_VW}, configurations of type $(4,4)$ must have one of the following \WV s:
$$
\begin{array}{c}
\left[2,1,1,1,1,1,1\right],\left[1,1,1,1,1,1,1,1\right].\\
\end{array}
$$
\begin{itemize}
\item $\left[2,1,1,1,1,1,1\right]$ cannot generate valid
  configurations. In fact  there are  four d.c.'s  and four d.r.'s,
  for any of this column the weight must be at least $2$.
  This implies that for each column
  and row the weight is exactly $2$ otherwise the total sum will exceed
  $8$. Wherever the 2-element lies that row and column is completed (every row and column has weight two), hence it is isolated,  but this cannot be for Lemma \ref{lem:isolation}.
\item $\left[1,1,1,1,1,1,1,1\right]$ can only generate configuration \ref{item:g8-3,4a}. 
It is only a case by case analysis to prove that any possible
distribution of ones, such that form a cycle, can be re-conduced to configuration \ref{item:g8-4,4a} by a combination of rows and column permutations.
.
\end{itemize}

\end{proof}	

\end{theorem}

\subsection*{Proof of theorem \ref{the:QC_g8}}
\begin{theorem}[\ref{the:QC_g8}]
 Let be $M \in \C_{m,\alpha,\beta,\gamma}$. The configurations in $M$
that may contain a cycles of length $8$,  are the
following \footnote{For brevity the transposes are omitted}
\begin{enumerate}
    \item $$
            \left| C-8 \right|,
          $$
    \item $$
            \left| C-6 \quad C-2 \right|,
          $$
    \item $$
            \left| C-4 \quad C-4 \right|,
          $$

    \item $$
            \left| C-4 \quad C-2 \quad C-2  \right| \,.
          $$
    \item $$
            \left| C-2 \quad C-2 \quad C-2 \quad C-2 \right| \,.
          $$
    \item $$
        \left|\begin{array}{cc}
        C-5      & \Delta-1 \\
        \Delta-1 & \Delta-1 \\
       \end{array}\right|\,,
        $$
    \item $$
        \left|\begin{array}{cc}
        C-4      & \Delta-2 \\
        0        & \C-2 \\
       \end{array}\right|\,,
        $$
    \item $$
        \left|\begin{array}{cc}
        C-4 & C-2 \\
        C-2 & 0 \\
       \end{array}\right|\,,
        $$
    \item $$
        \left|\begin{array}{cc}
        C-3      & \Delta-1 \\
        C-3 & \Delta-1 \\
       \end{array}\right|\,,
        $$
    \item $$
        \left|\begin{array}{cc}
        C-3      & \Delta-1 \\
        \Delta-1 & C-3 \\
       \end{array}\right|\,,
        $$
    \item $$
        \left|\begin{array}{cc}
        \Delta-2      & \Delta-2 \\
        \Delta-2 & \Delta-2 \\
       \end{array}\right|\,,
        $$
\item $$
    \left|\begin{array}{ccc}
        C-4 & \Delta-1 & \Delta-1 \\
         O  & \Delta-1    & \Delta-1 \\
       \end{array}\right|,
    $$
\item $$
    \left|\begin{array}{ccc}
        C-3     & C-2   & \Delta-1 \\
      \Delta-1  &  O    & \Delta-1 \\
       \end{array}\right|,
    $$
\item $$
    \left|\begin{array}{ccc}
        C-3     & \Delta-1 & O   \\
      \Delta-1  & \Delta-1 &  C-2  \\
       \end{array}\right|,
    $$
\item $$
    \left|\begin{array}{ccc}
       \Delta-2 & C-2   & C-2 \\
         C-2    & O & O \\
       \end{array}\right|,
    $$
\item $$
    \left|\begin{array}{ccc}
       \Delta-2 & \Delta-1   & \Delta-1 \\
       \Delta-2 & \Delta-1   & \Delta-1 \\
       \end{array}\right|,
    $$
\item $$
    \left|\begin{array}{ccc}
       C-2 & \Delta-2   &  O\\
       O        & \Delta-2   & C-2 \\
       \end{array}\right|,
    $$

\item $$
    \left|\begin{array}{cccc}
       C-2 & C-2  & \Delta-1 & \Delta-1 \\
       O   &  O   & \Delta-1 & \Delta-1 \\
       \end{array}\right|,
    $$
\item $$
    \left|\begin{array}{cccc}
       C-2 & \Delta-1 & \Delta-1  &  O \\
       O    & \Delta-1 & \Delta-1 & C-2 \\
       \end{array}\right|,
    $$
\item $$
    \left|\begin{array}{cccc}
       \Delta-1 & \Delta-1  & \Delta-1 & \Delta-1 \\
       \Delta-1 & \Delta-1  & \Delta-1 & \Delta-1 \\
       \end{array}\right|,
    $$

    \item $$
    \left|\begin{array}{ccc}
           C-3 & \Delta-1 &    O     \\
        \Delta-1 &    O     & \Delta-1 \\
            O    & \Delta-1 & \Delta-1 \\
       \end{array}\right|\,.
    $$
    \item $$
    \left|\begin{array}{ccc}
          \Delta-2 & \Delta-1 &    \Delta-1     \\
        C-2 &    O     & O \\
            O    & \Delta-1 & \Delta-1 \\
       \end{array}\right|\,.
    $$
    \item $$
    \left|\begin{array}{ccc}
        \Delta-2 & \Delta-1 & \Delta-1     \\
        \Delta-1 & \Delta-1 & O \\
        \Delta-1 &  O       & \Delta-1 \\
       \end{array}\right|\,.
    $$
    \item $$
    \left|\begin{array}{ccc}
            C-2 &    O     & O \\
         \Delta-1 & \Delta-1 & O\\
          \Delta-1 & \Delta-1 &    C-2     \\
       \end{array}\right|\,.
    $$

    \item $$
    \left|\begin{array}{cccc}
        C-2 & \Delta-1 & \Delta-1 & O\\
           O     & \Delta-1 &     O    & \Delta-1\\
           O     &   O      & \Delta-1 & \Delta-1\\
       \end{array}\right|\,.
    $$
\item $$
    \left|\begin{array}{cccc}
        \Delta-1 & \Delta-1 & \Delta-1 & \Delta-1\\
        \Delta-1 & \Delta-1 &     O    & 0 \\
           O     &   O      & \Delta-1 & \Delta-1\\
       \end{array}\right|\,.
    $$
    \item $$
    \left|\begin{array}{ccccc}
        \Delta-1 & \Delta-1 &     O    & O\\
        \Delta-1 &    O     & \Delta-1 & O\\
           O     & \Delta-1 &     O    & \Delta-1\\
           O     &   O      & \Delta-1 & \Delta-1\\
       \end{array}\right|\,.
    $$

\end{enumerate}
\begin{proof}
All the configurations appeared in Theorem \ref{the:gencase8} are consider and the relative cycle configurations are provided.
\vskip 3mm
\hrule
\vskip 2mm
\noindent
{\bf Configuration \ref{item:g8-1,1a}} gives\\
$$
\left| C-8 \right| \,.
$$
Cycle configuration $|J-8|$ may be discarded (Lemma \ref{lem:lemmino5}).
\vskip 3mm
\hrule
\vskip 2mm
\noindent
{\bf Configuration \ref{item:g8-1,2a}} \\
\begin{enumerate}
\item Configuration \ref{item:g8-1,2a}.1 gives 
\begin{align*}
\left|\begin{array}{cc}
        C-6 & C-2 \\
       \end{array}\right| \,.
\end{align*}
\item Configuration \ref{item:g8-1,2a}.2 gives 
\begin{align*}
\left|\begin{array}{cc}
        C-4 & C-4 \\
       \end{array}\right| \,.
\end{align*}
\end{enumerate}
In fact other cycle configurations $|C-6\quad J-2|$, $|J-6\quad J-2|$ , $|J-6\quad C-2|$ and
$|C-4\quad J-4|$, $|J-4\quad J-4|$ , $|J-4\quad C-4|$ may be discarded because in $J-2$ and in $J-4$
there is no cycle columns (Lemma \ref{lem:lemmino1}-1).
\vskip 3mm
\hrule
\vskip 2mm
\noindent
{\bf Configuration \ref{item:g8-1,3a}} gives\\
\begin{align*}
\left|\begin{array}{ccc}
        C-6 & C-2 & C-2 \\
       \end{array}\right| \,.
\end{align*}
Other cycle configurations may be discarded because they contain a d.s.
$J-2$ or $J-6$ and in them there is no cycle columns (Lemma \ref{lem:lemmino1}-1).
\vskip 3mm
\hrule
\vskip 2mm
\noindent
{\bf Configuration \ref{item:g8-1,4a}} gives\\
\begin{align*}
\left|\begin{array}{cccc}
        C-2 & C-2 & C-2 & C-2 \\
       \end{array}\right| \,.
\end{align*}
Other cycle configurations  may be discarded because they contain a d.s.
$J-2$ and in it there is no cycle column (Lemma \ref{lem:lemmino1}-1).
\vskip 3mm
\hrule
\vskip 2mm
\noindent
{\bf Configuration \ref{item:g8-2,2a}}  \\
\begin{enumerate}
\item Configuration \ref{item:g8-2,2a}.1 gives
\begin{align*}
\left|\begin{array}{cc}
        C-5 & \Delta-1 \\
	\Delta-1 & \Delta-1 \\
       \end{array}\right| \,.
\end{align*}
Other cycle configurations  may be discarded because they contain a d.s.
$J-5$ (Lemma \ref{lem:lemmino5}).

\item Configuration \ref{item:g8-2,2a}.2 gives
\begin{align*}
\left|\begin{array}{cc}
        C-4 & C-2 \\
	C-2 & 0 \\
       \end{array}\right| \,.
\end{align*}
       Other cycle configurations  may be discarded because they contain a d.s.
$J-2$ as the only non-zero matrix in a d.r./d.c. or a $J-4$ and that cannot be (Lemma \ref{lem:lemmino5}).
\item Configuration \ref{item:g8-2,2a}.3 gives
\begin{align*}
\left|\begin{array}{cc}
        C-4 & \Delta-2 \\
	 & C-2 \\
       \end{array}\right| \,.
\end{align*}
       Other cycle configurations  may be discarded because they contain a d.s.
$J-2$ or $J-4$  as the only non-zero matrix in a d.r. or d.c.

\item Configuration \ref{item:g8-2,2a}.4 gives
\begin{align*}
\left|\begin{array}{cc}
        C-3 & \Delta-1 \\
	C-3 & \Delta-1 \\
       \end{array}\right| \,.
\end{align*}
       Other cycle configurations  may be discarded because they contain a d.s.
$J-3$  (Lemma \ref{lem:lemmino5}).
\item Configuration \ref{item:g8-2,2a}.5 gives
\begin{align*}
\left|\begin{array}{cc}
        C-3 & \Delta-1 \\
	\Delta-1 &  C-3\\
       \end{array}\right| \,.
\end{align*}
       Other cycle configurations  may be discarded because they contain a d.s.
$J-3$  (Lemma \ref{lem:lemmino5}).
\item Configuration \ref{item:g8-2,2a}.6 gives
\begin{align*}
\left|\begin{array}{cc}
        \Delta-2 & \Delta-2 \\
	\Delta-2 &  \Delta-2\\
       \end{array}\right| \,.
\end{align*}
       This cover all the possibilities cycle configuration for the particular case.
\end{enumerate}
\vskip 3mm
\hrule
\newpage
\hrule
\vskip 2mm
\noindent
{\bf Configuration \ref{item:g8-2,3a}}\\
\begin{enumerate}
\item Configuration \ref{item:g8-2,3a}.1 gives
\begin{align*}
\left|\begin{array}{ccc}
        C-4 & \Delta-1& \Delta-1 \\
	0 & \Delta-1 & \Delta-1\\
       \end{array}\right| \,.
\end{align*}
Other cycle configurations  may be discarded because they contain a d.s.
$J-4$  (Lemma \ref{lem:lemmino5}).
\item Configuration \ref{item:g8-2,3a}.2 gives
\begin{align*}
\left|\begin{array}{ccc}
        C-3 & C-2& \Delta-1 \\
	\Delta-1 & 0 & \Delta-1\\
       \end{array}\right| \,.
\end{align*}
 Other cycle configurations  may be discarded because they contain a d.s.
$J-2$ as the only non-zero matrix in a d.c. or a  $J-3$ and this cannot be (Lemma \ref{lem:lemmino5}).
\item Configuration \ref{item:g8-2,3a}.3 gives
\begin{align*}
\left|\begin{array}{ccc}
        C-3 & \Delta-1 & 0 \\
	\Delta-1 & \Delta-1 &C-2 \\
       \end{array}\right| \,.
\end{align*}
 Other cycle configurations  may be discarded because they contain a d.s.
$J-2$ as the only non-zero matrix in a d.c. or  they contain a $J-3$ this cannot be (Lemma \ref{lem:lemmino5}).
\item Configuration \ref{item:g8-2,3a}.4 gives
\begin{align*}
\left|\begin{array}{ccc}
        \Delta-2 & C-2& C-2 \\
	C-2 & 0 & 0\\
       \end{array}\right| \,.
\end{align*}
 Other cycle configurations may be discarded  because they contain a d.s.
$J-2$ as the only non-zero matrix in a d.c. or d.r. .
\item Configuration \ref{item:g8-2,3a}.5 gives
\begin{align*}
\left|\begin{array}{ccc}
        \Delta-2 & \Delta-1 & \Delta-1 \\
	\Delta-2 & \Delta-1 & \Delta-1 \\
       \end{array}\right| \,.
\end{align*}
       This cover all the possibilities cycle configuration for the particular case.
\item Configuration \ref{item:g8-2,3a}.6 gives
\begin{align*}
\left|\begin{array}{ccc}
        \Delta-2 & C-2 &0 \\
	\Delta-2 & 0 &C-2\\
       \end{array}\right| \,.
\end{align*}
Other cycle configurations  may be discarded because they contain a d.s.
$J-2$ as the only non-zero matrix in a d.c. .
\end{enumerate}
\vskip 3mm
\hrule
\vskip 2mm
\noindent
{\bf Configuration \ref{item:g8-2,4a}}\\
\begin{enumerate}
\item Configuration \ref{item:g8-2,4a}.1 gives
\begin{align*}
\left|\begin{array}{cccc}
        C-2 & C-2 & \Delta-1 & \Delta-1 \\
	0 & 0 & \Delta-1 & \Delta-1\\
       \end{array}\right| \,.
\end{align*}
    Other cycle configurations  may be discarded because they contain a d.s.
$J-2$ as the only non-zero matrix in a d.c. .
\item Configuration \ref{item:g8-2,4a}.2 gives
\begin{align*}
\left|\begin{array}{cccc}
        C-2 &  \Delta-1 & \Delta-1 & 0\\
	0  & \Delta-1 & \Delta-1 & C-2\\
       \end{array}\right| \,.
\end{align*}
   Other cycle configurations  may be discarded because they contain a d.s.
$J-2$ as the only non-zero matrix in a d.c. .
\item Configuration \ref{item:g8-2,4a}.3 gives
\begin{align*}
\left|\begin{array}{cccc}
        \Delta-1 & \Delta-1 & \Delta-1 & \Delta-1 \\
	\Delta-1 & \Delta-1 & \Delta-1 & \Delta-1\\
       \end{array}\right| \,.
\end{align*}
This cover all the possibilities cycle configuration for the particular case.
\end{enumerate}
\vskip 3mm
\hrule
\vskip 2mm
\noindent
{\bf Configuration \ref{item:g8-3,3a}}\\
\begin{enumerate}
\item Configuration \ref{item:g8-3,3a}.1 gives
\begin{align*}
\left|\begin{array}{ccc}
        C-3 & \Delta-1 & 0  \\
	\Delta-1 & 0 & \Delta-1 \\
	0 & \Delta-1 & \Delta-1 \\
       \end{array}\right| \,.
\end{align*}
  Other cycle configurations  may be discarded because they contain a d.s.
$J-3$ (Lemma \ref{lem:lemmino5}) .
\item Configuration \ref{item:g8-3,3a}.2 gives
\begin{align*}
\left|\begin{array}{ccc}
        \Delta-2 & \Delta-1 & \Delta-1  \\
	C-2 & 0 & 0 \\
	0 & \Delta-1 & \Delta-1 \\
       \end{array}\right| \,.
\end{align*}
 Other cycle configurations  may be discarded because they contain a d.s.
$J-2$ as the only non-zero matrix in a d.r. .
\item Configuration \ref{item:g8-3,3a}.3 gives
\begin{align*}
\left|\begin{array}{ccc}
        \Delta-2 & \Delta-1 & \Delta-1  \\
	\Delta-1 & \Delta-1 & 0 \\
	\Delta-1 & 0 & \Delta-1 \\
       \end{array}\right| \,.
\end{align*}
This cover all the possibilities cycle configuration for the particular case.
\item Configuration \ref{item:g8-3,3a}.4 gives
\begin{align*}
\left|\begin{array}{ccc}
        C-2 & 0 & 0  \\
	\Delta-1 & \Delta-1 & 0 \\
	\Delta-1 & \Delta-1 & C-2\\
       \end{array}\right| \,.
\end{align*}
Other cycle configurations  may be discarded because they contain a d.s.
$J-2$ as the only non-zero matrix in a d.r. or d.c. .
\end{enumerate}
\vskip 3mm
\hrule
\vskip 2mm
\noindent
{\bf Configuration \ref{item:g8-3,4a}}\\
\begin{enumerate}
\item Configuration \ref{item:g8-3,4a}.1 gives
\begin{align*}
\left|\begin{array}{cccc}
        \Delta-1 & \Delta-1 & \Delta-1 & \Delta-1 \\
	\Delta-1 & \Delta-1 & 0 & 0\\
	0 & 0 & \Delta-1 & \Delta-1\\
       \end{array}\right| \,.
\end{align*}
This cover all the possibilities cycle configuration for the particular case.
\item  Configuration \ref{item:g8-3,4a}.2 gives
\begin{align*}
\left|\begin{array}{cccc}
        \C-2 & \Delta-1 & \Delta-1 & 0  \\
	0 & \Delta-1 & 0 & \Delta-1\\
	0 & 0 & \Delta-1 & \Delta-1\\
       \end{array}\right| \,.
\end{align*}
Other cycle configurations  may be discarded because they contain a d.s.
$J-2$ as the only non-zero matrix in a d.c. .
\end{enumerate}
\vskip 3mm
\hrule
\vskip 2mm
\noindent
{\bf Configuration \ref{item:g8-4,4a}} \\
\begin{align*}
\left|\begin{array}{cccc}
        \Delta-1 & \Delta-1 & 0 & 0\\
	\Delta-1 & 0 & \Delta-1 & 0\\
	0 & \Delta-1 & 0 & \Delta-1\\
	0 & 0 & \Delta-1 & \Delta-1\\
       \end{array}\right| \,.
\end{align*}
This cover all the possibilities cycle configuration for the particular case.

For Remark~\ref{remark:g8-transpose} it is not necessary to study the cycle configurations that are transposed of the one considered.
Hence it has been proved that the listed configurations are the only valid.
\end{proof}
\end{theorem}

\subsection*{Proof of theorem \ref{the:conditions8}}
\begin{theorem}[\ref{the:conditions8}]
Let be $M \in \C_{m,\alpha,\beta,\gamma}$. The configurations in $M$
that may contain a cycles of length exactly $8$,  are the following 
\footnote{ Configurations with two or more weight-$2$ circulants in
the same row or column are  n
not listed since they always contain a cycle of at most $8$
(Lemma \ref{lemma:2C=g8}), 
hence they do not add any information.
}
\begin{enumerate}

    \item \label{item:g8-1a}
    \begin{align*} 
            &\left| C-8 \right|,& s(p)=m/4
          \end{align*}
    \item \label{item:g8-6a}
    \begin{gather*}
            \left|\begin{array}{cc}
        C^1-5      & J^2-1 \\
        J^3-1 & \Delta^4-1 \\
       \end{array}\right|,\\
       \epsilon(p^1)-\epsilon(p^2)-\epsilon(p^3)+\epsilon(p^4)\equiv\pm 2s(p^1),
        \end{gather*}
    \item \label{item:g8-7a}
    \begin{gather*}
            \left|\begin{array}{cc}
        C^1-4      & J^2-2 \\
        0        & \C^3-2 \\
       \end{array}\right|,\\
       \pm s(p^3) \equiv 2s(p^1) 
        \end{gather*}
    \item\label{item:g8-10a}
    	\begin{gather*}
            \left|\begin{array}{cc}
        C^1-3      & J^2-1 \\
        J^3-1 & C^4-3 \\
       \end{array}\right|,\\
\epsilon(p^1)-\epsilon(p^2)-\epsilon(p^3)+\epsilon(p^4)\equiv \pm s(p^1) \pm s(p^3),
        \end{gather*}
    \item\label{item:g8-11a}
     \begin{gather*}	
        \left|\begin{array}{cc}
        \Delta^1-2      & \Delta^2-2 \\
        \Delta^3-2 & \Delta^4-2 \\
       \end{array}\right|,\\
\epsilon(p^1)+\epsilon(p^1)-\epsilon(p^2)-\epsilon(p^2)-\epsilon(p^3)-\epsilon(p^3)+\epsilon(p^4)+\epsilon(p^4)\equiv 0\,,
\end{gather*}
\item \label{item:g8-13a}
\begin{gather*}
            \left|\begin{array}{ccc}
        C^1-4 & J^2-1 & J^3-1 \\
         O  & \Delta^4-1    & \Delta^5-1 \\
       \end{array}\right|,\\
\epsilon(p^2)-\epsilon(p^3)-\epsilon(p^4)+\epsilon(p^5)\equiv \pm 2s(p^1),
    \end{gather*}
\item \label{item:g8-15a}
\begin{gather*}
            \left|\begin{array}{ccc}
        C^1-3     & O   & J^2-1 \\
      J^3-1  &  C^4-2    & J^5-1 \\
       \end{array}\right|,\\
\epsilon(p^1)-\epsilon(p^2)-\epsilon(p^3)+\epsilon(p^5)\equiv \pm s(p^1) \pm s(p^4),
    \end{gather*}
\item \label{item:g8-17a}
\begin{gather*}
            \left|\begin{array}{ccc}
       \Delta^1-2 & \Delta^2-1   & \Delta^3-1 \\
       \Delta^4-2 & \Delta^5-1   & \Delta^6-1 \\
       \end{array}\right|,\\
\epsilon(p^1)+\epsilon(p^1)-\epsilon(p^2)-\epsilon(p^3)-\epsilon(p^4)-\epsilon(p^4)+\epsilon(p^5)+\epsilon(p^6)=0\,,
\mbox{ or }\\
\epsilon(p^2)-\epsilon(p^3)+\ep(p^4)-\ep(p^4)-\epsilon(p^5)+\epsilon(p^6)=\pm s(p^1)\,,
\mbox{ or }\\
\epsilon(p^2)-\epsilon(p^3)+\ep(p^1)-\ep(p^1)-\epsilon(p^5)+\epsilon(p^6)=\pm s(p^4)\,.
    \end{gather*}
\item \label{item:g8-18a}
\begin{gather*}
            \left|\begin{array}{ccc}
       C^1-2 & J^2-2   &  O\\
       O        & J^3-2   & C^4-2 \\
       \end{array}\right|,\\
  s(p^1)\equiv  s(p^4),
    \end{gather*}
\item \label{item:g8-20a}
\begin{gather*}
            \left|\begin{array}{cccc}
       C^1-2 & O  & J^3-1 & J^4-1 \\
         O   &  C^2-2   & J^5-1 & J^6-1 \\
       \end{array}\right|,\\
\epsilon(p^3)-\epsilon(p^4)-\epsilon(p^5)+\epsilon(p^6)\equiv \pm s(p^1) \pm s(p^2),
    \end{gather*}
\item \label{item:g8-21a}
\begin{gather*}
            \left|\begin{array}{cccc}
       \Delta^1-1 & \Delta^2-1  & \Delta^3-1 & \Delta^4-1 \\
       \Delta^5-1 & \Delta^6-1  & \Delta^7-1 & \Delta^8-1 \\
       \end{array}\right|,\\
\epsilon(p^1)+\epsilon(p^2)-\epsilon(p^3)-\epsilon(p^4)-\epsilon(p^5)-\epsilon(p^6)+\epsilon(p^7)+\epsilon(p^8)=0\,, \mbox{ or}\\
\epsilon(p^1)-\epsilon(p^2)+\epsilon(p^3)-\epsilon(p^4)-\epsilon(p^5)+\epsilon(p^6)-\epsilon(p^7)+\epsilon(p^8)=0\,, \mbox{ or}\\
\epsilon(p^1)-\epsilon(p^2)-\epsilon(p^3)+\epsilon(p^4)-\epsilon(p^5)+\epsilon(p^6)+\epsilon(p^7)-\epsilon(p^8)=0\,.
    \end{gather*}
    \item \label{item:g8-22a}
    \begin{gather*}
            \left|\begin{array}{ccc}
           C^1-3 & J^2-1 &    O     \\
           J^3-1 &    O     & \Delta^4-1 \\
            O    & \Delta^5-1 & \Delta^6-1 \\
       \end{array}\right|,\\
\epsilon(p^1)-\epsilon(p^2)-\epsilon(p^3)+\epsilon(p^4)+\epsilon(p^5)-\epsilon(p^6)\equiv \pm s(p^1),
    \end{gather*}
    \item \label{item:g8-23a}
    \begin{gather*}
            \left|\begin{array}{ccc}
          J^1-2 & \Delta^2-1 &    \Delta^3-1     \\
        C^4-2 &    O     & O \\
            O    & \Delta^5-1 & \Delta^6-1 \\
       \end{array}\right|,\\
\epsilon(p^2)-\epsilon(p^3)-\epsilon(p^5)+\epsilon(p^6)\equiv \pm s(p^4)
    \end{gather*}

    \item \label{item:g8-24a}
    \begin{gather*}
            \left|\begin{array}{ccc}
        \Delta^1-2 & \Delta^2-1 & \Delta^3-1     \\
        \Delta^4-1 & \Delta^5-1 & O \\
        \Delta^6-1 &  O       & \Delta^7-1 \\
       \end{array}\right|,\\
\epsilon(p^1)+\epsilon(p^1)-\epsilon(p^2)-\epsilon(p^3)-\epsilon(p^4)+\epsilon(p^5)-\epsilon(p^6)+\epsilon(p^7)=0\,, \mbox{ or}\\
\epsilon(p^2)-\epsilon(p^3)+\epsilon(p^4)-\epsilon(p^5)-\epsilon(p^6)+\epsilon(p^7)=\pm s(p^1),
    \end{gather*}
 \item \label{item:g8-25a}
 \begin{gather*}
            \left|\begin{array}{ccc}
            C^1-2 &    O     & O \\
            J^2-1 & \Delta^3-1 & O\\
            J^4-1 & J^5-1 &    \C^6-2     \\
       \end{array}\right|,\\
\epsilon(p^2)-\epsilon(p^3)-\epsilon(p^4)+\epsilon(p^5 )\equiv \pm s(p^1) \pm s(p^6),
    \end{gather*}
\item \label{item:g8-26-aa}
    \begin{gather*}
            \left|\begin{array}{cccc}
        \Delta^1-2 & \Delta^2-1 & \Delta^3-1 & \Delta^4-1\\
        \Delta^5-1 & \Delta^6-1 &     O    & 0 \\
           O     &   O      & \Delta^7-1 & \Delta^8-1\\
       \end{array}\right|,\\
\epsilon(p^2)-\epsilon(p^3)-\epsilon(p^4)+\epsilon(p^5)+\epsilon(p^6)-\epsilon(p^7)=\pm s(p^1),
    \end{gather*}
\item \label{item:g8-26a}
    \begin{gather*}
            \left|\begin{array}{cccc}
        C^1-2 & J^2-1 & J^3-1 & O\\
           O     & \Delta^4-1 &     O    & \Delta^5-1\\
           O     &   O      & \Delta^6-1 & \Delta^7-1\\
       \end{array}\right|,\\
\epsilon(p^2)-\epsilon(p^3)-\epsilon(p^4)+\epsilon(p^5)+\epsilon(p^6)-\epsilon(p^7)=\pm s(p^1),
    \end{gather*}
    \item \label{item:g8-27a}
    \begin{gather*}
            \left|\begin{array}{ccccc}
        \Delta^1-1 & \Delta^2-1 &     O    & O\\
        \Delta^3-1 &    O     & \Delta^4-1 & O\\
           O     & \Delta^5-1 &     O    & \Delta^6-1\\
           O     &   O      & \Delta^7-1 & \Delta^8-1\\
       \end{array}\right|,\\
\epsilon(p^1)-\epsilon(p^2)-\epsilon(p^3)+\epsilon(p^4)+\epsilon(p^5)-\epsilon(p^6)-\epsilon(p^7)+\epsilon(p^8)=0,
    \end{gather*}

\end{enumerate}

\begin{proof}

As for the theorems~\ref{the:condition4} and~\ref{the:condition6} the
proof is once more divided in smaller lemmas each considering a
particular case.\\
An $8$-cycle is formed by four  cycle comumns called $y,z,v,u$ and four
cycle rows called $x,t,w,l$. 
In every $8$-cycle there are $8$ cycle points, they take the name of
the column and row where they lie:
$(x,y),(x,z),(t,y),(t,v),(w,u),(w,z),(l,u),(l,v)$.\\The notation is used to compute
the conditions.
\newpage

\begin{lemma}\label{lem:g8-1} 
There is a $8$-cycle and there are no $6$-cycles or $4$-cycles  in case  \ref{item:g8-1a} if and only if
$$
   4|m  \mbox{ and } \quad
    \quad s(p)=m/4 \,.
$$

\begin{proof}

There is a $8$-cycle and no $6$-cycle or $4$-cycle if and only if $g=6$.
Applying Prop. \ref{prop:girth-circulant} to the case $g=8$ and
$M=C$, this is equivalent to
$$
   8= g =2 \frac{m}{\gcd(m,s)}
$$
i.e. $m= 4 \gcd(m,s)$. In particular, $4|m$ and
$m/4 | s$, but $s\leq m/2$, so that $s=m/4$. On the other hand,
if $m$ is divisible by $4$ and $s=m/4$ then $\gcd(m,s)=s=m/4$.
\end{proof}

\end{lemma}
\begin{lemma}\label{lem:g8-6} 
There is a $8$-cycle in case \ref{item:g8-6a}  if and only if
$$
\ep(p^1)-\ep(p^2)-\ep(p^3)+\ep(p^4) \equiv \pm 2s(p^1)
$$
\begin{proof}
It can be assumed that there is a $8$-cycle if and only if, simultaneously,
 cycle columns $y$ and $z$ lie in $C^1$,
 cycle point $(t,v)$ lies in $C^1$,
 cycle point $(w,u)$ lies in $\Delta^2$,
 cycle point $(l,v)$ lies in $\Delta^3$ and
 cycle point $(l,u)$ lies in $\Delta^4$.
Applying Proposition \ref{prop:cas}-4 to cycle column $y$ and cycle column $z$ yields
\begin{equation} \label{eq:dim_g8-6-1}
x-t \equiv \pm s^1 \,,
\end{equation}
\begin{equation} \label{eq:dim_g8-6-2}
x-w \equiv \mp s^1 \,.
\end{equation}
Since cycle point $(t,v)$ lies in $\C^1$, applying Proposition \ref{prop:cas}-1
\begin{equation} \label{eq:dim_g8-6-3}
v \equiv t +\ep(p^1) \,.
 \end{equation}
Since cycle point $(w,u)$ lies in $\Delta^2$, applying Proposition \ref{prop:cas}-1
\begin{equation} \label{eq:dim_g8-6-4}
u \equiv w +\ep(p^2) \,.
 \end{equation}
Since cycle point $(l,v)$ lies in $\Delta^3$, applying Lemma \ref{lem:id}
\begin{equation} \label{eq:dim_g8-6-5}
v \equiv l +\ep(p^3)\,.
 \end{equation}
Since cycle point $(l,u)$ lies in $\Delta^4$, applying Lemma \ref{lem:id}
\begin{equation} \label{eq:dim_g8-6-6}
u \equiv l +\ep(p^4) \,.
 \end{equation}
From (\ref{eq:dim_g8-6-3}), (\ref{eq:dim_g8-6-5}), and
(\ref{eq:dim_g8-6-4}), (\ref{eq:dim_g8-6-6}) : 
\begin{equation} \label{eq:dim_g8-6-7}
 l +\ep(p^3)\ \equiv t +\ep(p^1) \,.
 \end{equation}
 \begin{equation} \label{eq:dim_g8-6-8}
 l +\ep(p^4)\ \equiv w +\ep(p^2) \,.
 \end{equation}
Subtracting (\ref{eq:dim_g8-6-7}) from  (\ref{eq:dim_g8-6-8}) : 
\begin{align} \label{eq:dim_g8-6-9}
&\ep(p^4)-\ep(p^3)\equiv w-t +\ep(p^2)-\ep(p^1) \\
&\ep(p^2)-\ep(p^1)-\ep(p^3)+ \ep(p^4)\equiv w-t
 \end{align}
but $(w-t) = (x-t)-(x-w)$ hence from (\ref{eq:dim_g8-6-1}) and (\ref{eq:dim_g8-6-2}),
 \begin{equation} \label{eq:dim_g8-6-10}
 w-t=\pm 2 s(p^1)
 \end{equation}
The condition can eb obtained from (\ref{eq:dim_g8-6-9}) and (\ref{eq:dim_g8-6-10}).
\end{proof}

\end{lemma}
\begin{lemma}\label{lem:g8-7} 
 There is an $8$-cycle in case \ref{item:g8-7a} if and only if
 \begin{align*}
   \ep(p^2)-\ep(p^2)\equiv \pm s(p^3) \pm 2s(p^1)
\end{align*}
\begin{proof}
It can be assumed that there is a $8$-cycle if and only if 
columns $y$ and $z$ lie in $C^1$,
cycle row $l$ lies in $C^3$, 
 cycle point $(t,v)$ lies in $\Delta^2$, and
 cycle point $(w,u)$ lies in $\Delta^2$.
Applying Proposition \ref{prop:cas}-4 to cycle column $y$ and cycle column $z$ yields
\begin{equation} \label{eq:dim_g8-7-1}
x-t \equiv \pm s^1 \,,
\end{equation}
\begin{equation} \label{eq:dim_g8-7-2}
x-w \equiv \mp s^1 \,.
\end{equation}
Since cycle point $(t,v)$ lies in $\Delta^2$, applying Lemma \ref{lem:id}
\begin{equation} \label{eq:dim_g8-7-3}
v \equiv t+\ep(p^2)\,.
 \end{equation}
Since cycle point $(w,u)$ lies in $\Delta^2$, applying Lemma \ref{lem:id}
\begin{equation} \label{eq:dim_g8-7-4}
u \equiv w +\ep(p^2)\,.
 \end{equation}
Since cycle row $l$ lies in $C^3$, applying Proposition \ref{prop:cas}-3
\begin{equation} \label{eq:dim_g8-7-5}
v-u \equiv \pm s^3.
 \end{equation}
Subtracting (\ref{eq:dim_g8-7-4}) to (\ref{eq:dim_g8-7-3}) and substituting the result
into (\ref{eq:dim_g8-7-5}) and using (\ref{eq:dim_g8-7-1}),
 (\ref{eq:dim_g8-7-2})  
the  desired result is obtained.
\end{proof}
\end{lemma}
\begin{lemma}\label{lem:g8-10} 
 There is a $8$-cycle in case \ref{item:g8-10a} if and only if
 \begin{align*}
    \ep(p^1)-\ep(p^2)-\ep(p^3)+\ep(p^4)\equiv \pm s(p^1) \pm s(p^4)
\end{align*}
\begin{proof}
It can be assumed that there is a $8$-cycle if and only if 
cycle row $x$ lies in $C^1$,
cycle row $l$ lies in $C^4$,
cycle point $(w,z)$ lies in $C^1$,
cycle point $(w,u)$ lies in $J^2$,
cycle point $(t,y)$ lies in $J^3$, and
cycle point $(t,v)$ lies in $C^4$.
Since cycle row $x$ lies in $C^1$, applying Proposition \ref{prop:cas}-3
\begin{equation} \label{eq:dim_g8-10-1}
z-y \equiv \pm s^1\,.
 \end{equation}
Since cycle row $l$ lies in $C^4$, applying Proposition \ref{prop:cas}-3
\begin{equation} \label{eq:dim_g8-10-2}
u-v \equiv \pm s^4\,.
 \end{equation}
Since cycle point $(w,z)$ lies in $C^1$, applying Lemma \ref{lem:id}
\begin{equation} \label{eq:dim_g8-10-3}
z \equiv w+\ep(p^1)\,.
 \end{equation}
Since cycle point $(w,u)$ lies in $J^2$, applying Lemma \ref{lem:id}
\begin{equation} \label{eq:dim_g8-10-4}
u \equiv w+\ep(p^2)\,.
 \end{equation}
 Since cycle point $(t,y)$ lies in $J^3$, applying Lemma \ref{lem:id}
\begin{equation} \label{eq:dim_g8-10-5}
y \equiv t+\ep(p^3)\,.
 \end{equation}
Since cycle point $(t,v)$ lies in $C^4$, applying Lemma \ref{lem:id}
\begin{equation} \label{eq:dim_g8-10-6}
v \equiv t+\ep(p^4)\,.
 \end{equation}

The desired result can be computed form the previous formulas.
\end{proof}

\end{lemma}
\begin{lemma}\label{lem:g8-11} 
 There is a $8$-cycle in case \ref{item:g8-11a} if and only if
 \begin{align*}
\epsilon(p^1)+\epsilon(p^1)-\epsilon(p^2)-\epsilon(p^2)-\epsilon(p^3)-\epsilon(p^3)+\epsilon(p^4)+\epsilon(p^4)\equiv 0\,,
\end{align*}
\begin{proof}
Two cases are considered, the first when there is a circulant column
or row in $\Delta^1$, the other when there is not.
The aim is now to prove that in the first case no $8$ cycle can be
formed for the class of \QC~codes considered.
Thanks to the equivalence of the configurations over row column
permutation it is possible to suppose that the cycle column/row lies
in  $\Delta^1$, moreover only the case of cycle column need to be
considered since the case with $\Delta^1$ containing a cycle row is the transpose of this.\\
To have a cycle column in $\Delta^1$ it must be a weight $2$ circulant.
For lemma \ref{lemma:2C=g8} $\Delta^2$ and $\Delta^3$ cannot
be weight-2 circulants.
The  two cycle points lying in $\Delta^3$ cannot form a cycle column
because  $\Delta^3$ is a  weight-1 circulant, hence 
they must be part of two different cycle columns
but this is not possible because in $C^1$ there are no other points to
form cycle columns.
This shows how cycles exist in this configuration if and only if  none
of the four circulants contain a cycle column or a cycle row.

If there are no cycle columns or row in any circulant it can be assumed that there is a $8$-cycle if and only if 
cycle point $(x,y)$ lies in $\Delta^1$,
cycle point $(l,u)$ lies in $\Delta^1$,
cycle point $(x,z)$ lies in $\Delta^2$,
cycle point $(l,v)$ lies in $\Delta^2$,
cycle point $(w,u)$ lies in $\Delta^3$,
cycle point $(t,y)$ lies in $\Delta^3$,
cycle point $(w,z)$ lies in $\Delta^4$,
cycle point $(t,v)$ lies in $\Delta^4$,
since cycle point $(x,y)$ lies in $\Delta^1$, applying Lemma \ref{lem:id}
\begin{equation} \label{eq:dim_g8-11-1}
y \equiv x+\ep(p^1)\,.
 \end{equation}
 Since cycle point $(l,u)$ lies in $\Delta^1$, applying Lemma \ref{lem:id}
\begin{equation} \label{eq:dim_g8-11-2}
u \equiv l+\ep(p^1)\,.
 \end{equation}
Since cycle point $(x,z)$ lies in $\Delta^2$, applying Lemma \ref{lem:id}
\begin{equation} \label{eq:dim_g8-11-3}
z \equiv x+\ep(p^2)\,.
 \end{equation}
 Since cycle point $(l,v)$ lies in $\Delta^2$, applying Lemma \ref{lem:id}
\begin{equation} \label{eq:dim_g8-11-4}
v \equiv l+\ep(p^2)\,.
 \end{equation}
  Since cycle point $(w,u)$ lies in $\Delta^3$, applying Lemma \ref{lem:id}
\begin{equation} \label{eq:dim_g8-11-5}
u \equiv w+\ep(p^3)\,.
 \end{equation}
Since cycle point $(t,y)$ lies in $\Delta^3$, applying Lemma \ref{lem:id}
\begin{equation} \label{eq:dim_g8-11-6}
y \equiv t+\ep(p^3)\,.
 \end{equation}
 Since cycle point $(w,z)$ lies in $\Delta^4$, applying Lemma \ref{lem:id}
\begin{equation} \label{eq:dim_g8-11-7}
z \equiv w+\ep(p^4)\,.
 \end{equation}
 Since cycle point $(t,v)$ lies in $\Delta^4$, applying Lemma \ref{lem:id}
\begin{equation} \label{eq:dim_g8-11-8}
v \equiv t+\ep(p^4)\,.
 \end{equation}
From (\ref{eq:dim_g8-11-2}) and (\ref{eq:dim_g8-11-5}) eq.~\ref{eq:new1} is obtained, substituting (in order)
(\ref{eq:dim_g8-11-4}), (\ref{eq:dim_g8-11-8}),
 (\ref{eq:dim_g8-11-6}), (\ref{eq:dim_g8-11-1}),
 (\ref{eq:dim_g8-11-3}) and (\ref{eq:dim_g8-11-7}) the desired
condition is found.
\begin{equation}\label{eq:new1}
l-w+\ep(p^1)-\ep(p^3) = 0
\end{equation}

\begin{remark}x
In the case when all the circulants in the configuration are weight-1 circulants then this condition is redundant. In fact is such case the condition became:
$$
2\epsilon(p^1)-2\epsilon(p^2)-2\epsilon(p^3)+2\epsilon(p^4)\equiv 0\\
$$
that is equivalent to case \ref{item:g4-3} of Theorem~\ref{the:condition4} hence this condition assure the existence of a $4$-cycle and not of an $8$-cycle.

\end{remark}

\end{proof}
\end{lemma}
\begin{lemma}\label{lem:g8-13} 
 There is a $8$-cycle in case \ref{item:g8-13a} if and only if
 \begin{align*}
   \ep(p^2)-\ep(p^3)-\ep(p^4)+\ep(p^5) \equiv \pm 2 s(p^1) 
\end{align*}
\begin{proof}
It can be assumed that there is a $8$-cycle if and only if 
cycle columns $y$ and $z$  lie in $C^1$,
cycle point $(t,v)$ lies in $\Delta^2$,
cycle point $(w,u)$ lies in $\Delta^3$,
cycle point $(l,v)$ lies in $\Delta^4$, and
cycle point $(l,u)$ lies in $\Delta^5$.
Applying Proposition \ref{prop:cas}-4 to cycle column $y$ and cycle column $z$ yields
\begin{equation} \label{eq:dim_g8-13-1}
x-t \equiv \pm s^1 \,,
\end{equation}
\begin{equation} \label{eq:dim_g8-13-2}
x-w \equiv \mp s^1 \,.
\end{equation}
Since cycle point $(t,v)$ lies in $\Delta^2$, applying Lemma \ref{lem:id}
\begin{equation} \label{eq:dim_g8-13-3}
v \equiv t+\ep(p^2)\,.
 \end{equation}
 Since cycle point $(w,u)$ lies in $\Delta^3$, applying Lemma \ref{lem:id}
\begin{equation} \label{eq:dim_g8-13-4}
u \equiv w+\ep(p^3)\,.
 \end{equation}
Since cycle point $(l,v)$ lies in $\Delta^4$, applying Lemma \ref{lem:id}
\begin{equation} \label{eq:dim_g8-13-5}
v \equiv l+\ep(p^4)\,.
 \end{equation}
 Since cycle point $(l,u)$ lies in $\Delta^5$, applying Lemma \ref{lem:id}
\begin{equation} \label{eq:dim_g8-13-6}
u \equiv l+\ep(p^5)\,.
 \end{equation}
 
Substituting  (\ref{eq:dim_g8-13-2}) into (\ref{eq:dim_g8-13-1}) : 
\begin{equation} \label{eq:dim_g8-13-7}
w -t \equiv \pm 2s^1\,.
 \end{equation}
 Substituting (\ref{eq:dim_g8-13-3}),
 (\ref{eq:dim_g8-13-4}), (\ref{eq:dim_g8-13-5}) and
 (\ref{eq:dim_g8-13-6}) in  (\ref{eq:dim_g8-13-7})  the desired condition is obtained.
\end{proof}

\end{lemma}
\begin{lemma}\label{lem:g8-15} 
 There is a $8$-cycle in case \ref{item:g8-15a} if and only if
  \begin{align*}
    \ep(p^1)-\ep(p^2)-\ep(p^3)+\ep(p^5)\equiv \pm  s(p^1)\pm  s(p^4) 
\end{align*}
\begin{proof}

It can be assumed that there is a $8$-cycle if and only if 
cycle columns $y$  lies in $C^1$,
cycle columns $u$  lies in $C^4$,
cycle point $(t,v)$ lies in $C^1$,
cycle point $(x,z)$ lies in $J^2$,
cycle point $(l,v)$ lies in $J^3$, and 
cycle point $(w,z)$ lies in $J^5$.
Applying Proposition \ref{prop:cas}-2 to cycle column $y$ 
\begin{equation} \label{eq:dim_g8-15-1}
x-t \equiv \pm s^1 \,.
\end{equation}
Applying Proposition \ref{prop:cas}-2 to cycle column $z$ 
\begin{equation} \label{eq:dim_g8-15-2}
w-l \equiv \pm s^2 \,.
\end{equation}
Since cycle point $(t,v)$ lies in $C^1$, applying Lemma \ref{lem:id}
\begin{equation} \label{eq:dim_g8-15-3}
v \equiv t+\ep(p^1)\,.
 \end{equation}
 Since cycle point $(x,z)$ lies in $J^2$, applying Lemma \ref{lem:id}
\begin{equation} \label{eq:dim_g8-15-4}
z \equiv x+\ep(p^2)\,.
 \end{equation}
Since cycle point $(l,v)$ lies in $J^3$, applying Lemma \ref{lem:id}
\begin{equation} \label{eq:dim_g8-15-5}
v \equiv l+\ep(p^3)\,.
 \end{equation}
 Since cycle point $(w,z)$ lies in $J^5$, applying Lemma \ref{lem:id}
\begin{equation} \label{eq:dim_g8-14-6}
z \equiv w+\ep(p^5)\,.
 \end{equation}
 
Substituting  in the proper order  the desired result is obtained.
\end{proof}

\end{lemma}
\begin{lemma}\label{lem:g8-17} 
 There is a $8$-cycle in case \ref{item:g8-17a} if and only if
\begin{align*}
&\epsilon(p^1)+\epsilon(p^1)-\epsilon(p^2)-\epsilon(p^3)-\epsilon(p^4)-\epsilon(p^4)+\epsilon(p^5)+\epsilon(p^6)=0\,,
\mbox{ or }\\
&\epsilon(p^2)-\epsilon(p^3)+\ep(p^4)-\ep(p^4)-\epsilon(p^5)+\epsilon(p^6)=\pm s(p^1)\,,\mbox{ or }\\
&\epsilon(p^2)-\epsilon(p^3)+\ep(p^1)-\ep(p^1)-\epsilon(p^5)+\epsilon(p^6)=\pm s(p^3)\,.
\end{align*}

\begin{proof}
Two cases must be considered, the first when nor $\Delta^1$ neither
$\Delta^4$ contain a cycle column or a cycle row, the second when one
of them contains a cycle row. 
Note how for lemma \ref{lemma:2C=g8} $\Delta^1$ and $\Delta^4$ cannot
be weight-2 circulants at the same time hence they cannot both contain a cycle row or cycle column.
The situation with only one cycle columns in $\Delta^1$ or in
$\Delta^4$ can be dismissed, in fact  it can be assumed the cycle column to lie in $\Delta^1$. Then the two cycle points in $\Delta^4$ must be in two different columns but these columns cannot exist because there are not other point in $\Delta^1$ to complete them.

\begin{itemize}
\item If there are no cycle columns or row in any circulant it can be assumed that there is a $8$-cycle if and only if 
cycle point $(x,y)$ lies in $\Delta^1$,
cycle point $(l,u)$ lies in $\Delta^1$,
cycle point $(x,z)$ lies in $\Delta^3$,
cycle point $(l,v)$ lies in $\Delta^2$,
cycle point $(w,u)$ lies in $\Delta^4$,
cycle point $(t,y)$ lies in $\Delta^4$,
cycle point $(w,z)$ lies in $\Delta^6$, and
cycle point $(t,v)$ lies in $\Delta^5$.
Since cycle point $(x,y)$ lies in $\Delta^1$, applying Lemma \ref{lem:id}
\begin{equation} \label{eq:dim_g8-17-1}
y \equiv x+\ep(p^1)\,.
 \end{equation}
 Since cycle point $(l,u)$ lies in $\Delta^1$, applying Lemma \ref{lem:id}
\begin{equation} \label{eq:dim_g8-17-2}
u \equiv l+\ep(p^1)\,.
 \end{equation}
Since cycle point $(x,z)$ lies in $\Delta^3$, applying Lemma \ref{lem:id}
\begin{equation} \label{eq:dim_g8-17-3}
z \equiv x+\ep(p^3)\,.
 \end{equation}
 Since cycle point $(l,v)$ lies in $\Delta^2$, applying Lemma \ref{lem:id}
\begin{equation} \label{eq:dim_g8-17-4}
v \equiv l+\ep(p^2)\,.
 \end{equation}
  Since cycle point $(w,u)$ lies in $\Delta^4$, applying Lemma \ref{lem:id}
\begin{equation} \label{eq:dim_g8-17-5}
u \equiv w+\ep(p^4)\,.
 \end{equation}
Since cycle point $(t,y)$ lies in $\Delta^4$, applying Lemma \ref{lem:id}
\begin{equation} \label{eq:dim_g8-17-6}
y \equiv t+\ep(p^4)\,.
 \end{equation}
 Since cycle point $(w,z)$ lies in $\Delta^6$, applying Lemma \ref{lem:id}
\begin{equation} \label{eq:dim_g8-17-7}
z \equiv w+\ep(p^6)\,.
 \end{equation}
 Since cycle point $(t,v)$ lies in $\Delta^5$, applying Lemma \ref{lem:id}
\begin{equation} \label{eq:dim_g8-17-8}
v \equiv t+\ep(p^5)\,.
 \end{equation}
From (\ref{eq:dim_g8-17-2}) and (\ref{eq:dim_g8-17-5}) eq.~\ref{eq:new2} is obtained and substituting (in order)
(\ref{eq:dim_g8-17-4}), (\ref{eq:dim_g8-17-8}),
 (\ref{eq:dim_g8-17-6}), (\ref{eq:dim_g8-17-1}),
 (\ref{eq:dim_g8-17-3}) and (\ref{eq:dim_g8-17-7}) 
 the listed condition is obtained.
\begin{equation}\label{eq:new2}
l-w+\ep(p^1)-\ep(p^4) = 0
\end{equation}

\item If there is a cycle row in a circulant it can be assumed that it is in $C^1$.
Then there is not any other cycle row or cycle column in $C^4$, hence it can be assumed that there is a $8$-cycle if and only if 
cycle row $x$ lies in $\C^1$,
cycle point $(w,z)$ lies in $\Delta^4$,
cycle point $(t,y)$ lies in $\Delta^4$,
cycle point $(t,v)$ lies in $\Delta^6$,
cycle point $(w,u)$ lies in $\Delta^5$,
cycle point $(l,u)$ lies in $\Delta^2$, and
cycle point $(l,v)$ lies in $\Delta^3$.
Since cycle row $x$ lies in $C^1$, applying Proposition \ref{prop:cas}-3
\begin{equation} \label{eq:dim_g8-17-15}
y-z \equiv \pm s^1\,.
 \end{equation}
Since cycle point $(w,z)$ lies in $\Delta^4$, applying Lemma \ref{lem:id}
\begin{equation} \label{eq:dim_g8-17-16}
z \equiv w+\ep(p^4)\,.
 \end{equation}
Since cycle point $(t,y)$ lies in $\Delta^4$, applying Lemma \ref{lem:id}
\begin{equation} \label{eq:dim_g8-17-17}
y \equiv t+\ep(p^4)\,.
 \end{equation}
Since cycle point $(t,v)$ lies in $\Delta^6$, applying Lemma \ref{lem:id}
\begin{equation} \label{eq:dim_g8-17-18}
v \equiv t+\ep(p^6)\,.
 \end{equation}
Since cycle point $(w,u)$ lies in $\Delta^5$, applying Lemma \ref{lem:id}
\begin{equation} \label{eq:dim_g8-17-19}
u \equiv w+\ep(p^5)\,.
 \end{equation}
Since cycle point $(l,u)$ lies in $\Delta^2$, applying Lemma \ref{lem:id}
\begin{equation} \label{eq:dim_g8-17-20}
u \equiv l+\ep(p^2)\,.
 \end{equation}
 Since cycle point $(l,v)$ lies in $\Delta^3$, applying Lemma \ref{lem:id}
\begin{equation} \label{eq:dim_g8-17-21}
v \equiv l+\ep(p^3)\,.
 \end{equation}

Staring from (\ref{eq:dim_g8-17-15}) and substituting in order
(\ref{eq:dim_g8-17-16}), (\ref{eq:dim_g8-17-17}),
(\ref{eq:dim_g8-17-18}), (\ref{eq:dim_g8-17-19}),
(\ref{eq:dim_g8-17-20}) and (\ref{eq:dim_g8-17-21}) 
the  result is obtained:
$$
 \epsilon(p^2)-\epsilon(p^3)+\ep(p^4)-\ep(p^4)-\epsilon(p^5)+\epsilon(p^6)=\pm s(p^1)
$$

The last condition
$$
\epsilon(p^2)-\epsilon(p^3)+\ep(p^1)-\ep(p^1)-\epsilon(p^5)+\epsilon(p^6)=\pm s(p^3)\,
$$
Is equivalent to this case but with cycle row laying in $C^4$ instead of $C^1$.
\end{itemize}

\end{proof}

\end{lemma}
\begin{lemma}\label{lem:g8-18} 
 There is a $8$-cycle in case \ref{item:g8-18a} if and only if
  \begin{align*}
    \ep(p^2)-\ep(p^2)+ \ep(p^3)-\ep(p^3)\,  \equiv\pm  s(p^1)\pm s(p^4).
\end{align*}
\begin{proof}
It can be assumed that there is a $8$-cycle if and only if 
cycle column $y$  lies in $C^1$,
cycle column $u$  lies in $C^4$,
cycle point $(x,z)$ lies in $\Delta^2$,
cycle point $(t,v)$ lies in $\Delta^2$,
cycle point $(w,z)$ lies in $\Delta^3$, and
cycle point $(l,v)$ lies in $\Delta^3$.
Applying Proposition \ref{prop:cas}-2 to cycle column $y$ 
\begin{equation} \label{eq:dim_g8-18-1}
x-t \equiv \pm s^1 \,.
\end{equation}
Applying Proposition \ref{prop:cas}-2 to cycle column $u$ 
\begin{equation} \label{eq:dim_g8-18-2}
w-l \equiv \pm s^4 \,.
\end{equation}
Since cycle point $(x,z)$ lies in $\Delta^2$, applying Lemma \ref{lem:id}
\begin{equation} \label{eq:dim_g8-18-3}
z \equiv x+\ep(p^2)\,.
 \end{equation}
Since cycle point $(t,v)$ lies in $\Delta^2$, applying Lemma \ref{lem:id}
\begin{equation} \label{eq:dim_g8-18-4}
v \equiv t+\ep(p^2)\,.
 \end{equation}
 Since cycle point $(w,z)$ lies in $\Delta^3$, applying Lemma \ref{lem:id}
\begin{equation} \label{eq:dim_g8-18-5}
z \equiv w+\ep(p^3)\,.
 \end{equation}
 Since cycle point $(l,v)$ lies in $\Delta^3$, applying Lemma \ref{lem:id}
\begin{equation} \label{eq:dim_g8-18-6}
v \equiv l+\ep(p^3)\,.
 \end{equation}
 Substituting  (\ref{eq:dim_g8-18-3}) and (\ref{eq:dim_g8-18-4}) into
 (\ref{eq:dim_g8-18-1}) and then  (\ref{eq:dim_g8-18-2}),
 (\ref{eq:dim_g8-18-5}) and (\ref{eq:dim_g8-18-6}),  
the desired result is obtained.
\end{proof}

\end{lemma}
\begin{lemma}\label{lem:g8-20} 
 There is a $8$-cycle in case \ref{item:g8-20a} if and only if
  \begin{align*}
   \ep(p^3)-\ep(p^4)-\ep(p^5)+\ep(p^6) \equiv  \pm  s(p^1)\pm  s(p^2) \,.
\end{align*}
\begin{proof}
This can be prove with identical reasoning used to prove Lemma \ref{lem:g8-20}
\end{proof}
\end{lemma}
\begin{lemma}\label{lem:g8-21} 
 There is a $8$-cycle in case \ref{item:g8-21a} if and only if
 \begin{align*}
 &\epsilon(p^1)+\epsilon(p^2)-\epsilon(p^3)-\epsilon(p^4)-\epsilon(p^5)-\epsilon(p^6)+\epsilon(p^7)+\epsilon(p^8)=0
 \,, \mbox { or}\\
&\epsilon(p^1)-\epsilon(p^2)+\epsilon(p^3)-\epsilon(p^4)-\epsilon(p^5)+\epsilon(p^6)-\epsilon(p^7)+\epsilon(p^8)=0
\,, \mbox { or}\\
&\epsilon(p^1)-\epsilon(p^2)-\epsilon(p^3)+\epsilon(p^4)-\epsilon(p^5)+\epsilon(p^6)+\epsilon(p^7)-\epsilon(p^8)=0\,.
  \end{align*}
\begin{proof}
To prove this lemma it is sufficient to consider that there is one
cycle point for each circulant in the configuration, hence even if the
circulant may have weight $2$ they act as weight one.
Such situation is identical to the class of codes used by
Fossorier  in~\cite{Fossorier:04}. Theorem~\ref{th:fossorier}
presented in section~\ref{sec:QC_case} can be applied directly to this
class. The three listed conditions are the expansion of
the theorem apply to the only possible cycles in such configuration.\\
In particular fixed the first cycle column to be fixed between
$\Delta_1$ and $\Delta_5$ there are three possible column that the
second row can lie in. For each of this possibilities there are two
possible positions of the remaining two cycle columns.
A total of six possible cycles are obtained but they are 
two by two equivalent by row swap (see~\ref{fig:ap_hard}). 
Since there are only two rows a row swap changes the sign  to all the
terms hence only three conditions are required.

\begin{figure}[htbp]
\begin{center}
\psfrag{a}{$\equiv$} 
\includegraphics[width=0.9\columnwidth]{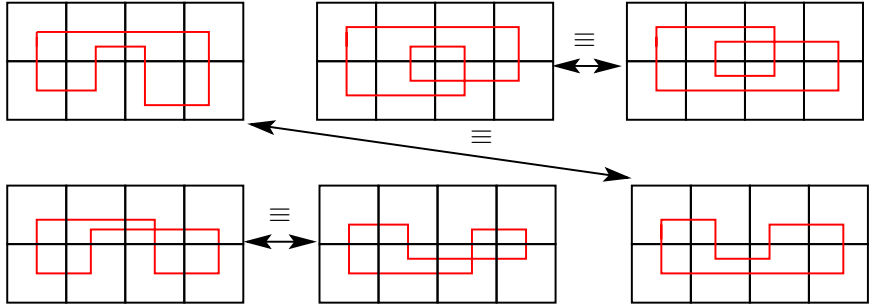}
\end{center}
\caption{All possible $8$-cycles for configuration $11$}
\label{fig:ap_hard}
\end{figure}

\end{proof}
\end{lemma}
\begin{lemma}\label{lem:g8-22} 
 There is a $8$-cycle in case \ref{item:g8-22a} if and only if
  \begin{align*}
   \ep(p^1)-\ep(p^2)-\ep(p^3)+\ep(p^4)+\ep(p^5)-\ep(p^6) \equiv  \pm  s(p^1)\,.
\end{align*}
\begin{proof}
It can be assumed that there is a $8$-cycle if and only if 
cycle column $y$  lies in $C^1$,
cycle point $(t,v)$ lies in $C^1$,
cycle point $(x,z)$ lies in $\Delta^2$,
cycle point $(l,v)$ lies in $\Delta^3$,
cycle point $(l,u)$ lies in $\Delta^4$,
cycle point $(w,z)$ lies in $\Delta^5$, and
cycle point $(w,u)$ lies in $\Delta^6$.
Applying Proposition \ref{prop:cas}-2 to cycle column $y$ 
\begin{equation} \label{eq:dim_g8-22-1}
x-t \equiv \pm s^1 \,.
\end{equation}
Since cycle point $(t,v)$ lies in $C^1$, applying Lemma \ref{lem:id}
\begin{equation} \label{eq:dim_g8-22-2}
v \equiv t+\ep(p^1)\,.
 \end{equation}
Since cycle point $(x,z)$ lies in $\Delta^2$, applying Lemma \ref{lem:id}
\begin{equation} \label{eq:dim_g8-22-3}
z \equiv x+\ep(p^2)\,.
 \end{equation}
Since cycle point $(l,v)$ lies in $\Delta^3$, applying Lemma \ref{lem:id}
\begin{equation} \label{eq:dim_g8-22-4}
v \equiv l+\ep(p^3)\,.
 \end{equation}
Since cycle point $(l,u)$ lies in $\Delta^4$, applying Lemma \ref{lem:id}
\begin{equation} \label{eq:dim_g8-22-5}lem:
u \equiv l+\ep(p^4)\,.
 \end{equation} 
 Since cycle point $(w,z)$ lies in $\Delta^5$, applying Lemma \ref{lem:id}
\begin{equation} \label{eq:dim_g8-22-6}
z \equiv w+\ep(p^5)\,.
 \end{equation}
  Since cycle point $(w,u)$ lies in $\Delta^6$, applying Lemma \ref{lem:id}
\begin{equation} \label{eq:dim_g8-22-7}
u \equiv w+\ep(p^6)\,.
 \end{equation}
 Starting from (\ref{eq:dim_g8-22-1}) and substituting the other equations it is easy to  find the desired result.
\end{proof}
\end{lemma}
\begin{lemma}\label{lem:g8-23} 
  There is a $8$-cycle in case \ref{item:g8-23a} if and only if
  \begin{align*}
   \ep(p^1)-\ep(p^1)+\ep(p^2)-\ep(p^3)-\ep(p^5)+\ep(p^6) \equiv  \pm  s(p^4)\,.
\end{align*}
\begin{proof}
It can be assumed that there is a $8$-cycle if and only if 
cycle row $x$  lies in $C^4$,
cycle point $(w,z)$ lies in $\Delta^1$,
cycle point $(t,y)$ lies in $\Delta^1$,
cycle point $(t,v)$ lies in $\Delta^2$,
cycle point $(w,u)$ lies in $\Delta^3$,
cycle point $(l,v)$ lies in $\Delta^5$, and
cycle point $(l,u)$ lies in $\Delta^6$.
Since cycle row $x$ lies in $C^4$, applying Proposition \ref{prop:cas}-3
\begin{equation} \label{eq:dim_g8-23-1}
z-y \equiv \pm s^4\,.
 \end{equation}
Since cycle point $(w,z)$ lies in $\Delta^1$, applying Lemma \ref{lem:id}
\begin{equation} \label{eq:dim_g8-23-2}
z \equiv w+\ep(p^1)\,.
 \end{equation}
Since cycle point $(t,y)$ lies in $\Delta^1$, applying Lemma \ref{lem:id}
\begin{equation} \label{eq:dim_g8-23-3}
y \equiv t+\ep(p^1)\,.
 \end{equation}
Since cycle point $(t,v)$ lies in $\Delta^2$, applying Lemma \ref{lem:id}
\begin{equation} \label{eq:dim_g8-23-4}
v \equiv t+\ep(p^2)\,.
 \end{equation}
 Since cycle point $(w,u)$ lies in $\Delta^3$, applying Lemma \ref{lem:id}
\begin{equation} \label{eq:dim_g8-23-5}
u \equiv w+\ep(p^3)\,.
 \end{equation}
 Since cycle point $(l,v)$ lies in $\Delta^5$, applying Lemma \ref{lem:id}
\begin{equation} \label{eq:dim_g8-23-6}
v \equiv l+\ep(p^5)\,.
 \end{equation}
 Since cycle point $(l,u)$ lies in $\Delta^6$, applying Lemma \ref{lem:id}
\begin{equation} \label{eq:dim_g8-23-7}
u \equiv l+\ep(p^6)\,.
 \end{equation}  
 Starting from (\ref{eq:dim_g8-23-1}) and substituting the other equations it is easy to  find the desired result.
\end{proof}

\end{lemma}
\begin{lemma}\label{lem:g8-24} 
There is a $8$-cycle in case \ref{item:g8-24a} if and only if
  \begin{align*}
   &\ep(p^1)+\ep(p^1)-\ep(p^2)-\ep(p^3)-\ep(p^4)+\ep(p^5)-\ep(p^6)+\ep(p^7) \equiv  0\,, \mbox{ or }\\
   &\ep(p^2)-\ep(p^3)+\ep(p^4)-\ep(p^5)-\ep(p^6)+\ep(p^7) \equiv  \pm s(p^1)
\end{align*}
\begin{proof}
Two possible cases must be considered. The first is when the two point laying in $\Delta^1$ form a cycle column (or row) the second when they do not.

\begin{itemize}
\item If there is a cycle column in $\Delta^1$ it can be assumed that there is a $8$-cycle if and only if 
cycle column $y$  lies in $\Delta^1$,
cycle point $(x,z)$ lies in $\Delta^3$,
cycle point $(t,v)$ lies in $\Delta^2$,
cycle point $(l,v)$ lies in $\Delta^5$,
cycle point $(l,u)$ lies in $\Delta^4$,
cycle point $(w,z)$ lies in $\Delta^7$, and
cycle point $(w,u)$ lies in $\Delta^6$.
Applying Proposition \ref{prop:cas}-2 to cycle column $y$ 
\begin{equation} \label{eq:dim_g8-24-1}
x-t \equiv \pm s^1 \,.
\end{equation}
Since cycle point $(x,z)$ lies in $\Delta^3$, applying Lemma \ref{lem:id}
\begin{equation} \label{eq:dim_g8-24-2}
z \equiv x+\ep(p^3)\,.
 \end{equation}
Since cycle point $(t,v)$ lies in $\Delta^2$, applying Lemma \ref{lem:id}
\begin{equation} \label{eq:dim_g8-24-3}
v \equiv t+\ep(p^2)\,.
 \end{equation}
Since cycle point $(l,v)$ lies in $\Delta^5$, applying Lemma \ref{lem:id}
\begin{equation} \label{eq:dim_g8-24-4}
v \equiv l+\ep(p^5)\,.
 \end{equation}
Since cycle point $(l,u)$ lies in $\Delta^4$, applying Lemma \ref{lem:id}
\begin{equation} \label{eq:dim_g8-24-5}
u \equiv l+\ep(p^4)\,.
 \end{equation}  
Since cycle point $(w,z)$ lies in $\Delta^7$, applying Lemma \ref{lem:id}
\begin{equation} \label{eq:dim_g8-24-6}
z \equiv w+\ep(p^7)\,.
 \end{equation}
 Since cycle point $(w,u)$ lies in $\Delta^6$, applying Lemma \ref{lem:id}
\begin{equation} \label{eq:dim_g8-24-7}
u \equiv w+\ep(p^6)\,.
 \end{equation}
 Starting from (\ref{eq:dim_g8-24-1}) and substituting the other
 equations it is easy to  find the condition in the statement.\\
If the two points in $\Delta^1$ form a cycle row and not a cycle
column 
it is sufficient to transpose the matrix (it is a symmetric 3x3
matrix) and the same condition is valid.

\item If there is no cycle column or row in $\Delta^1$ it can be assumed that there is a $8$-cycle if and only if 
cycle point $(x,y)$ lies in $\Delta^1$,
cycle point $(l,u)$ lies in $\Delta^1$,
cycle point $(x,z)$ lies in $\Delta^2$,
cycle point $(l,v)$ lies in $\Delta^3$,
cycle point $(w,u)$ lies in $\Delta^4$,
cycle point $(w,z)$ lies in $\Delta^5$,
cycle point $(t,y)$ lies in $\Delta^6$, and
cycle point $(t,v)$ lies in $\Delta^7$.
Since cycle point $(x,y)$ lies in $\Delta^1$, applying Lemma \ref{lem:id}
\begin{equation} \label{eq:dim_g8-24-8}
y \equiv x+\ep(p^1)\,.
 \end{equation}
Since cycle point $(l,u)$ lies in $\Delta^1$, applying Lemma \ref{lem:id}
\begin{equation} \label{eq:dim_g8-24-9}
u \equiv l+\ep(p^1)\,.
 \end{equation}
Since cycle point $(x,z)$ lies in $\Delta^2$, applying Lemma \ref{lem:id}
\begin{equation} \label{eq:dim_g8-24-10}
z \equiv x+\ep(p^2)\,.
 \end{equation}
Since cycle point $(l,v)$ lies in $\Delta^3$, applying Lemma \ref{lem:id}
\begin{equation} \label{eq:dim_g8-24-11}
v \equiv l+\ep(p^3)\,.
 \end{equation}
Since cycle point $(w,u)$ lies in $\Delta^4$, applying Lemma \ref{lem:id}
\begin{equation} \label{eq:dim_g8-24-12}
u \equiv w+\ep(p^4)\,.
 \end{equation}
Since cycle point $(w,z)$ lies in $\Delta^5$, applying Lemma \ref{lem:id}
\begin{equation} \label{eq:dim_g8-24-13}
z \equiv w+\ep(p^5)\,.
 \end{equation}
Since cycle point $(t,y)$ lies in $\Delta^6$, applying Lemma \ref{lem:id}
\begin{equation} \label{eq:dim_g8-24-14}
y \equiv t+\ep(p^6)\,.
 \end{equation}  
Since cycle point $(t,v)$ lies in $\Delta^7$, applying Lemma \ref{lem:id}
\begin{equation} \label{eq:dim_g8-24-15}
v \equiv t+\ep(p^7)\,.
 \end{equation}

 Starting from (\ref{eq:dim_g8-24-9}) and (\ref{eq:dim_g8-24-12}) eq.~\ref{eq:new4} is obtained and substituting the other
 equations into this the desired condition  is found.
\begin{equation}\label{eq:new4}
 l+\ep(p^1)\equiv w+\ep(p^4).
\end{equation}

\end{itemize}

\end{proof}
\end{lemma}
\begin{lemma}\label{lem:g8-25} 
There is a $8$-cycle in case \ref{item:g8-25a} if and only if
  \begin{align*}
   \ep(p^2)-\ep(p^3)-\ep(p^4)+\ep(p^5)\equiv  \pm  s(p^1) \pm s(p^6)\,.
\end{align*}
\begin{proof}
It can be assumed that there is a $8$-cycle if and only if 
cycle row $x$  lies in $C^1$,
cycle column $v$  lies in $C^6$,
cycle point $(w,z)$ lies in $\Delta^2$,
cycle point $(w,u)$ lies in $\Delta^3$,
cycle point $(t,y)$ lies in $\Delta^4$, and
cycle point $(l,u)$ lies in $\Delta^5$.
Since cycle row $x$ lies in $C^1$, applying Proposition \ref{prop:cas}-3
\begin{equation} \label{eq:dim_g8-25-1}
z-y \equiv \pm s^1\,.
 \end{equation}
Applying Proposition \ref{prop:cas}-2 to cycle column $v$ 
\begin{equation} \label{eq:dim_g8-25-2}
t-l \equiv \pm s^6 \,.
\end{equation}
Since cycle point $(w,z)$ lies in $\Delta^2$, applying Lemma \ref{lem:id}
\begin{equation} \label{eq:dim_g8-25-3}
z \equiv w+\ep(p^2)\,.
 \end{equation}
 Since cycle point $(w,u)$ lies in $\Delta^3$, applying Lemma \ref{lem:id}
\begin{equation} \label{eq:dim_g8-25-4}
u \equiv w+\ep(p^3)\,.
 \end{equation}
 Since cycle point $(t,y)$ lies in $\Delta^4$, applying Lemma \ref{lem:id}
\begin{equation} \label{eq:dim_g8-25-5}
y \equiv t+\ep(p^4)\,.
 \end{equation}
 Since cycle point $(l,u)$ lies in $\Delta^5$, applying Lemma \ref{lem:id}
\begin{equation} \label{eq:dim_g8-25-6}
u \equiv l+\ep(p^5)\,.
 \end{equation} 
 Starting from (\ref{eq:dim_g8-25-1}) and substituting the other equations it is easy to  find the desired result.
\end{proof}
\end{lemma}
\begin{lemma}\label{lem:g8-26-a} 
 There is a $8$-cycle in case \ref{item:g8-26-aa} if and only if
  \begin{align*}
   \epsilon(p^2)-\epsilon(p^3)-\epsilon(p^4)+\epsilon(p^5)+\epsilon(p^6)-\epsilon(p^7)=\pm s(p^1),
\end{align*}
\begin{proof}
It can be assumed that there is a $8$-cycle if and only if 
cycle point $(x,y)$ lies in $\Delta^1$,
cycle point $(t,y)$ lies in $\Delta^5$,
cycle point $(x,z)$ lies in $\Delta^4$,
cycle point $(t,v)$ lies in $\Delta^6$,
cycle point $(w,z)$ lies in $\Delta^8$,
cycle point $(w,u)$ lies in $\Delta^7$,
cycle point $(l,v)$ lies in $\Delta^2$, and
cycle point $(l,u)$ lies in $\Delta^3$.

Since cycle point $(x,y)$ lies in $\Delta^1$, applying Lemma \ref{lem:id}
\begin{equation} \label{eq:dim_g8-26-1-a}
y \equiv x+\ep(p^1)\,.
 \end{equation}
Since cycle point $(t,y)$ lies in $\Delta^5$, applying Lemma \ref{lem:id}
\begin{equation} \label{eq:dim_g8-26-2-a}
y \equiv t+\ep(p^5)\,.
 \end{equation}
Since cycle point $(x,z)$ lies in $\Delta^4$, applying Lemma \ref{lem:id}
\begin{equation} \label{eq:dim_g8-26-3-a}
z \equiv x+\ep(p^4)\,.
 \end{equation}
Since cycle point $(t,v)$ lies in $\Delta^6$, applying Lemma \ref{lem:id}
\begin{equation} \label{eq:dim_g8-26-4-a}
v \equiv t+\ep(p^6)\,.
 \end{equation}
 Since cycle point $(w,z)$ lies in $\Delta^8$, applying Lemma \ref{lem:id}
\begin{equation} \label{eq:dim_g8-26-5-a}
z \equiv w+\ep(p^8)\,.
 \end{equation}
  Since cycle point $(w,u)$ lies in $\Delta^7$, applying Lemma \ref{lem:id}
\begin{equation} \label{eq:dim_g8-26-6-a}
u \equiv w+\ep(p^7)\,.
 \end{equation}
 Since cycle point $(l,v)$ lies in $\Delta^2$, applying Lemma \ref{lem:id}
\begin{equation} \label{eq:dim_g8-26-7-a}
v \equiv l+\ep(p^2)\,.
 \end{equation}
Since cycle point $(l,u)$ lies in $\Delta^3$, applying Lemma \ref{lem:id}
\begin{equation} \label{eq:dim_g8-26-8-a}
u \equiv l+\ep(p^3)\,.
 \end{equation} 
 Starting from (\ref{eq:dim_g8-26-1-a}) and substituting the other equations it is easy to  find the desired result.
\end{proof}

\end{lemma}
\begin{lemma}\label{lem:g8-26} 
 There is a $8$-cycle in case \ref{item:g8-26a} if and only if
  \begin{align*}
   \ep(p^2)-\ep(p^3)-\ep(p^4)+\ep(p^5)+\ep(p^6)-\ep(p^7) \equiv  \pm  s(p^1)\,.
\end{align*}
\begin{proof}
It can be assumed that there is a $8$-cycle if and only if 
cycle column $y$  lies in $C^1$,
cycle point $(x,z)$ lies in $\Delta^2$,
cycle point $(t,v)$ lies in $\Delta^3$,
cycle point $(w,z)$ lies in $\Delta^4$,
cycle point $(w,u)$ lies in $\Delta^5$,
cycle point $(l,v)$ lies in $\Delta^6$, and 
cycle point $(l,u)$ lies in $\Delta^7$.
Applying Proposition \ref{prop:cas}-2 to cycle column $y$ 
\begin{equation} \label{eq:dim_g8-26-1}
x-t \equiv \pm s^1 \,.
\end{equation}
Since cycle point $(x,z)$ lies in $\Delta^2$, applying Lemma \ref{lem:id}
\begin{equation} \label{eq:dim_g8-26-2}
z \equiv x+\ep(p^2)\,.
 \end{equation}
Since cycle point $(t,v)$ lies in $\Delta^3$, applying Lemma \ref{lem:id}
\begin{equation} \label{eq:dim_g8-26-3}
v \equiv t+\ep(p^3)\,.
 \end{equation}
 Since cycle point $(w,z)$ lies in $\Delta^4$, applying Lemma \ref{lem:id}
\begin{equation} \label{eq:dim_g8-26-4}
z \equiv w+\ep(p^4)\,.
 \end{equation}
  Since cycle point $(w,u)$ lies in $\Delta^5$, applying Lemma \ref{lem:id}
\begin{equation} \label{eq:dim_g8-26-5}
u \equiv w+\ep(p^5)\,.
 \end{equation}
 Since cycle point $(l,v)$ lies in $\Delta^6$, applying Lemma \ref{lem:id}
\begin{equation} \label{eq:dim_g8-26-6}
v \equiv l+\ep(p^6)\,.
 \end{equation}
Since cycle point $(l,u)$ lies in $\Delta^7$, applying Lemma \ref{lem:id}
\begin{equation} \label{eq:dim_g8-26-7}
u \equiv l+\ep(p^7)\,.
 \end{equation} 
 Starting from (\ref{eq:dim_g8-26-1}) and substituting the other equations it is easy to  find the desired result.
\end{proof}

\end{lemma}
\begin{lemma}\label{lem:g8-27} 
There is a $8$-cycle in case \ref{item:g8-27a} if and only if
  \begin{align*}
   \ep(p^1)-\ep(p^2)-\ep(p^3)+\ep(p^4)+\ep(p^5)-\ep(p^6)-\ep(p^7)+\ep(p^8) \equiv 0\,.
\end{align*}
\begin{proof}
It can be assumed that there is a $8$-cycle if and only if 
cycle point $(x,y)$ lies in $\Delta^1$,
cycle point $(x,z)$ lies in $\Delta^2$,
cycle point $(t,y)$ lies in $\Delta^3$,
cycle point $(t,v)$ lies in $\Delta^4$,
cycle point $(w,z)$ lies in $\Delta^5$,
cycle point $(w,u)$ lies in $\Delta^6$,
cycle point $(l,v)$ lies in $\Delta^7$, and
cycle point $(l,u)$ lies in $\Delta^8$.
Since cycle point $(x,y)$ lies in $\Delta^1$, applying Lemma \ref{lem:id}
\begin{equation} \label{eq:dim_g8-27-1}
y \equiv x+\ep(p^1)\,.
 \end{equation}
Since cycle point $(x,z)$ lies in $\Delta^2$, applying Lemma \ref{lem:id}
\begin{equation} \label{eq:dim_g8-27-2}
z \equiv x+\ep(p^2)\,.
 \end{equation}
Since cycle point $(t,y)$ lies in $\Delta^3$, applying Lemma \ref{lem:id}
\begin{equation} \label{eq:dim_g8-27-3}
y \equiv t+\ep(p^3)\,.
 \end{equation}
Since cycle point $(t,v)$ lies in $\Delta^4$, applying Lemma \ref{lem:id}
\begin{equation} \label{eq:dim_g8-27-4}
v \equiv t+\ep(p^4)\,.
 \end{equation}
 Since cycle point $(w,z)$ lies in $\Delta^5$, applying Lemma \ref{lem:id}
\begin{equation} \label{eq:dim_g8-27-5}
z \equiv w+\ep(p^5)\,.
 \end{equation}
  Since cycle point $(w,u)$ lies in $\Delta^6$, applying Lemma \ref{lem:id}
\begin{equation} \label{eq:dim_g8-27-6}
u \equiv w+\ep(p^6)\,.
 \end{equation}
 Since cycle point $(l,v)$ lies in $\Delta^7$, applying Lemma \ref{lem:id}
\begin{equation} \label{eq:dim_g8-27-7}
v \equiv l+\ep(p^7)\,.
 \end{equation}
Since cycle point $(l,u)$ lies in $\Delta^8$, applying Lemma \ref{lem:id}
\begin{equation} \label{eq:dim_g8-27-8}
u \equiv l+\ep(p^8)\,.
 \end{equation} 
 From (\ref{eq:dim_g8-27-1}) and (\ref{eq:dim_g8-27-3}) eq.~\ref{eq:new5}
 is obtained and substituting the other
 equations into this  the desired result is found.
\begin{equation}\label{eq:new5}
x+\ep(p^1) \equiv t+\ep(p^3).
\end{equation}
\end{proof}

\end{lemma}

It has hence been proved that the conditions listed on the statement
considers all the possible $8$-cycles that can exist 
on the studied \QC~matrices.

\end{proof}
\end{theorem}

\end{document}